\documentclass[reqno]{amsart}
\usepackage[dvipsnames]{xcolor}
\usepackage{amsmath}
\usepackage{graphicx, color}
\usepackage{chngcntr}
\usepackage{apptools}
\usepackage{tikz}
\usetikzlibrary{backgrounds}
\usepackage{amsmath,amsfonts, amssymb}
\usepackage[normalem]{ulem}
\usepackage{wrapfig}
\usepackage{chngcntr}
\usepackage{paralist}
\usepackage{graphics} 
  \usepackage{epsfig} 
 \usepackage[colorlinks=true]{hyperref}
 \usepackage{textcomp}
\usepackage{subcaption}
\usepackage{epstopdf}
\usepackage[UKenglish]{babel}
\usepackage{syntonly}
\usepackage{fancyhdr}
\usepackage{esint}
\usepackage[title]{appendix}
\usepackage{algorithm}
\usepackage[noend]{algpseudocode}
\usepackage{tabularx}

 \usepackage{tikz-3dplot}
 \usepackage{tikz}
 \usepackage{pgfplots}
 \usepackage{bm}
 \usepgfplotslibrary{fillbetween}
 \usetikzlibrary{backgrounds}
 \usetikzlibrary{patterns}
 \usetikzlibrary{scopes}
 \usetikzlibrary{shapes,arrows}
 \usetikzlibrary{plotmarks,intersections}
 \usetikzlibrary{calc}
 \usetikzlibrary{positioning}
 
 \usepackage{float}

 \usepackage{multirow}
 \usepackage{array}
 
\usetikzlibrary{calc}
\usepackage{pgfplots}
 
 \usepackage{bbm}
\hypersetup{urlcolor=blue, citecolor=red}

\algrenewcommand\algorithmicrequire{\textbf{Precondition:}}

\makeatletter
\newcommand{\multiline}[1]{%
	\begin{tabularx}{\dimexpr\linewidth-\ALG@thistlm}[t]{@{}X@{}}
		#1
	\end{tabularx}
}
\makeatother

\usepackage{array}
\newcolumntype{N}{@{}m{0pt}@{}}
\makeatletter
\newcommand*{\centerfloat}{%
  \parindent \z@
  \leftskip \z@ \@plus 1fil \@minus \textwidth
  \rightskip\leftskip
  \parfillskip \z@skip}
\makeatother

\def\mathrlap{\mathpalette\mathrlapinternal}
\def\mathrlapinternal#1#2{%
	\rlap{$\mathsurround=0pt#1{#2}$}}

\numberwithin{equation}{section}

\title[]{Entropy Decay Rates for Conservative Spectral Schemes Modeling Fokker-Planck-Landau Type Flows in the Mean Field Limit}
\author[Clark A. Pennie  and Irene M. Gamba]{}

\begin{document}
\maketitle

\centerline{\scshape  Clark A. Pennie  and Irene M. Gamba}
\medskip
{\footnotesize
	\centerline{Department of Mathematics and Oden Institute }
	\centerline{University of Texas at Austin}
	\centerline{2515 Speedway Stop C1200
		Austin, Texas, 78712-1202, USA}

}

\date{}
\setcounter{tocdepth}{2}
\setcounter{secnumdepth}{2}

	
	
	
	
	
	
	\begin{abstract}
	
	The focus of this work is to create benchmark simulations of decay rates to statistical equilibrium in transport plasma models for Coulomb particle interactions given by a coupled Vlasov-Poisson Fokker-Planck-Landau equation, as well as with Maxwell type and hard sphere interactions.  The qualitative decay to the equilibrium Maxwell-Boltzmann distribution through relative entropy is studied in detail for all three types of particle interactions by means of a conservative hybrid spectral and discontinuous Galerkin scheme adapted from previous work.  More precisely, the Coulomb case shows that there is a degenerate spectrum, with a decay rate close to the law of two thirds predicted by upper estimates in a work of Strain and Guo in 2006, while the Maxwell type and hard sphere examples both exhibit a spectral gap as predicted by Desvillettes and Villani in 2000.  Such decay rate  behavior indicates that the analytical estimates for the Coulomb case is sharp while, still  to this  date, there is no analytical proof of sharp degenerate spectral behaviour for the Fokker-Planck-Landau operator.
	
	Simulations are presented, both for the space-homogeneous case of just particle potential interactions and the space-inhomogeneous case for the mean field coupling through the Poisson equation for total charges in periodic domains.  New explicit derivations of spectral collisional weights are presented in the case of Maxwell type and hard sphere interactions and the stability of all three scenarios, including Coulomb interactions, is investigated.
	
\medskip
	
{\textbf {Kewords:}}
		Fokker-Planck-Landau Type Equations, Vlasov-Poisson Equation, Boltzmann Equation, Mean Field Limit - Numerical Schemes, Conservative Spectral Methods, Equilibrium Decay Rate, Entropy Decay.
	\end{abstract}



\section{Introduction}

An important model for plasmas is the Landau equation, which results from the grazing collision limit of the Boltzmann equation.  This limit, first derived by Landau \cite{LandauEq}, assumes that colliding particles are travelling almost parallel to each other due to repulsive Coulomb forces.  

A more mathematical description of the limit was detailed by Degond and Lucquin-Desreux \cite{Degond&LD}, Desvillettes \cite{Desvillettes92, Desvillettes15}, Villani \cite{VillaniLandau} and Desvillettes and Villani \cite{DesvillettesVillani}, even for extended potential rates higher than Coulomb interactions and up to hard spheres.  When rates different to Coulomb interactions are used, the equation is referred to as being of Fokker-Planck-Landau type.  Computationally, the limiting problem has been studied by Bobylev and Potapenko \cite{BobylevPotapenko}, using Monte Carlo methods, and in Fourier space by Haack and Gamba \cite{GambaHaack1, GambaHaack2}.

The Landau equation is rather difficult to model, either analytically or numerically, due to the high dimensionality, non-linearity and non-locality.  For numerical simulations, a deterministic scheme can be used, such as the conservative spectral method, developed by Zhang and Gamba \cite{Chenglong}, which is the model of choice for the current work.  The method described in \cite{Chenglong} is in fact a solver for the space-inhomogeneous Landau equation, coupled to Poisson's equation, where the advection is modeled by a discontinuous Galerkin scheme.  Some results for the space-homogeneous version of the equation have already been described by the present authors \cite{RGDPaper}.  

The current work improves upon the results of \cite{RGDPaper} and extends them to the space-inhomogeneous case.  This produces benchmark computations of accurate dynamics for long time approximations to the Maxwell-Boltzmann equilibrium distribution determined by moments of the initial state.  As in \cite{RGDPaper}, the calculations are also included for Fokker-Planck-Landau type equations associated to Maxwell type interactions and hard spheres, expanding upon the previous work of \cite{Chenglong}.

The version of the spectral method in this work exploits the weak form of the Landau equation in order to calculate the Fourier transform of the collision operator.  It does so in just $\mathcal{O}(N^3 \log{N})$ operations, where the number of Fourier modes $N$ in each velocity dimension can be small, thanks to the conservation enforcement with just a further $\mathcal{O}(N^3)$ operations.  For computational purposes, a cut-off domain in velocity space is used, within which the majority of the solution's mass should be supported, based on a result by Gamba et al.\ \cite{BoltzmannSupport} for the Boltzmann equation.  This general construction of a spectral method was first applied to the Boltzmann equation by Gamba and Tharkabhushaman \cite{Harsha} and further details for the derivation of the Landau equation scheme can be found in  Zhang and Gamba \cite{Chenglong}.

Spectral methods  as an approximating  model for the space-homogeneous Landau equation were first considered by Pareschi et al.\ \cite{Pareschi_Spectral}, and later by Filbet and Pareschi\cite{Filbet_InhomSpectral} and Crouseilles and Filbet \cite{Crouseilles_Spectral}, but did not preserve the conservation properties of the Landau equation.  The lack of enforcing a conservation correction by minimization distance enforcing the collision invariants associate to the Landau operator  limited the ability of these schemes to compute accurate dynamics for long time approximations to the Maxwell-Boltzmann equilibrium , either in the scalar setting  or  in a system of multicomponent plasmas.    In fact,  the work of \cite{GambaHaack2} have shown the numerical conservation in the space homogeneous form  of Landau equation as a limit of the grazing collisions, both simulated by a conservative spectral scheme, for fairly   . More recently the work   Zhang and Gamba  \cite [Section 7.1.2]{Chenglong}, inspired in the implementation of the conservative spectral methods for a system of Boltzmann equations in  Munafo, Haack, Gamba, and Magin, \cite{MHGM-jcp2014},    has shown that  a system of Landau equations whose temperature evolution  solution matches  the explicit example of a system for  electro-neutral hydrogen plasma \cite{Bobylev_Karpov_Potapenko-2012}. Such numerical verification of  matching the approximate and analytical solutions would not be possible if  the spectral scheme does not conserve the system collision invariants  corresponding to the total energy.

A particular attraction to our current method is its ability to yield the correct decay of entropy up to 400 time units, both in the space homogeneous case and in the one physical space and three velocity  dimensional space inhomogeneous Landau flow model.  The conservation enforcement is essential in the proof of convergence of the spectral method by rigorous analysis of semi-discrete error estimates  for the conservative spectral scheme applied to the Boltzmann equation \cite{BoltzmannConvergence} and the same is true for Fokker-Planck-Landau type equations \cite{Pennie_gamba_LandauConv}.  The entropy decay rate is also a consequence of this fact.  A recent manuscript by Carrillo, Hu, Wang and Wu \cite{Carrillo_etal_2020} proposes a particle method for the evolution of the homogeneous Landau equation. However the conservation of energy fails for  500 time steps for five units of time, while our  scheme preserves the conservation of energy passed 40.000 times steps for 400 times units of times, and arrives to the a neighborhood of equilibrium in 20 units of time whose with the time rate of $e^{k t^{2/3}}$ as analytically predicted in \cite{Strain&Guo}.

To the best of the authors' knowledge, this is the first time that the relative entropy convergence rate of two thirds, proven analytically by Strain and Guo \cite{Strain&Guo}, has been seen through a numerical approximation of the relative entropy. It further shows that the upper bound analytically calculated in \cite{Strain&Guo} is very sharp. This sharp numerical output  can play a role in the validation and verification of a numerical scheme for the Landau equation.

The computational scheme has been parallelized and the computational efficiency is fully discussed in Section~6. 

Finally, it should also be mentioned that work has been undertaken by the current authors to produce $L^2$ error estimates for the approximations produced by this numerical scheme by the authors in  \cite{Pennie_gamba_LandauConv}.  In particular, a proof has been constructed to show that the conservative spectral method for Fokker-Planck-Landau type equations associated to hard potentials has a unique solution with moments, $L^2$-norm and even $L^2$-norm of its derivatives remaining bounded for all time, under certain conditions.  This then allows the estimate to be produced which shows that the approximation does indeed converge to the true solution in $L^2$-norm and that, as time increases, also converges to the correct equilibrium Maxwellian distribution associated to the initial data.  This is the first time that asemi-discrete error estimate has been produced for any numerical method which approximates Fokker-Planck-Landau type equations associated to any range of potentials and complements the numerical evidence produced here.


The layout of this work is as follows.  First, the set up of the problem is described in Section \ref{Description}, along with any required definitions.  The expressions for the Fourier transform of the Fokker-Planck-Landau type operators corresponding to Coulomb, Maxwell type and hard sphere interactions are derived in Section \ref{Fourier} and the stability results given in Section \ref{Stability}.  Finally, Section \ref{Results} contains the numerical results.  In that section, the correct decay rate to equilibrium is demonstrated for the space-homogeneous problem associated to Coulomb, Maxwell type and hard sphere interactions, as well as the space-inhomogeneous Landau equation.  All work here is part of a PhD thesis by the first author, under advisorship of the second.

\section{Description of Problem} \label{Description}
\subsection{Fokker-Planck-Landau Type Equations}
A space-inhomogeneous Fokker-Planck-Landau type equation for the probability density function (pdf) $f(t, x, \boldsymbol{v})$, where $(t, x, \boldsymbol{v}) \in (\mathbb{R}^+, \Omega_{x}, \mathbb{R}^3)$, with $\Omega_{x} \subseteq \mathbb{R}$, is of the form
\begin{equation}
f_t(t, x, \boldsymbol{v}) + \boldsymbol{v}\cdot\nabla_{x}f(t, x, \boldsymbol{v}) - \boldsymbol{E}(t, x)\cdot\nabla_{\boldsymbol{v}}f(t, x, \boldsymbol{v}) = \frac{1}{\varepsilon}Q(f,f)(t, x, \boldsymbol{v}), \label{Landau}
\end{equation}
where $\varepsilon$ is the Knudsen number and $Q(f,f)$ is the collision operator given by
\begin{flalign*}
&& Q(f,f) = \nabla_{\boldsymbol{v}}\cdot \int_{\mathbb{R}^3} & S(\boldsymbol{v} - \boldsymbol{v}_*) (f_*\nabla_{\boldsymbol{v}}f - f \nabla_{\boldsymbol{v}_*}f_*)~\textrm{d} \boldsymbol{v}_*, && \\
\textrm{for} && S(\boldsymbol{u}) &= |\boldsymbol{u}|^{\lambda + 2}\left(\textrm{I} - \frac{\boldsymbol{uu}^T}{|\boldsymbol{u}|^2}\right), &&
\end{flalign*}
with $-3 \leq \lambda \leq 1$, $I \in \mathbb{R}^{3 \times 3}$ the identity matrix and the subscript notation $f_*$ meaning evaluation at $\boldsymbol{v_*}$ (the velocity of a colliding particle).  In general, $\lambda > 0$  corresponds to hard potentials and $\lambda < 0$ to soft potentials.  More precisely, $\lambda = 1$ models hard sphere interactions; $\lambda = 0$ is known as a Maxwell type interaction; and $\lambda = -3$ models Coulomb interactions between particles.

In addition, $\boldsymbol{E}$ is the electric field found by solving Poisson's equation, namely
\begin{equation*}
\boldsymbol{E}(t, x) = - \nabla_{x}\Phi(t, x), 
\end{equation*}
where $\Phi$ is the potential solved from
\begin{equation}
-\Delta_{x}\Phi(t, x) = 1 - \int_{\mathbb{R}^3} f(t,x,\boldsymbol{v}) ~\textrm{d}\boldsymbol{v}. \label{Poisson}
\end{equation}
Note that the right-hand side of (\ref{Poisson}) is the density of positively charged ions (assumed here to be a constant background density) minus the density of electrons (due to the negative charge).  Also, in this context where $\boldsymbol{v}$ is a vector but $x$ is a scalar, $\boldsymbol{E}(t, x) = (E(t, x), 0, 0)$ and the gradient in $x$ is treated as $\nabla_x = \left(\frac{\partial}{\partial x}, 0, 0 \right)$.

In the current work, boundary conditions for both the Fokker-Planck-Landau type and Poisson equations are taken as periodic in space.  Furthermore, since the Poisson equation is an ordinary differential equation for any given $t \geq 0$ with periodic boundary conditions, if $\Omega_x = [0, L_x]$ then it has explicit solution given by
\begin{flalign*}
&& \Phi(t, x) &= \int_{0}^{x} \int_{0}^{s} \int_{\mathbb{R}^3} f(t, z, \boldsymbol{v}) ~\textrm{d}\boldsymbol{v}\textrm{d}z\textrm{d}s - \frac{1}{2} x^2 - C_E x + \Phi(t, 0), && && \\
\textrm{where } && & C_E = -\frac{1}{2}L_x + \frac{1}{L_x}\int_{0}^{L_x}\int_{0}^{s} \int_{\mathbb{R}^3} f(t, z, \boldsymbol{v}) ~\textrm{d}\boldsymbol{v}\textrm{d}z\textrm{d}s. && &&
\end{flalign*}
The potential $\Phi$ is never explicitly used, however, and it is in fact the derivative that is more relevant for the Landau equation.  For this reason, the value of $\Phi(t, 0)$ is irrelevant and is chosen as $\Phi(t, 0) = 0$ for convenience.

It should also be noted here that the space-homogeneous version of the Fokker-Planck-Landau type equation (\ref{Landau}) is simply to find the pdf $f(t, \boldsymbol{v})$, where $(t, \boldsymbol{v}) \in (\mathbb{R}^+, \mathbb{R}^3)$, such that
\begin{equation}
f_t(t, \boldsymbol{v}) = \frac{1}{\varepsilon}Q(f,f)(t, \boldsymbol{v}). \label{Landau_homo}
\end{equation}

\subsection{Properties of Fokker-Planck-Landau Type Equations}
Since Fokker-Planck-Landau type equations are a limit of the Boltzmann equation, they enjoy the same conservation laws.  In particular, for the set of collision invariants $\left\{\phi_k(\boldsymbol{v})\right\}_{k=0}^4 = \left\{1, v_1, v_2, v_3, |\boldsymbol{v}|^2\right\}$,
\begin{equation}
\int_{\mathbb{R}^3} Q(f, f)(\boldsymbol{v}) \phi_k (\boldsymbol{v}) ~\textrm{d}\boldsymbol{v} = 0, ~~~~~\textrm{for } k = 0, 1, \ldots, 4. \label{conservation}
\end{equation}
This is important because it leads to the conservation of mass $\rho$, average velocity $\boldsymbol{V}$ and total energy $T^{tot}$, where each of these quantities are found via
\begin{equation*}
\rho = \int_{\Omega_x} \int_{\mathbb{R}^3} f(t, x, \boldsymbol{v}) ~\textrm{d}\boldsymbol{v} \textrm{d}x, ~~~\boldsymbol{V} = \frac{1}{\rho} \int_{\Omega_x}\int_{\mathbb{R}^3} f(t, x, \boldsymbol{v}) \boldsymbol{v} ~\textrm{d}\boldsymbol{v} \textrm{d}x
\end{equation*}
\begin{flalign}
\textrm{and} &&T^{tot}(t) = \frac{3}{2} \rho T^K(t) + T^E(t), && \label{T_tot}
\end{flalign}
\begin{flalign}
\textrm{where} && T^K = \frac{1}{3\rho} \int_{\Omega_x} \int_{\mathbb{R}^3} f(t, x, \boldsymbol{v}) |\boldsymbol{v}|^2 ~\textrm{d}\boldsymbol{v} \textrm{d}x \textrm{~~and~~} T^E = \frac{1}{2} \int_{\Omega_x} |\Phi'(t, x)|^2 ~\textrm{d}x && \label{T_E}
\end{flalign}
are the kinetic energy $T^K$ and the electric energy $T^E$.

These moments will always be conserved for the single-species space- \hspace{-5pt} inhomogeneous Landau equation (\ref{Landau}) when solved with appropriate boundary conditions (including the periodic ones considered here).  If the initial mass, average velocity and total energy are denoted by $\rho_0$, $\boldsymbol{V}_0$ and $T^{tot}_0$, respectively, the equilibrium solution of the Landau equation is a Gaussian distribution with the same moments.  This is referred to as the equilibrium Maxwellian, denoted $\mathcal{M}_{eq}$, and is the specific Maxwellian distribution with moments equal to those of the initial condition, given by
\begin{equation}
\mathcal{M}_{eq}(x, \boldsymbol{v}) = \frac{\rho_0}{(2\pi T_{eq})^{\frac{3}{2}} \int_{0}^{L_x} e^{\frac{\Phi_{eq}(x)}{T_{eq}}} ~\textrm{d}x} e^{\frac{\Phi_{eq}(x)}{T_{eq}}} e^{-\frac{|\boldsymbol{v} - \boldsymbol{V}_0|^2}{2 T_{eq}}}, \label{M_eq_inhom}
\end{equation}
where $\Phi_{eq}$ is the equilibrium potential and $T_{eq}$ is such that using $\mathcal{M}_{eq}$ in expression (\ref{T_tot}) returns $T^{tot} = T^{tot}_0$.  

In the space-homogeneous setting there is no integration with respect to $x$ to evaluate the moments $\rho$, $\boldsymbol{V}$ and $T^K$; $T^{tot} = T^K$; there is no field $\Phi$; and the equilibrium Maxwellian reduces to
\begin{equation}
\mathcal{M}_{eq}(\boldsymbol{v}) = \frac{\rho_0}{(2 \pi T_0)^{\frac{3}{2}}} e^{-\frac{|\boldsymbol{v} - \boldsymbol{V}_0|^2}{2T_0}}, \label{M_eq_hom} 
\end{equation} 
where $T_0 = T^K(0)$.

The H-theorem also holds for Fokker-Planck-Landau type equations, which states that the entropy decays throughout time.  The entropy is defined as
\begin{equation*}
\mathcal{H}[f](t) = \int_{\Omega_x} \int_{\mathbb{R}^3} f \ln (f) ~\textrm{d}\boldsymbol{v} \textrm{d}x
\end{equation*}
\begin{flalign*}
\textrm{and so the H-theorem gives that}&& \frac{\textrm{d}}{\textrm{d}t}\left(\mathcal{H}[f]\right) \leq 0. && && && &&
\end{flalign*}

At this point it is also useful to define the entropy relative to the equilibrium Maxwellian $\mathcal{M}_{eq}$ as
\begin{align}
\mathcal{H}[f|\mathcal{M}_{eq}](t) &= \int_{\Omega_x} \int_{\mathbb{R}^3} f \ln (f) ~\textrm{d}\boldsymbol{v} \textrm{d}x - \int_{\Omega_x} \int_{\mathbb{R}^3} \mathcal{M}_{eq} \ln (\mathcal{M}_{eq}) ~\textrm{d}\boldsymbol{v} \textrm{d}x \nonumber \\
&= \int_{\Omega_x} \int_{\mathbb{R}^3} f \ln \left(\frac{f}{\mathcal{M}_{eq}}\right) ~\textrm{d}\boldsymbol{v} \textrm{d}x. \label{RelEnt}
\end{align}
Again, in the space-homogeneous case, there is no integration with respect to $x$ when considering the entropy.

\subsection{Choosing a Computational Domain}
Initially $f(0, x, \boldsymbol{v}) = f_0(x, \boldsymbol{v})$ and it is assumed that $\textrm{supp}f \Subset \Omega_{\boldsymbol{v}}$, for some domain $\Omega_{\boldsymbol{v}} \subset \mathbb{R}^3$, since $f$ should have sufficient decay in velocity-space \cite{BoltzmannSupport} and $\Omega_{\boldsymbol{v}}$ is chosen depending on the initial data (see \cite{BoltzmannConvergence}, Section 2).  In fact, $\boldsymbol{v} \in \mathbb{R}^3$ but values of $f$ are negligible outside a sufficiently large ball.  The initial data is then extended by zero outside the computational domain, which means it can be controlled by $e^{-c|\boldsymbol{v}|^2}$, for $c > 0 $ depending on the moments of $f_0$.  Under such conditions, it is expected that the computational solution will remain supported on $\Omega_{\boldsymbol{v}}$ up to a fixed small error that depends on the initial data (more details can be seen in the proof for the conservative spectral method applied to the Boltzmann equation in \cite{BoltzmannConvergence}).

More precisely, assume that the support of $f$ is in fact contained in $B_R(\boldsymbol{0})$, for $R > 0$ large enough, and choose the approximate velocity domain as $\Omega_{\boldsymbol{v}} = [-L_v, L_v]^3$, for $L_v > R$.  Then, to match up with the required reciprocity relation for the discrete Fourier transform that is used by the FFTW3 package \cite{fftw3} in the code, there is a corresponding transformed Fourier space $\Omega_{\boldsymbol{\xi}}$.  This is given by $\Omega_{\boldsymbol{\xi}} = [-L_{\xi}, L_{\xi}]^3$, for $L_{\xi} = \frac{N \pi}{2 L_v}$ when $N$ Fourier modes are used in each dimension of velocity.

\subsection{Time Splitting}
For computational purposes, the space-inhomogeneous Fokker-Planck-Landau type equation (\ref{Landau}) is broken down into two smaller problems in a process known as time splitting.  To describe this, let time be discretised by $t_n = t_0 + n \Delta t$, for some time-step $\Delta t$, and let $f_n(x, \boldsymbol{v}) = f(t_n, x, \boldsymbol{v})$.  First, given the solution $f_n$, a collisionless advection problem is solved for $g$, namely
\begin{equation}
g_t(t, x, \boldsymbol{v}) + \boldsymbol{v}\cdot\nabla_{x}g(t, x, \boldsymbol{v}) - \boldsymbol{E}(t, x)\cdot\nabla_{\boldsymbol{v}}g(t, x, \boldsymbol{v}) = 0, \label{Vlasov1}
\end{equation}
along with Poisson's equation (\ref{Poisson}), with $g(0, x, \boldsymbol{v}) = f_n(x, \boldsymbol{v})$.  Then a space-homogeneous collision problem is solved for $\tilde{f}$ at each $x \in \Omega_x$, namely
\begin{equation}
\tilde{f}_t(t, x, \boldsymbol{v}) = \frac{1}{\varepsilon} Q(\tilde{f},\tilde{f})(t, x, \boldsymbol{v}), \label{Landau_homogeneous}
\end{equation}
with $\tilde{f}(0, x, \boldsymbol{v}) = g(\Delta t, x, \boldsymbol{v})$.  Finally, the solution at time $t = t_{n+1}$ is given by
\begin{equation*}
f_{n+1}(x, \boldsymbol{v}) = \tilde{f}(\Delta t, x, \boldsymbol{v}).
\end{equation*}

Equation (\ref{Vlasov1}) is solved by a discontinuous Galerkin (D.G.) method, with piecewise linear polynomials in $x$ and piecewise quadratic polynomials in $\boldsymbol{v}$, and third order Runge-Kutta in time.  Proofs of how the choice of quadratic basis functions in velocity space ensure moment conservation at this stage are given in \cite{Chenglong}.

Then, equation (\ref{Landau_homogeneous}) is solved by the conservative spectral method with fourth order Runge-Kutta for time-stepping.  Conservation is enforced by considering a constrained minimisation problem, which will be described in Section \ref{ConservationRoutine}.  The spectral method will be described in Section \ref{Fourier} and is extended from the Landau equation with Coulomb interactions to Fokker-Planck-Landau type equations with Maxwell type and hard sphere interactions.

\section{The Fourier Transform of the Collision Operator} \label{Fourier}
As is shown in \cite{Chenglong}, when the pdf $f$ is supported in a ball of radius $R > 0$, the Fourier transform of the collision operator $Q$ is
\begin{equation}
\hat{Q}(\hat{f}, \hat{f})\left(\boldsymbol{\xi}\right) = \int_{\Omega_{\boldsymbol{\xi}}}\hat{f}\left(\boldsymbol{\xi} - \boldsymbol{\omega}\right)\hat{f} (\boldsymbol{\omega})\Bigl({\boldsymbol{\omega}}^T\hat{S}~\left(\boldsymbol{\omega}\right)\boldsymbol{\omega} ~-~ {\left(\boldsymbol{\xi} - \boldsymbol{\omega}\right)}^T\hat{S}~\left(\boldsymbol{\omega}\right)\left(\boldsymbol{\xi} - \boldsymbol{\omega}\right)\Bigr)~\textrm{d}\boldsymbol{\omega}, \label{qHat}
\end{equation}
for $\boldsymbol{\xi} \in \Omega_{\boldsymbol{\xi}}$, the Fourier space domain described in the previous section, where
\begin{equation*}
\hat{S}\left(\boldsymbol{\omega }\right) = (2\pi)^{-\frac{3}{2}} \int_{B_R(\boldsymbol{0})}{S\left(\boldsymbol{u}\right)}e^{-i\boldsymbol{\omega } \cdot \boldsymbol{u}}\mathrm{d}\boldsymbol{u},
\end{equation*}
\begin{flalign*}
\textrm{for} && S(\boldsymbol{u}) = |\boldsymbol{u}|^{\lambda + 2}\left(\textrm{I} - \frac{\boldsymbol{uu}^T}{|\boldsymbol{u}|^2}\right), ~~~\textrm{with } -3 \leq \lambda \leq 1. &&
\end{flalign*}

This means that evaluating $\hat{Q}$ is performed by a fast Fourier transform (F.F.T.) of the pdf $f$ and then a weighted convolution with itself.  The F.F.T.\ requires $\mathcal{O}(N^3 \log{N})$ operations and multiplication by the weight and quadrature to calculate the convolution requires $\mathcal{O}(N^3)$ operations.  The weights can also be pre-computed and stored at the beginning of the code run, where the bulk of the calculation is in evaluation of $\hat{S}$.  This has different forms depending on the value of $\lambda$ but the results are found through the same general method.

First, the entries of $\hat{S}$ can be decomposed as $\hat{S}_{i, j}(\boldsymbol{\omega}) = \hat{S}^1_{i, j}(\boldsymbol{\omega}) - \hat{S}^2_{i, j}(\boldsymbol{\omega})$, for $i, j = 1, 2, 3$, with
\begin{flalign*}
&& \hat{S}^1_{i, j}(\boldsymbol{\omega}) &= (2\pi)^{-\frac{3}{2}} \int_{B_R(\boldsymbol{0})} |\boldsymbol{u}|^{\lambda + 2} \delta_{i,j} e^{-i \boldsymbol{\omega}\cdot\boldsymbol{u}} ~\textrm{d}\boldsymbol{u} && \\
\textrm{and} &&\hat{S}^2_{i, j}(\boldsymbol{\omega}) &= (2\pi)^{-\frac{3}{2}} \int_{B_R(\boldsymbol{0})} |\boldsymbol{u}|^{\lambda}u_i u_j e^{-i \boldsymbol{\omega}\cdot\boldsymbol{u}} ~\textrm{d}\boldsymbol{u}. &&
\end{flalign*}
Then, for a given $\boldsymbol{\omega} = (\omega_1, \omega_2, \omega_3)$, it should be noted that when $j = i$, there is only one value of $\hat{S}^1_{i, i}(\boldsymbol{\omega})$, for each $i = 1, 2, 3$, and that $\hat{S}^1_{i, j}(\boldsymbol{\omega}) = 0$ when $i \neq j$ (thanks to the Kronecker delta).  Also note that, for $i = j$,
\begin{equation*}
\hat{S}^2_{1, 1}(\omega_1, \omega_2, \omega_3) = \hat{S}^2_{3, 3}(\omega_2, \omega_3, \omega_1) ~~~\textrm{and} ~~~\hat{S}^2_{2, 2}(\omega_1, \omega_2, \omega_3) = \hat{S}^2_{3, 3}(\omega_1, \omega_3, \omega_2)
\end{equation*}
and, for $i \neq j$,
\begin{equation*}
\hat{S}^2_{1, 2}(\omega_1, \omega_2, \omega_3) = \hat{S}^2_{1, 3}(\omega_1, \omega_3, \omega_2) ~~~\textrm{and} ~~~\hat{S}^2_{2, 3}(\omega_1, \omega_2, \omega_3) = \hat{S}^2_{1, 3}(\omega_2, \omega_1, \omega_3).
\end{equation*}
The sub-diagonal entries are then also known since $\hat{S}$ is a symmetric matrix because $S$ is too.  This means that only $\hat{S}^1_{1, 1}$, $\hat{S}^2_{3, 3}$ and $\hat{S}^2_{1, 3}$ need to be calculated, the explicit formulae for which are found to be, evaluated at $\boldsymbol{\omega}$ such that $|\boldsymbol{\omega}| \neq 0$,
\begin{equation*}
\hat{S}^1_{1, 1} (\boldsymbol{\omega}) =
\left\{
\begin{aligned}
\displaystyle
\sqrt{\frac{2}{\pi}} \frac{1}{|\boldsymbol{\omega}|^2}&\Bigl(1 - \cos(R |\boldsymbol{\omega}|)\Bigr), &&\textrm{when } \lambda = -3, \\
\displaystyle
\sqrt{\frac{2}{\pi}} \frac{1}{|\boldsymbol{\omega}|^5}&\Bigl(-(R|\boldsymbol{\omega}|)^3 \cos(R|\boldsymbol{\omega}|) + \mathrlap{3 (R|\boldsymbol{\omega}|)^2 \sin(R|\boldsymbol{\omega}|)}\\
&~~~ + 6 (R|\boldsymbol{\omega}|) \cos(R|\boldsymbol{\omega}|) - 6 \sin(R|\boldsymbol{\omega}|) \Bigr), &&\textrm{when } \lambda = 0, \\
\displaystyle
\sqrt{\frac{2}{\pi}} \frac{1}{|\boldsymbol{\omega}|^6}&\Bigl(-(R|\boldsymbol{\omega}|)^4 \cos(R|\boldsymbol{\omega}|) + \mathrlap{4 (R|\boldsymbol{\omega}|)^3 \sin(R|\boldsymbol{\omega}|)} \\
&~~~ + \mathrlap{12 (R|\boldsymbol{\omega}|)^2 \cos(R|\boldsymbol{\omega}|) - 24 (R|\boldsymbol{\omega}|) \sin(R|\boldsymbol{\omega}|)} \\
&~~~~~~ - 24 \cos(R|\boldsymbol{\omega}|) + 24\Bigr), &&\textrm{when } \lambda = 1,
\end{aligned}
\right.
\end{equation*}
\begin{equation*}
\hat{S}^2_{3, 3} (\boldsymbol{\omega}) =
\left\{
\begin{aligned}
\displaystyle
\sqrt{\frac{2}{\pi}} \frac{1}{|\boldsymbol{\omega}|^4}&\biggl(\Bigl(\omega_1^2 + \omega_2^2\Bigr) \frac{R|\boldsymbol{\omega}| - \sin(R|\boldsymbol{\omega}|)}{R|\boldsymbol{\omega}|} \\
&~~~- \mathrlap{\omega_3^2 \frac{R|\boldsymbol{\omega}| + (R|\boldsymbol{\omega}|) \cos(R|\boldsymbol{\omega}|) - 2 \sin(R|\boldsymbol{\omega}|)}{R|\boldsymbol{\omega}|}\biggr),} \\
&~~~~~~~~~~~~~~~~~~~~~~~~~~~~~~~~~~~~~~~~~~~~~~~~~~~~ &&\textrm{when } \lambda = -3, \\
\displaystyle
\sqrt{\frac{2}{\pi}} \frac{1}{|\boldsymbol{\omega}|^7}&\biggl(\Bigl(\omega_1^2 + \omega_2^2\Bigr)\Bigl(- (R|\boldsymbol{\omega}|)^2 \sin(R|\boldsymbol{\omega}|) - \mathrlap{3 (R|\boldsymbol{\omega}|) \cos(R|\boldsymbol{\omega}|)} \\
&~~~~~~~~~~~~~~~~~~ + 3 \sin(R|\boldsymbol{\omega}|)\Bigr) \\
&~~~+ \omega_3^2 \Bigl(- (R|\boldsymbol{\omega}|)^3 \cos(R|\boldsymbol{\omega}|) + \mathrlap{5 (R|\boldsymbol{\omega}|)^2 \sin(R|\boldsymbol{\omega}|)} \\
&~~~~~~~~~~~~~ + 12 (R|\boldsymbol{\omega}|) \cos(R|\boldsymbol{\omega}|) - \mathrlap{12 \sin(R|\boldsymbol{\omega}|)\Bigr)\biggr),} \\
&~~~~~~~~~~~~~~~~~~~~~~~~~~~~~~~~~~~~~~~~~~~~~~~~~~~~ &&\textrm{when } \lambda = 0, \\
\displaystyle
\sqrt{\frac{2}{\pi}} \frac{1}{|\boldsymbol{\omega}|^8}&\biggl(\Bigl(\omega_1^2 + \omega_2^2\Bigr)\Bigl(- (R|\boldsymbol{\omega}|)^3 \sin(R|\boldsymbol{\omega}|) - \mathrlap{4 (R|\boldsymbol{\omega}|)^2 \cos(R|\boldsymbol{\omega}|)} \\
&~~~~~~~~~~~~~~~~~~ + 8 (R|\boldsymbol{\omega}|) \sin(R|\boldsymbol{\omega}|) + \mathrlap{8 \cos(R|\boldsymbol{\omega}|) - 8\Bigr)} \\
&~~~+ \omega_3^2 \Bigl(- (R|\boldsymbol{\omega}|)^4 \cos(R|\boldsymbol{\omega}|) + \mathrlap{6 (R|\boldsymbol{\omega}|)^3 \sin(R|\boldsymbol{\omega}|)} \\
&~~~~~~~~~~~~~ + 20 (R|\boldsymbol{\omega}|)^2 \cos(R|\boldsymbol{\omega}|) - \mathrlap{40 (R|\boldsymbol{\omega}|) \sin(R|\boldsymbol{\omega}|)} \\
&~~~~~~~~~~~~~ - 40 \cos(R|\boldsymbol{\omega}|) + 40\Bigr)\biggr), &&\textrm{when } \lambda = 1
\end{aligned}
\right.
\end{equation*}
and
\begin{equation*}
\hat{S}^2_{1, 3} (\boldsymbol{\omega}) =
\left\{
\begin{aligned}
\displaystyle
-\sqrt{\frac{2}{\pi}} \frac{\omega_1 \omega_3}{|\boldsymbol{\omega}|^4} &\frac{2R|\boldsymbol{\omega}| + R|\boldsymbol{\omega}| \cos(R|\boldsymbol{\omega}|) - 3\sin(R|\boldsymbol{\omega}|)}{R|\boldsymbol{\omega}|}, &&\textrm{when } \lambda = -3, \\
\displaystyle
\sqrt{\frac{2}{\pi}} \frac{\omega_1 \omega_3}{|\boldsymbol{\omega}|^7}&\Bigl(- (R|\boldsymbol{\omega}|)^3 \cos(R|\boldsymbol{\omega}|) + \mathrlap{6 (R|\boldsymbol{\omega}|)^2 \sin(R|\boldsymbol{\omega}|)} \\
&~ + 15 (R|\boldsymbol{\omega}|) \cos(R|\boldsymbol{\omega}|) - 15 \sin(R|\boldsymbol{\omega}|)\Bigr), && \textrm{when } \lambda = 0, \\
\displaystyle
\sqrt{\frac{2}{\pi}} \frac{\omega_1 \omega_3}{|\boldsymbol{\omega}|^8}&\Bigl(- (R|\boldsymbol{\omega}|)^4 \cos(R|\boldsymbol{\omega}|) + \mathrlap{7 (R|\boldsymbol{\omega}|)^3 \sin(R|\boldsymbol{\omega}|)} \\
&~ + 24 (R|\boldsymbol{\omega}|)^2 \cos(R|\boldsymbol{\omega}|) - \mathrlap{48 (R|\boldsymbol{\omega}|) \sin(R|\boldsymbol{\omega}|)} \\
&~ - 48 \cos(R|\boldsymbol{\omega}|) + 48\Bigr), && \textrm{when } \lambda = 1.
\end{aligned}
\right.
\end{equation*}
The details leading to these expressions can be found in appendix \ref{S_calculations}.  In addition, by substituting $\boldsymbol{\omega} = \boldsymbol{0}$ into the integrands found in $\hat{S}^1_{1, 1}$, $\hat{S}^2_{3, 3}$ and $\hat{S}^2_{1, 3}$ and evaluating directly (noting that the exponential evaluated at $\boldsymbol{\omega} = \boldsymbol{0}$ is equal to one),
\begin{equation*}
\hat{S}^1_{1, 1} (\boldsymbol{0}) =
\begin{cases}
\displaystyle
~~\sqrt{\frac{1}{2\pi}} R^2, &~~~~~\textrm{when } \lambda = -3, \\
\displaystyle
\frac{2}{5}\sqrt{\frac{1}{2 \pi}}R^5, &~~~~~\textrm{when } \lambda = 0, \\
\displaystyle
\frac{1}{3}\sqrt{\frac{1}{2 \pi}}R^6, &~~~~~\textrm{when } \lambda = 1,
\end{cases}
\end{equation*}
\begin{equation*}
\hat{S}^2_{3, 3} (\boldsymbol{0}) =
\begin{cases}
\displaystyle
\frac{1}{3 \sqrt{2\pi}} R^2, &~~~~~\textrm{when } \lambda = -3, \\
\displaystyle
\frac{2}{15 \sqrt{2\pi}} R^5, &~~~~~\textrm{when } \lambda = 0, \\
\displaystyle
\frac{1}{9 \sqrt{2\pi}} R^6, &~~~~~\textrm{when } \lambda = 1
\end{cases}
\end{equation*}
\begin{flalign*}
\textrm{and} && \hat{S}^2_{1, 3} (\boldsymbol{0}) = 0, ~~~~~\textrm{for all } \lambda. &&
\end{flalign*}

\section{The Conservation Routine} \label{ConservationRoutine}
Even though the approximated pdf $f$ is assumed to have support inside $B_R(0)$, the true solution does still take values outside this ball, albeit negligible.  In general, a larger choice of $B_R(0)$ and  $\Omega_{\boldsymbol{\xi}}$ will give a more accurate approximation to $\hat{Q}_L$, and therefore $Q$, but some amount of error is unavoidable whenever truncating the velocity domain and its associated Fourier domain.  This is because any collision operator defined on a truncated domain cannot hope to conserve moments of the solution, since the property satisfied by the collision invariants is defined by integrals over all of $\mathbb{R}^3$, as in expression (\ref{conservation}).  Conservation can be enforced, however, by considering a constrained minimisation problem.

\subsection{Conserving in Velocity Space}
Given a collection of discrete values of the collision operator $Q$ resulting from the spectral method, say $\{\tilde{Q}_n\}_{n=1}^{N^3}$, a new set of values $\{Q_n\}_{n=1}^{N^3}$ must be found which are as close as possible to the original values in $\ell^2$-norm but satisfy the discrete form of (\ref{conservation}).  This discrete form replaces the integrals in (\ref{conservation}) with quadrature sums and can be written as
\begin{flalign*}
&& && \sum_{n=1}^{N^3} Q_n (\phi_k)_n \omega_n = 0, && \textrm{for } k = 0, 1, \ldots, 4,
\end{flalign*}
where $\{(\phi_k)_n\}_{k=0}^{4}$ are evaluations of the collision invariants at the same discrete point where $Q_n$ is evaluated and $\omega_n$ is the corresponding quadrature weight for that point.  If the discrete values of $Q$ are stored in the vector $\boldsymbol{Q}$ of length $N^3$, this discrete conservation can be written as 
\begin{flalign}
A \boldsymbol{Q} = \boldsymbol{0}, ~~\textrm{where } A_{k,n} = (\phi_{k-1})_n \omega_n, ~~\textrm{for } k = 1,2\ldots,5, ~n = 1,2,\ldots N^3. \label{A_entries}
\end{flalign}
Then, given $\tilde{\boldsymbol{Q}} = (\tilde{Q}_1, \tilde{Q}_2, \ldots, \tilde{Q}_{N^3})$, the least squares problem is to find the vector $\boldsymbol{Q} = (Q_1, Q_2, \ldots, Q_{N^3})$ of conserved evaluations of the collision operator which solves
\begin{flalign}
&& \min_{\boldsymbol{Q} \in \mathbb{R}^{N^3}} \bigl|\bigl|\tilde{\boldsymbol{Q}} - \boldsymbol{Q} \bigr|\bigr|^2_{\ell^2} && \textrm{such that } A \boldsymbol{Q} = \boldsymbol{0}. && \label{least_squares_v}
\end{flalign}
This can then be solved as a $5$-dimensional Lagrange multiplier problem by defining the operator
\begin{equation*}
L(\boldsymbol{Q},\boldsymbol{\gamma}) = \sum_{n=1}^{N^3} \bigl(\tilde{Q}_n - Q_n \bigr)^2 - \boldsymbol{\gamma}^T A \boldsymbol{Q}.
\end{equation*}
By solving $\nabla_{\boldsymbol{Q}} = \boldsymbol{0}$ for the Lagrange multiplier $\boldsymbol{\gamma}$, the discrete values of the conserved collision operator are found to be
\begin{flalign}
&& \boldsymbol{Q} = \Lambda(A) \tilde{\boldsymbol{Q}} && \textrm{where } \Lambda(A) = I - A^T(A A^T)^{-1} A. && \label{conserve_velocity}
\end{flalign}

This means that the conservation is simply matrix-vector multiplication.  The details of the derivation of $\Lambda(A)$ can be found in \cite{Harsha} and \cite{Chenglong} for the Boltzmann and Landau equations, respectively, but it should be noted that $\Lambda(A)$ is identical for both equations.  The full algorithm of the conservative spectral method for solving the space-homogeneous Fokker-Planck-Landau type equation (\ref{Landau_homogeneous}) when conserving in velocity space is then given in Algorithm \ref{conservation_algorithm_velocity}.

\begin{algorithm}
	\caption{The conservative spectral method for solving the space-homogeneous Fokker-Planck-Landau type equation (\ref{Landau_homogeneous}) when conserving in velocity space
		\label{conservation_algorithm_velocity}}
	\begin{algorithmic}[1]
		\Require{$\boldsymbol{F}$ contains evaluations of $f$ on the uniform velocity grid at a given time-step $t_n$}
		\Statex
		\For{each step in Runge-Kutta}
			\State Calculate the F.F.T.\ of $\boldsymbol{F}$ and store the values in $\hat{\boldsymbol{F}}$ \Comment{$\mathcal{O}(N^3 \log{N})$}
			\State \multiline{Calculate $\hat{Q}(\hat{\boldsymbol{F}})$ at each point in the uniform Fourier space grid using identity (\ref{qHat}) and store the values in $\hat{\boldsymbol{Q}}$ \Comment{$\mathcal{O}(N^3)$}}
			\State Calculate the I.F.F.T.\ of $\hat{\boldsymbol{Q}}$ and store the values in $\tilde{\boldsymbol{Q}}$ \Comment{$\mathcal{O}(N^3 \log{N})$}
			\State Set $\boldsymbol{Q} = \Lambda(A) \hat{\boldsymbol{Q}}$ as in (\ref{conserve_velocity}), with $A$ given in (\ref{A_entries}) \Comment{$\mathcal{O}(N^3)$}
			\State Perform the iteration of Runge-Kutta to update $\boldsymbol{F}$ \Comment{$\mathcal{O}(N^3)$}
		\EndFor
	\end{algorithmic}
\end{algorithm}	

\subsection{Conserving in Fourier Space}
The method of conservation just described is the one which is used in deriving error estimates for the spectral method for the Boltzmann equation, as in \cite{BoltzmannConvergence}, and in work currently undergo for Fokker-Planck-Landau equations as well.  In practice, however, for all simulations in Section \ref{Results}, conservation is actually enforced in Fourier space.  

To describe the conservation in Fourier space, first consider the partial Fourier series reconstruction of $Q$.  If the velocity domain is $\Omega_{\boldsymbol{v}} = [-L_{\boldsymbol{v}}, L_{\boldsymbol{v}}]^3$, for large enough $L_{\boldsymbol{v}} > 0$ and the Fourier modes at which the F.F.T.\ is evaluated are denoted by $\{\boldsymbol{\xi}_{n}\}_{n = 1,2,\ldots,N^3}$
\begin{equation*}
Q(f,f) \approx \frac{(2 \pi)^{\frac{3}{2}}}{2 L_{\boldsymbol{v}}} \sum_{n=1}^{N^3} \hat{Q}_L(\boldsymbol{\xi}_{n})e^{i \boldsymbol{\xi}_{n} \cdot \boldsymbol{v}}. 
\end{equation*}
Then, using this approximation to $Q$ in the integrals (\ref{conservation}) which enforce conservation gives
\begin{equation*}
\int_{\Omega_{\boldsymbol{v}}} \frac{(2 \pi)^{\frac{3}{2}}}{2 L_{\boldsymbol{v}}} \left(\sum_{n=1}^{N^3} \hat{Q}_L(\boldsymbol{\xi}_{n}) \right)e^{i \boldsymbol{\xi}_{n} \cdot \boldsymbol{v}} \phi_k (\boldsymbol{v}) ~\textrm{d}\boldsymbol{v} = 0, ~~~~~\textrm{for } k = 0, 1, \ldots, d,
\end{equation*}
which is equivalent to
\begin{equation*}
\sum_{n=1}^{N^3} \left(\int_{\Omega_{\boldsymbol{v}}}e^{i \boldsymbol{\xi}_{n} \cdot \boldsymbol{v}} \phi_k (\boldsymbol{v}) ~\textrm{d}\boldsymbol{v} \right) \hat{Q}_L(\boldsymbol{\xi}_{n}) = 0, ~~~~~\textrm{for } k = 0, 1, \ldots, d,
\end{equation*}
The left-hand side here is another matrix vector multiplication.  So, if $\tilde{\hat{\boldsymbol{Q}}} = (\tilde{\hat{Q}}_1, \tilde{\hat{Q}}_2, \ldots, \tilde{\hat{Q}}_{N^3})$, where $\tilde{\hat{Q}}_n = \hat{Q}_L(\boldsymbol{\xi})_{\boldsymbol{n}}$ for $n = 1,2,\ldots,N^3$,  the least squares problem in Fourier space is to find the vector $\hat{\boldsymbol{Q}} = (\hat{Q}_1, \hat{Q}_2, \ldots, \hat{Q}_{N^3})$ which solves
\begin{flalign}
&& \min_{\hat{\boldsymbol{Q}} \in \mathbb{R}^{N^3}} \bigl|\bigl|\tilde{\hat{\boldsymbol{Q}}} - \hat{\boldsymbol{Q}} \bigr|\bigr|^2_{\ell^2} && \textrm{such that } C \hat{\boldsymbol{Q}} = \boldsymbol{0}, && \label{least_squares_Fourier}
\end{flalign}
where $C$ is the matrix with entries
\begin{flalign}
&& C_{k,n} = \int_{\Omega_{\boldsymbol{v}}}e^{i \boldsymbol{\xi}_{n} \cdot \boldsymbol{v}} \phi_k (\boldsymbol{v}) ~\textrm{d}\boldsymbol{v}, && \textrm{for } k = 1,2\ldots,d+2, ~n = 1,2,\ldots N^3. \label{C_entries}
\end{flalign}

This least squares problem (\ref{least_squares_Fourier}) in Fourier space is the exact same form as the least squares problem (\ref{least_squares_v}) in velocity space, but with the matrix $C$ instead of $A$.  This means that it has the same solution $\hat{\boldsymbol{Q}} = \Lambda(C) \tilde{\hat{\boldsymbol{Q}}}$, for the same operator $\Lambda$ in (\ref{A_entries}), but evaluated with the matrix $C$.  

A couple of things should be mentioned here.  First, since the conservation is enforced in Fourier space, the I.F.F.T.\ must then be taken of the conserved vector $\hat{\boldsymbol{Q}}$ to obtain $\boldsymbol{Q}$.  This means that there may be a tiny amount of conservation lost in during the inverse Fourier transform.  On the other hand, the least squares problem for conserving in velocity space involved the quadrature matrix $A$.  This means that conservation in velocity space will always have some error as well, resulting from the choice of quadrature.  The values of the integrals in the matrix $C$ actually have explicit values which can be calculated by hand, as shown in \cite{Chenglong}.  In practice, there seems to be less of an error resulting from I.F.F.T.\ after conserving in Fourier space than there would be from conserving in velocity space using the quadrature matrix $A$.

The full algorithm of the conservative spectral method for solving the space-homogeneous Fokker-Planck-Landau type equation (\ref{Landau_homogeneous}) when conserving in Fourier space is then given in Algorithm \ref{conservation_algorithm_Fourier}.  Note that this is the algorithm which is used in all simulations in Section \ref{Results}.

\begin{algorithm}
	\caption{The conservative spectral method for solving the space-homogeneous Fokker-Planck-Landau type equation (\ref{Landau_homogeneous}) when conserving in Fourier space
		\label{conservation_algorithm_Fourier}}
	\begin{algorithmic}[1]
		\Require{$\boldsymbol{F}$ contains evaluations of $f$ on the uniform velocity grid at a given time-step $t_n$}
		\Statex
		\For{each step in Runge-Kutta}
		\State Calculate the F.F.T.\ of $\boldsymbol{F}$ and store the values in $\hat{\boldsymbol{F}}$ \Comment{$\mathcal{O}(N^3 \log{N})$}
		\State \multiline{Calculate $\hat{Q}(\hat{\boldsymbol{F}})$ at each point in the uniform Fourier space grid using identity (\ref{qHat}) and store the values in $\tilde{\hat{\boldsymbol{Q}}}$ \Comment{$\mathcal{O}(N^3)$}}
		\State Set $\hat{\boldsymbol{Q}} = \Lambda(C) \tilde{\hat{\boldsymbol{Q}}}$ for $\Lambda$ as in (\ref{A_entries}) and $C$ given in (\ref{C_entries})  \Comment{$\mathcal{O}(N^3)$}
		\State Calculate the I.F.F.T.\ of $\hat{\boldsymbol{Q}}$ and store the values in $\tilde{\boldsymbol{Q}}$ \Comment{$\mathcal{O}(N^3 \log{N})$}
		\State Perform the iteration of Runge-Kutta to update $\boldsymbol{F}$ \Comment{$\mathcal{O}(N^3)$}
		\EndFor
	\end{algorithmic}
\end{algorithm}	

\section{Stability of the Space-homogeneous Spectral Method} \label{Stability}
In order to consider the stability of the spectral method, first note that the integral (\ref{qHat}) to calculate $\hat{Q}$ is approximated using quadrature.  The current code uses the composite trapezoidal rule but, in general, for $M$ equally spaced quadrature nodes $\{\boldsymbol{\xi}_m\}_{m=1}^M \subset \Omega_{L_{\boldsymbol{\xi}}}$ in Fourier space, corresponding weights $\{w_m\}_{m=1}^M$ and Fourier space stepsize $h_{\xi} = \frac{2 L_{\xi}}{N} = \frac{\pi}{L_{v}}$, 
\begin{multline}
\hat{Q}\left(\boldsymbol{\xi}_k\right) = h_{\xi}^3\sum_{m = 1}^{M} w_m \hat{f}\left(\boldsymbol{\xi}_k - \boldsymbol{\xi}_m\right)\hat{f} (\boldsymbol{\xi}_m)\Bigl({\boldsymbol{\xi}_m}^T\hat{S}~\left(\boldsymbol{\xi}_m\right)\boldsymbol{\xi}_m \\
~-~ {\left(\boldsymbol{\xi}_k - \boldsymbol{\xi}_m\right)}^T\hat{S}~\left(\boldsymbol{\xi}_m\right)\left(\boldsymbol{\xi}_k - \boldsymbol{\xi}_m\right)\Bigr). \label{Q_quad}
\end{multline}

Now, according to Lebedev \cite{Lebedev}, the criterion for stability of a numerical method of the form
\begin{equation*}
\frac{\textrm{d}}{\textrm{d}t}\Bigl(\hat{f}(\boldsymbol{\xi}_k)\Bigr) = F(\hat{f}(\boldsymbol{\xi}_k))
\end{equation*}
is that the time-step $\Delta t$ must satisfy
\begin{equation*}
\Delta t \leq \frac{1}{\textrm{Lip}(F)},
\end{equation*}
for the Lipschitz norm of $F$, $\textrm{Lip}(F)$.  If an upper bound can be found on $\textrm{Lip}(F)$, this will in turn give a lower bound on $(\textrm{Lip}(F))^{-1}$, which $\Delta t$ must be below for the numerical method to remain stable.  To find the upper bound, note that
\begin{equation*}
\textrm{Lip}(F) \leq \max_{k,l = 1, \ldots M} |\mathcal{J}_{k,l}|,
\end{equation*}
for the Jacobian $\mathcal{J}$ of $F(\hat{f}(\boldsymbol{\xi}_k))$, with entries
\begin{flalign*}
&& \mathcal{J}_{k,l} = \frac{\partial}{\partial \hat{f}(\boldsymbol{\xi}_l)}\Bigl(F(\hat{f}(\boldsymbol{\xi}_k))\Bigr), ~~~~~~\textrm{for } k,l = 1,2,\ldots,M.
\end{flalign*}

Here, $F(\hat{f}(\boldsymbol{\xi}_k)) = \frac{1}{\varepsilon}\hat{Q}(\hat{f}, \hat{f})\left(\boldsymbol{\xi}_k\right)$ and, to calculate the derivative of $\hat{Q}(\hat{f}, \hat{f})\left(\boldsymbol{\xi}_k\right)$ with respect to $\hat{f}(\boldsymbol{\xi}_l)$, it should be noted that there are two chances for $\hat{f}(\boldsymbol{\xi}_l)$ to appear in the quadrature sum (\ref{Q_quad}).  These are when $m = l$ and in general (depending on the choice of quadrature nodes) at another index, say $m = n$, where $\boldsymbol{\xi}_k - \boldsymbol{\xi}_n = \boldsymbol{\xi}_l$.  Assuming that there are indeed two indices which give rise to non-zero derivatives in the sum, and considering that $\boldsymbol{\xi}_k - \boldsymbol{\xi}_n = \boldsymbol{\xi}_l$ is equivalent to $\boldsymbol{\xi}_n = \boldsymbol{\xi}_k - \boldsymbol{\xi}_l$, the derivative is given by
\begin{align}
&\frac{\partial}{\partial \hat{f}(\boldsymbol{\xi}_l)}\Bigl(\hat{Q}(\hat{f}, \hat{f})\left(\boldsymbol{\xi}_k\right)\Bigr) \nonumber \\
&~~~= ~h_{\xi}^3 w_l \hat{f}\left(\boldsymbol{\xi}_k - \boldsymbol{\xi}_l\right) \Bigl({\boldsymbol{\xi}_l}^T\hat{S}~\left(\boldsymbol{\xi}_l\right)\boldsymbol{\xi}_l \nonumber - {\left(\boldsymbol{\xi}_k - \boldsymbol{\xi}_l\right)}^T\hat{S}~\left(\boldsymbol{\xi}_l\right)\left(\boldsymbol{\xi}_k - \boldsymbol{\xi}_l\right)\Bigr) \nonumber \\
&~~~~~~+ h_{\xi}^3 w_n \hat{f}\left(\boldsymbol{\xi}_n\right) \Bigl({\boldsymbol{\xi}_n}^T\hat{S}~\left(\boldsymbol{\xi}_n\right)\boldsymbol{\xi}_n - {\left(\boldsymbol{\xi}_k - \boldsymbol{\xi}_n\right)}^T\hat{S}~\left(\boldsymbol{\xi}_n\right)\left(\boldsymbol{\xi}_k - \boldsymbol{\xi}_n\right)\Bigr) \nonumber \\
&~~~= ~h_{\xi}^3 w_l \hat{f}\left(\boldsymbol{\xi}_k - \boldsymbol{\xi}_l\right) \Bigl({\boldsymbol{\xi}_l}^T\hat{S}~\left(\boldsymbol{\xi}_l\right)\boldsymbol{\xi}_l - {\left(\boldsymbol{\xi}_k - \boldsymbol{\xi}_l\right)}^T\hat{S}~\left(\boldsymbol{\xi}_l\right)\left(\boldsymbol{\xi}_k - \boldsymbol{\xi}_l\right)\Bigr) \nonumber \\
&~~~~~~+ h_{\xi}^3 w_n \hat{f}\left(\boldsymbol{\xi}_k - \boldsymbol{\xi}_l\right) \Bigl({(\boldsymbol{\xi}_k - \boldsymbol{\xi}_l)}^T\hat{S}~\left(\boldsymbol{\xi}_k - \boldsymbol{\xi}_l\right)(\boldsymbol{\xi}_k - \boldsymbol{\xi}_l)  - \boldsymbol{\xi}_l^T\hat{S}~\left(\boldsymbol{\xi}_k - \boldsymbol{\xi}_l\right)\boldsymbol{\xi}_l\Bigr).
\end{align}

Then, since $h_{\xi} = \frac{\pi}{L_v}$ and $|w_l| \leq 1$ for any $l$, by the triangle inequality, 
\begin{multline*}
~~~\left|\frac{\partial}{\partial \hat{f}(\boldsymbol{\xi}_l)}\Bigl(\hat{Q}(\hat{f}, \hat{f})\left(\boldsymbol{\xi}_k\right)\Bigr) \right| \\
\leq \frac{\pi^3}{L_v^3} |\hat{f}\left(\boldsymbol{\xi}_k - \boldsymbol{\xi}_l\right)| \Bigl(|{\boldsymbol{\xi}_l}^T\hat{S}~\left(\boldsymbol{\xi}_l\right)\boldsymbol{\xi}_l| + |{\left(\boldsymbol{\xi}_k - \boldsymbol{\xi}_l\right)}^T\hat{S}~\left(\boldsymbol{\xi}_l\right)\left(\boldsymbol{\xi}_k - \boldsymbol{\xi}_l\right)|~~~~~~~~~~~~~~ \\
+ |{(\boldsymbol{\xi}_k - \boldsymbol{\xi}_l)}^T\hat{S}~\left(\boldsymbol{\xi}_k - \boldsymbol{\xi}_l\right)(\boldsymbol{\xi}_k - \boldsymbol{\xi}_l)|  + |\boldsymbol{\xi}_l^T\hat{S}~\left(\boldsymbol{\xi}_k - \boldsymbol{\xi}_l\right)\boldsymbol{\xi}_l|\Bigr).
\end{multline*}
Note that if there had been no such $\boldsymbol{\xi}_n$ then the final two terms would be omitted here and the bound would only be smaller.  This means the assumption that there are two appearances of $\hat{f}(\boldsymbol{\xi}_l)$ in the quadrature sum (\ref{Q_quad}) is more general.

Also, by definition of the Fourier transform,
\begin{equation*}
|\hat{f}\left(\boldsymbol{\xi}_k - \boldsymbol{\xi}_l\right)| \leq (2\pi)^{-\frac{3}{2}} \int_{B_R(\boldsymbol{0})} |f\left(\boldsymbol{u}\right)| |e^{-i(\boldsymbol{\xi}_k - \boldsymbol{\xi}_l) \cdot \boldsymbol{u}}| \mathrm{d}\boldsymbol{u} = (2\pi)^{-\frac{3}{2}} ||f||_{L_1(B_R(\boldsymbol{0}))},
\end{equation*}
since $|e^{-i(\boldsymbol{\xi}_k - \boldsymbol{\xi}_l) \cdot \boldsymbol{u}}| = 1$, and so
\begin{multline}
\left|\frac{\partial}{\partial \hat{f}(\boldsymbol{\xi}_l)}\Bigl(\hat{Q}(\hat{f}, \hat{f})\left(\boldsymbol{\xi}_k\right)\Bigr) \right| \\
\leq \frac{\pi^{\frac{3}{2}}}{2 \sqrt{2} L_v^3} ||f||_{L_1(B_R(\boldsymbol{0}))} \Bigl(|{\boldsymbol{\xi}_l}^T\hat{S}~\left(\boldsymbol{\xi}_l\right)\boldsymbol{\xi}_l| + |{\left(\boldsymbol{\xi}_k - \boldsymbol{\xi}_l\right)}^T\hat{S}~\left(\boldsymbol{\xi}_l\right)\left(\boldsymbol{\xi}_k - \boldsymbol{\xi}_l\right)|~~~~~~~~~~~~~~~~~~~ \\
+ |{(\boldsymbol{\xi}_k - \boldsymbol{\xi}_l)}^T\hat{S}~\left(\boldsymbol{\xi}_k - \boldsymbol{\xi}_l\right)(\boldsymbol{\xi}_k - \boldsymbol{\xi}_l)|  + |\boldsymbol{\xi}_l^T\hat{S}~\left(\boldsymbol{\xi}_k - \boldsymbol{\xi}_l\right)\boldsymbol{\xi}_l|\Bigr). \label{dQ_bound}
\end{multline}

Now, for the terms involving $\hat{S}$, note that for a general matrix $A \in \mathbb{R}^{3 \times 3}$ and vectors $\boldsymbol{y}, \boldsymbol{z} \in \mathbb{R}^3$, 
\begin{equation}
\boldsymbol{y}^T A \boldsymbol{z} = \sum_{i,j = 1}^{3} A_{i,j} y_i z_j ~~~\textrm{and so } ~~|\boldsymbol{y}^T A \boldsymbol{z}| \leq (3)^2 \max_{i,j = 1,2,3} |A_{i,j}| (\max_{i = 1,2,3} y_i) (\max_{i = 1,2,3} z_i). \label{zAz_bound}
\end{equation}
This means that a bound must be found on $|\hat{S}_{i,j}(\boldsymbol{\xi})|$, which is achieved by using the expressions in Section \ref{Fourier} for $\hat{S}^1_{1,1}$, $\hat{S}^2_{3,3}$ and $\hat{S}^2_{1,3}$, for $\lambda = -3$, $0$ and $1$.  As is shown in Appendix \ref{S_bounds}, for any $k = 1, 2, \ldots, M$,
\begin{align*}
|\hat{S}_{i,j} (\boldsymbol{\xi}_k)| &\leq
\left\{
\begin{aligned}
\displaystyle
&\Biggl(\sqrt{\frac{1}{2\pi}} + \frac{3}{\pi^3} \Bigl(\pi + 1\Bigr) \sqrt{\frac{2}{\pi}} \Biggr) L_v^2, &&\textrm{when } \lambda = -3, \\
\displaystyle
&\sqrt{\frac{2}{\pi}} \frac{1}{\pi^5} \Bigl(2 \pi^3 + 9 \pi^2 + 21 \pi + 21 \Bigr) L_v^5, && \textrm{when } \lambda = 0, \\
\displaystyle
&\sqrt{\frac{2}{\pi}} \frac{1}{\pi^6} \Bigl(2 \pi^6 + 11 \pi^3 + 36 \pi^2 + 72 \pi + 144 \Bigr) L_v^6, && \textrm{when } \lambda = 1
\end{aligned}
\right. \\
&\lesssim
\left\{
\begin{aligned}
\displaystyle
& L_v^2, &&\textrm{when } \lambda = -3, \\
\displaystyle
& L_v^5, && \textrm{when } \lambda = 0, \\
\displaystyle
& L_v^6, && \textrm{when } \lambda = 1.
\end{aligned}
\right.
\end{align*}

Then, by using the identity (\ref{zAz_bound}) and noting that $|(\boldsymbol{\xi}_k)_i| \leq L_{\xi} = \frac{\pi}{h_v}$, for any $k, l, n = 1, 2, \ldots, M$,
\begin{equation*}
|\boldsymbol{\xi}_k^T\hat{S}(\boldsymbol{\xi}_l) \boldsymbol{\xi}_n| \lesssim 9 \frac{\pi^2}{h_v^2} \times
\left\{
\begin{aligned}
\displaystyle
& L_v^2, &&\textrm{when } \lambda = -3, \\
\displaystyle
& L_v^5, && \textrm{when } \lambda = 0, \\
\displaystyle
& L_v^6, && \textrm{when } \lambda = 1.
\end{aligned}
\right.
\end{equation*}
Also, since $\boldsymbol{\xi}_k - \boldsymbol{\xi}_l = \boldsymbol{\xi}_n$, each mixed $\boldsymbol{\xi}_k - \boldsymbol{\xi}_l$ and $\boldsymbol{\xi}_l$ term in inequality (\ref{dQ_bound}) has the same upper bound.  This gives
\begin{equation*}
\left|\frac{\partial}{\partial \hat{f}(\boldsymbol{\xi}_l)}\Bigl(\hat{Q}(\hat{f}, \hat{f})\left(\boldsymbol{\xi}_k\right)\Bigr) \right| \lesssim 4 \left(9 \frac{\pi^{\frac{7}{2}}}{2 \sqrt{2} h_v^2 L_v^3} ||f||_{L_1(B_R(\boldsymbol{0}))}\right) \times
\left\{
\begin{aligned}
\displaystyle
& L_v^2, &&\textrm{when } \lambda = -3, \\
\displaystyle
& L_v^5, && \textrm{when } \lambda = 0, \\
\displaystyle
& L_v^6, && \textrm{when } \lambda = 1,
\end{aligned}
\right.
\end{equation*}
and so
\begin{equation*}
|\mathcal{J}_{k,l}| \leq \frac{1}{\varepsilon} \left|\frac{\partial}{\partial \hat{f}(\boldsymbol{\xi}_l)}\Bigl(\hat{Q}(\hat{f}, \hat{f})\left(\boldsymbol{\xi}_k\right)\Bigr) \right| \lesssim \frac{18 \pi^{\frac{7}{2}}}{\sqrt{2}\varepsilon h_v^2} ||f||_{L_1(B_R(\boldsymbol{0}))} \times
\left\{
\begin{aligned}
\displaystyle
&  \frac{1}{L_v}, &&\textrm{when } \lambda = -3, \\
\displaystyle
& L_v^2, && \textrm{when } \lambda = 0, \\
\displaystyle
& L_v^3, && \textrm{when } \lambda = 1,
\end{aligned}
\right.
\end{equation*}
which means
\begin{equation*}
\frac{1}{|\mathcal{J}_{k,l}|} \gtrsim 
\left\{
\begin{aligned}
\displaystyle
& \frac{\sqrt{2} \varepsilon L_v h_v^2}{18 \pi^{\frac{7}{2}} ||f||_{L_1(B_R(\boldsymbol{0}))}}, &&\textrm{when } \lambda = -3, \\
\displaystyle
& \frac{\sqrt{2} \varepsilon h_v^2}{18 \pi^{\frac{7}{2}} L_v^2 ||f||_{L_1(B_R(\boldsymbol{0}))}}, && \textrm{when } \lambda = 0, \\
\displaystyle
& \frac{\sqrt{2} \varepsilon h_v^2}{18 \pi^{\frac{7}{2}} L_v^3||f||_{L_1(B_R(\boldsymbol{0}))}}, && \textrm{when } \lambda = 1.
\end{aligned}
\right.
\end{equation*}

Therefore, to ensure that $\Delta t \leq \frac{1}{|\mathcal{J}_{k,l}|}$, choose $\Delta t$ such that
\begin{align*}
\Delta t &\leq
\left\{
\begin{aligned}
\displaystyle
& \frac{\sqrt{2} \varepsilon L_v h_v^2}{18 \pi^{\frac{7}{2}} ||f||_{L_1(B_R(\boldsymbol{0}))}}, &&\textrm{when } \lambda = -3, \\
\displaystyle
& \frac{\sqrt{2} \varepsilon h_v^2}{18 \pi^{\frac{7}{2}} L_v^2 ||f||_{L_1(B_R(\boldsymbol{0}))}}, && \textrm{when } \lambda = 0, \\
\displaystyle
& \frac{\sqrt{2} \varepsilon h_v^2}{18 \pi^{\frac{7}{2}} L_v^3||f||_{L_1(B_R(\boldsymbol{0}))}}, && \textrm{when } \lambda = 1
\end{aligned}
\right. \\
&=
\left\{
\begin{aligned}
\displaystyle
& \frac{2 \sqrt{2} \varepsilon L_v^3}{9 \pi^{\frac{7}{2}} N^2 ||f||_{L_1(B_R(\boldsymbol{0}))}}, &&\textrm{when } \lambda = -3, \\
\displaystyle
& \frac{2 \sqrt{2} \varepsilon}{9 \pi^{\frac{7}{2}} N^2 ||f||_{L_1(B_R(\boldsymbol{0}))}}, && \textrm{when } \lambda = 0, \\
\displaystyle
& \frac{2 \sqrt{2} \varepsilon}{9 \pi^{\frac{7}{2}} N^2 L_v ||f||_{L_1(B_R(\boldsymbol{0}))}}, && \textrm{when } \lambda = 1.
\end{aligned}
\right.
\end{align*}

\section{Numerical Results and Entropy Decay} \label{Results}
\subsection{Space Homogeneous Results} \label{Results_Homogeneous}
In the previous work by the current authors \cite{RGDPaper}, simulations were already run to demonstrate the entropy decay rates for both Coulomb and hard sphere interactions using only $N = 16$ Fourier modes in each velocity direction.  The results were satisfactory but it has since been discovered that the decay rates are even more convincing when increasing to $N = 32$.  In addition, simulations have now been run for Maxwell type interactions, which had caused some difficulty to produce at first.

\subsubsection{The Coulomb Case ($\lambda = -3$)} \label{Results_Homogeneous_Coulomb}
When $-3 \leq \lambda < 0$, there is no spectral gap for Fokker-Planck-Landau type equations.  This was proven analytically by Strain and Guo \cite{Strain&Guo} where they showed that, if the initial condition is bounded by $e^{-c|\boldsymbol{v}|^2}$, for some $c > 0$, the rate of convergence to a Maxwellian close to equilibrium is given by
\begin{equation}
	e^{\displaystyle -kt^p}, ~~~~~\textrm{with } p = -\frac{2}{\lambda} ~\textrm{and some } k > 0. \label{S&G_decay}
\end{equation} 
For Coulomb interactions, with $\lambda = -3$, this gives the law of two thirds.  In \cite{RGDPaper}, the current authors showed this rate of convergence to equilibrium numerically by plotting the natural log of the relative entropy on a ln-ln scale against time.  In particular, as the solution approaches equilibrium, it should be that
\begin{equation*}
	\ln\Bigl(\Bigl|\ln \bigl(\bigl|\mathcal{H}[f|\mathcal{M}_{eq}]\bigr|\bigr)\Bigr|\Bigr) \sim \frac{2}{3} \ln(t).
\end{equation*}

The rate was captured by choosing an initial condition far from equilibrium, which is a sum of four Maxwellians with shifted centers, namely
\begin{equation}
	f_0(\boldsymbol{v}) = \frac{1}{4} \sum_{l = 0}^{3} \mathcal{M}_v \left(\boldsymbol{v} + \left((-1)^{\lfloor \frac{l}{2} \rfloor}, (-1)^{l}, (-1)^{l}\right)\right), \label{IC}
\end{equation}
for the Maxwellian $\mathcal{M}_v (\boldsymbol{v}) = (2 \pi T)^{- \frac{3}{2}} e^{-\frac{|\boldsymbol{v}|^2}{2T}}$.  The temperature used was $T = 0.4$; the Knudsen number was $\varepsilon = 20$; the velocity domain had $L_v = 5.25$; $N = 16$ Fourier modes were chosen; and the time-stepsize used was $\Delta t = 0.01$.  With these parameters, the rate was seen to be $0.634$.  This result is improved in the current simulation, however, where the number of Fourier modes has been increased to $N = 32$ (so that $\Delta t = 0.01$ is still below the new upper bound of approximately 0.0162 calculated for stability with these parameters for $\lambda = -3$ in Section \ref{Stability}).  

\begin{figure}[!hbtp] 
	\resizebox{\textwidth}{!}{
	\begin{tabular}{cc}
		\includegraphics[width=75mm]{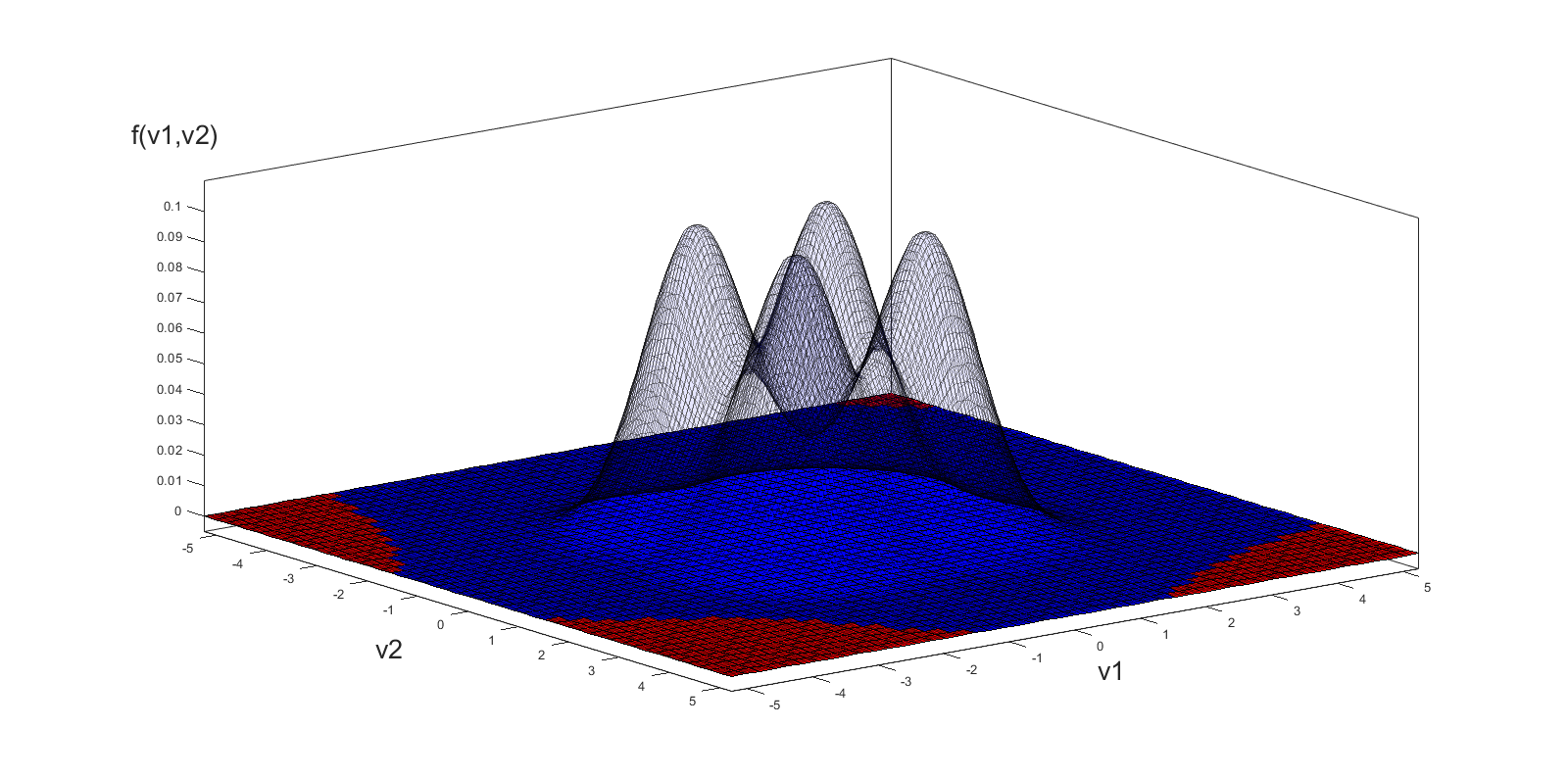} &   \includegraphics[width=75mm]{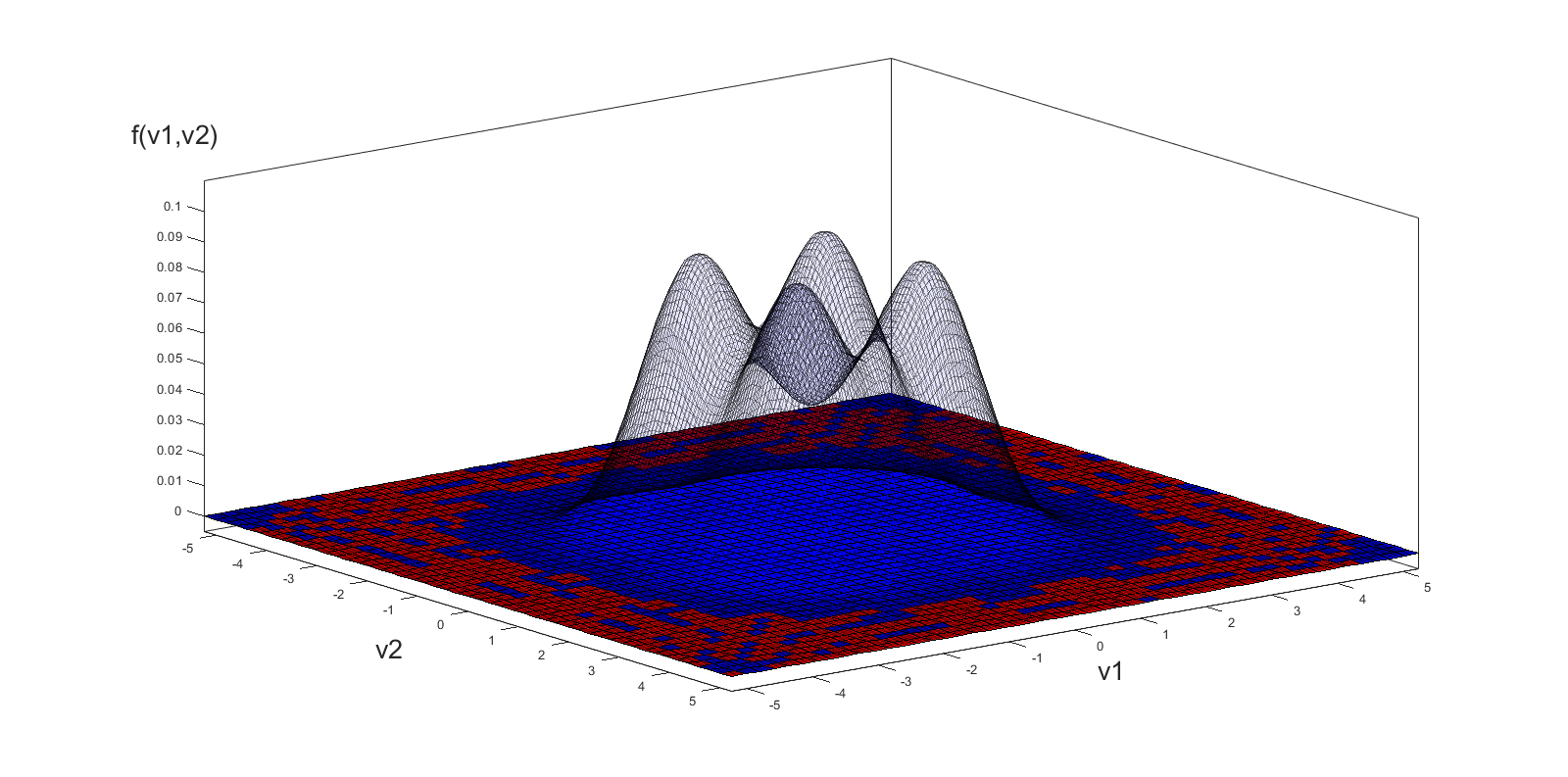} \\
		(a) t = 0 & (b) t = 2.8 \\[6pt]
		\includegraphics[width=75mm]{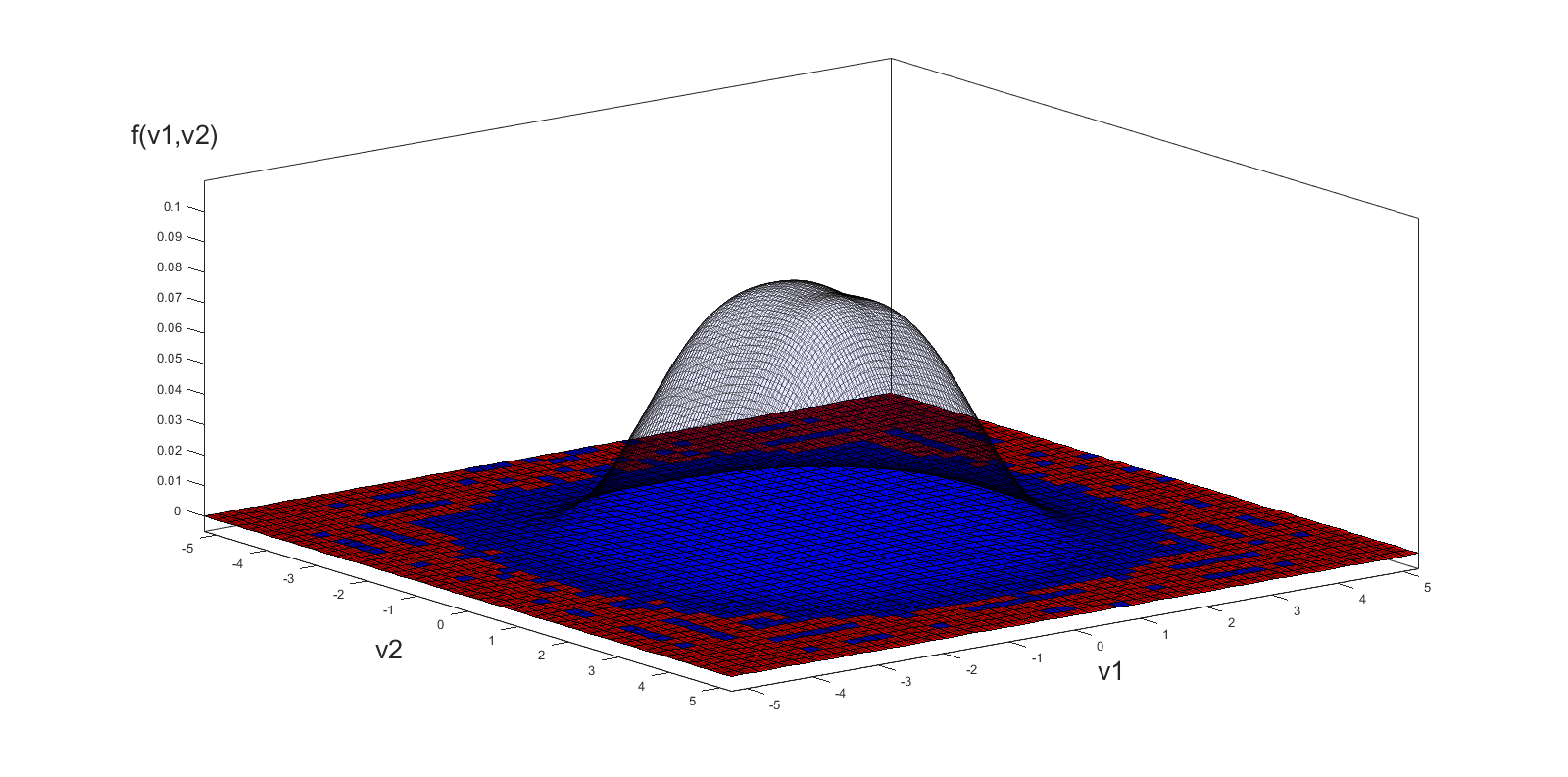} &   \includegraphics[width=75mm]{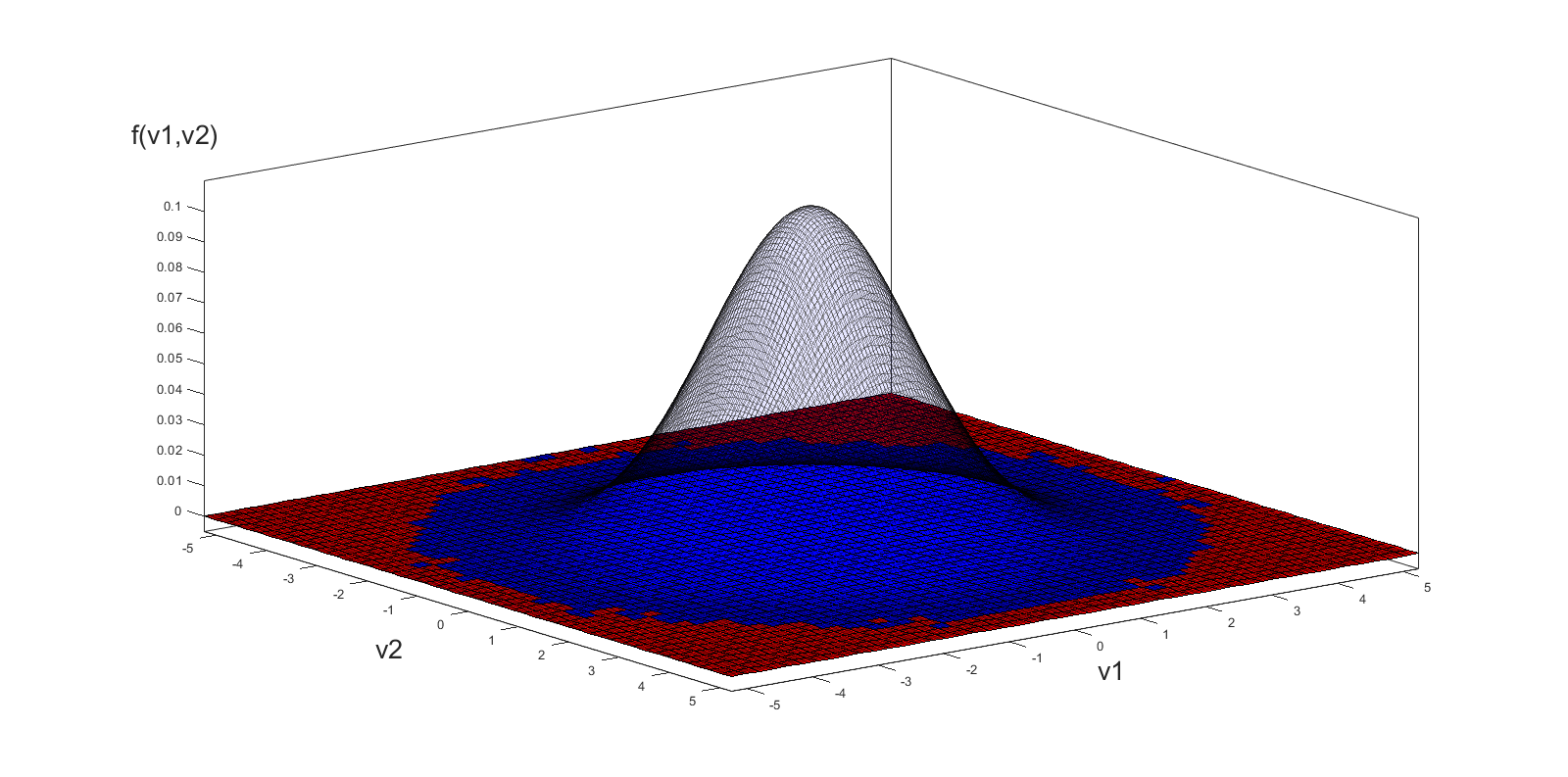} \\
		(c) t = 20 & (d) t = 100 \\[6pt]
	\end{tabular}
	}
	\caption{Marginals of $f$ in the variables $v_1$ and $v_2$ at various times during the simulation of the space-homogeneous Landau equation starting with the initial condition (\ref{IC}), with $T = 0.4$, $\varepsilon = 20$, $L_v = 5.25$, $N = 32$ and $\Delta t = 0.01$, showing cells in the domain where the solution is negative (in red, near the boundary) and positive (in blue, in the interior).}
	\label{Marginals}
\end{figure}

The marginal in ($v_1$, $v_2$)-space of the initial condition (\ref{IC}) is plotted in Fig.\ \ref{Marginals}(a), where it can be seen that this has the form of four humps.  Subsequent marginals of the approximation to the Landau equation starting at this initial condition are plotted at mean-free times $t = 2.8$, $20$ and $100$ in Fig.\ \ref{Marginals}(b)-(d).  This shows that the four humps merge together into one, before eventually taking shape as the space-homogeneous equilibrium Maxwellian (\ref{M_eq_hom}) (see Fig.\ \ref{Marginals}(d)) which, in this case with $T = 0.4$ in (\ref{IC}), has equilibrium temperature $T_{eq} = 1.4$ and is given by
\begin{equation}
\mathcal{M}_{eq}(\boldsymbol{v}) = \frac{1}{(2.8 \pi)^{\frac{3}{2}}} e^{-\frac{|\boldsymbol{v}|^2}{2.8}}. \label{M_eq}
\end{equation} 

\begin{figure}[!hbt]
	\centering
	\includegraphics[width=\linewidth]{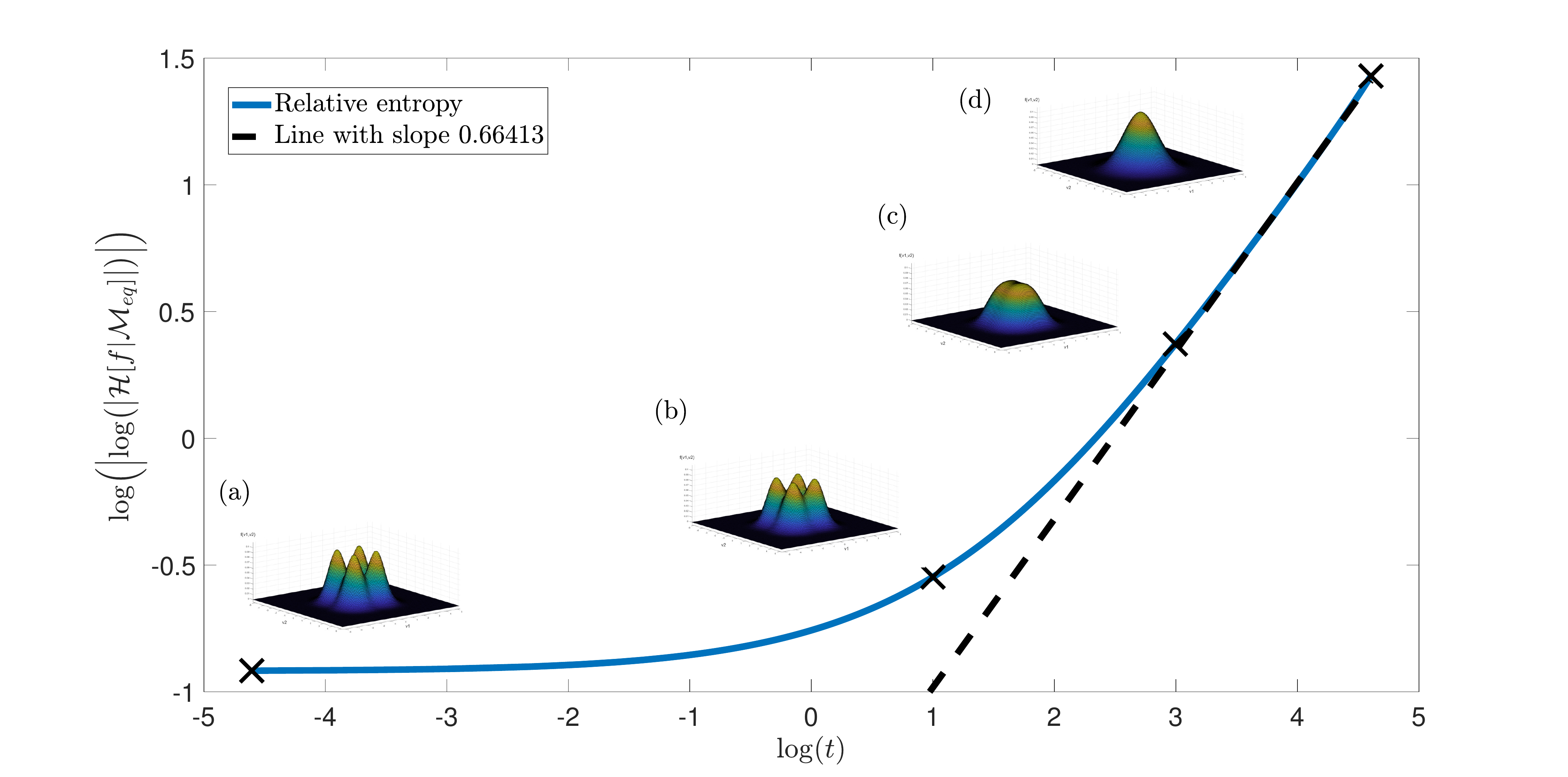} 
	\caption{Plot of $\ln\Bigl(\Bigl|\ln \bigl(\bigl|\mathcal{H}[f|\mathcal{M}_{eq}]\bigr|\bigr)\Bigr|\Bigr)$ against $\ln(t)$ for the numerical approximation $f$ to the space-homogeneous Landau equation, given initial condition (\ref{IC}), with $T = 0.4$, $\varepsilon = 20$, $L_v = 5.25$, $N = 32$ and $\Delta t = 0.01$, which has equilibrium solution $\mathcal{M}_{eq}$ given by (\ref{M_eq}).  A straight line has been added to show that the slope near equilibrium is close to two thirds, exhibiting the lack of spectral gap, but a degenerate spectrum corresponding to a stretch-time exponential decay given by $e^{\displaystyle -kt^p}$, with $p = \frac{2}{3}$ and some $k > 0$.  The labels correspond to the marginal plots in Fig.\ \ref{Marginals}.}
	\label{EntropyPlot_Coulomb}
\end{figure}

In Fig.\ \ref{EntropyPlot_Coulomb}, the relative entropy has been plotted.  When natural logarithms have been taken, the curve does indeed become a straight line when close to equilibrium.  It can be seen that, when $t = 2.8$ (corresponding to Fig.\ \ref{Marginals}(b)), the curve is not yet straight but that is because the solution is still far from a Maxwellian.  At around $t = 20$ (corresponding to Fig.\ \ref{Marginals}(c)), however, the four humps have disappeared and the solution is becoming close to that of a Maxwellian.  This is part of the entropy plot which is a straight line, with a slope of approximately 0.664.  It should be noted that this is much closer to two thirds than the value of 0.634 attained with the parameters in \cite{RGDPaper}.

\subsubsection{Lack of Positivity Preservation}
At this point, it should be mentioned that the numerical scheme does not preserve positivity.  There is potential for negativity to occur when conservation is enforced.  The good news, however, is that the negative parts of the solution only appear as a result of tiny oscillations near the tails.  The negative regions are shown underneath the marginal plots in Fig.\ \ref{Marginals}, on the $(v_1, v_2)$-axes, as red cells which are indeed next to the boundary near the tails.  In these regions, the solution is negligible anyway and so the effects of the negative values are not noticed.  On the other hand, calculating the natural log in expression (\ref{RelEnt}) for the relative entropy requires only positive values.  Since the negative values are so tiny though (and the parts of the solution so close to zero give negligible influence on any bulk quantities anyway), these are just discarded when calculating the entropy.  More precisely, the entropy is calculated through a quadrature method and any point for which $f$ has a negative value is considered a zero contribution to the overall sum.

\subsubsection{Parallel Computing Discussion}
The simulations are carried out with C++ code run on the Texas Advanced Computing Center's Stampede2 supercomputer \cite{TACC}, utilising all sixty eight cores on 24 of the Intel Xeon Phi 7250 1.4GHz Knights Landing processors using hybrid OpenMP \cite{OpenMP} and MPI \cite{MPI}.  Any procedure that requires a loop over the grid-cells in velocity space distributes the cells amongst the OpenMP threads then recombines the individual values calculated at the end of the loop.  In addition, when calculating the Fourier transform of $Q$, the evaluations at the $N^3$ many Fourier modes are evenly distributed across the MPI nodes.  This means that only the values of $\hat{Q}$ are calculated on the modes associated to the current MPI node and so time is saved by evaluating at multiple Fourier modes concurrently across MPI nodes.

In \cite{RGDPaper}, there was a table to show the performance increase when using more OpenMP threads was almost linear.  In this work, where MPI has also been added to the space-homogeneous code, the performance increase with more MPI processes is recorded and it also appears to be close to linear.  Table \ref{MPI_Table} records the times taken for $100$ time-steps of the current simulation with various number of MPI processes, each with 68 OpenMP threads (averaged over three runs).
\begin{table}[!hbt]
	\renewcommand{\arraystretch}{2}
	\resizebox{\textwidth}{!}{
	\begin{tabular}{| r | c  c  c  c  c  c | N} 
		\cline{1-7}
		No. of MPI processes & 1 & 2 & 4 & 8 & 16 & 24 & \\ 
		\cline{1-7}
		Average time for 100 time-steps (s) & 18,643 & 9,391 & 4,757 & 2,439 & 1,276 & 890 & \\
		\cline{1-7}
	\end{tabular}
	}
	\renewcommand{\arraystretch}{1}
	\caption{Average times after three runs of 100 time-steps with various number Intel Xeon Phi 7250 1.4GHz Knights Landing processors, each running one MPI task with 68 OpenMP threads in TACC's Stampede2 supercomputer}
	\label{MPI_Table}
\end{table}

\subsubsection{The Hard Sphere Case ($\lambda = 1$)}
Unlike when $\lambda < 0$, there is a spectral gap when $\lambda = 1$.  This means the rate of convergence to a Maxwellian close to equilibrium is in fact exponential, of the form $e^{-kt}$, for some $k > 0$.  Similar to the previous example, when close to equilibrium, the relative entropy should behave like $\ln\Bigl(\Bigl|\ln \bigl(\bigl|\mathcal{H}[f|\mathcal{M}_{eq}]\bigr|\bigr)\Bigr|\Bigr) \sim \ln(t)$.

Trying to simulate hard spheres introduced a fair amount of difficulty, which shed light on an issue that should be considered for modeling hard potentials with the current spectral method.  In particular, when choosing an initial condition for which the bulk of the mass is supported in too small a region near the center of the domain, the tails of the solution start to ripple after a small number of time-steps, causing an instability which leads to a blow-up.  It is believed that this problem stems from the fact that collisions are more significant for hard potentials than soft ones, with more weight being given to larger relative velocities.  The relative velocity becomes larger when closer to the tails in velocity-space.

At first, it may seem like a more compactly supported initial solution could help.  The problem, however, is that collisions are computed in Fourier space.  The Fourier transform will take a solution with small support in the original space and spread it out in the Fourier domain (consider, for example, that a Gaussian with large peak and small variance has a Gaussian with small peak and large variance as its Fourier transform).  This means that the Fourier transform of such an initial condition actually has tails with rather large magnitude near the boundaries.  When multiplied by the hard sphere weights calculated in Section \ref{Fourier}, this causes a problem computationally.  This issue did not exist for $\lambda = -3$ as the weights near the tails for Coulomb interactions are smaller in magnitude.  As a result, any part of the solution that turns negative is emphasized, which introduces the ripples as the conservation routine attempts to compensate.

This logic was followed for the simulations in \cite{RGDPaper} and a larger variance relative to the computational domain was chosen to fix the problem.  This worked to combat the instabilities but as mass started to spread out of the domain, any bulk quantities calculated were affected.  A better approach to this problem, which has been used in the current work, is to simply reduce the mass of the initial condition.  This has the same result of reducing the magnitude of the tails but allows the variance to be reduced in the process.

For the hard sphere simulation, a very similar initial condition is chosen to (\ref{IC}), namely
\begin{equation}
	f_0(\boldsymbol{v}) = \frac{\rho_0}{4} \sum_{l = 0}^{3} \mathcal{M}_v \left(\boldsymbol{v} + 0.016\left((-1)^{\lfloor \frac{l}{2} \rfloor}, (-1)^{l}, (-1)^{l}\right)\right), ~~~\textrm{for } \boldsymbol{v} \in \Omega_{\boldsymbol{v}}, \label{IC_Hardsphere}
\end{equation}
for the Maxwellian $\mathcal{M}_v (\boldsymbol{v}) = (2 \pi T)^{- \frac{3}{2}} e^{-\frac{|\boldsymbol{v}|^2}{2T}}$, with a smaller temperature of $T = 0.00015$ than for the Coulomb interactions example.  Also, in the initial condition (\ref{IC}), there was no $\rho_0$ factor but here $\rho_0 = 0.01$, which reduces the mass.  Again, the Knudsen number is $\varepsilon = 20$ and $N = 32$ Fourier modes are used, but a much smaller velocity domain is chosen here, with boundary $L_v = 0.1$.  This allows the time-stepsize to be increased slightly, as the stability results from Section \ref{Stability} show that a smaller value of $L_v$ and smaller mass is less restrictive.  In particular, the time-stepsize chosen is $\Delta t = 0.1$ (below the upper bound of approximately 0.1117 calculated for stability with these parameters for $\lambda = 1$ in Section \ref{Stability}).

The increased time-stepsize helps because when the mass is reduced there are fewer collisions and so simulations are slower on this time-scale.  In order for the results to be comparable to those from the Coulomb interaction simulations in Sub-section \ref{Results_Homogeneous_Coulomb}, the time-scale should be adjusted to match that used for solutions with larger mass.  An explanation of how the timescales differ for two simulations with different masses is given in Appendix \ref{Timescales}.  In particular, in those calculations, let $t^a$ = $t^C$ be the time-scale from the Coulomb interaction simulations; $t^b = t^H$ the time-scale from the current hard sphere simulations; and $\tau = \rho_0 = 0.01$ the mass ratio.  Then, the entropy results in this section are plotted on the scales $t = t^C = \rho_0 t^H$ and $\mathcal{H}[f](t) = \mathcal{H}^C[f] = \frac{1}{\rho_0}\mathcal{H}^H[f]$.  As implied here, the superscripts are dropped in any plots.

\begin{figure}[!b]
	\centering
	\includegraphics[width=\linewidth]{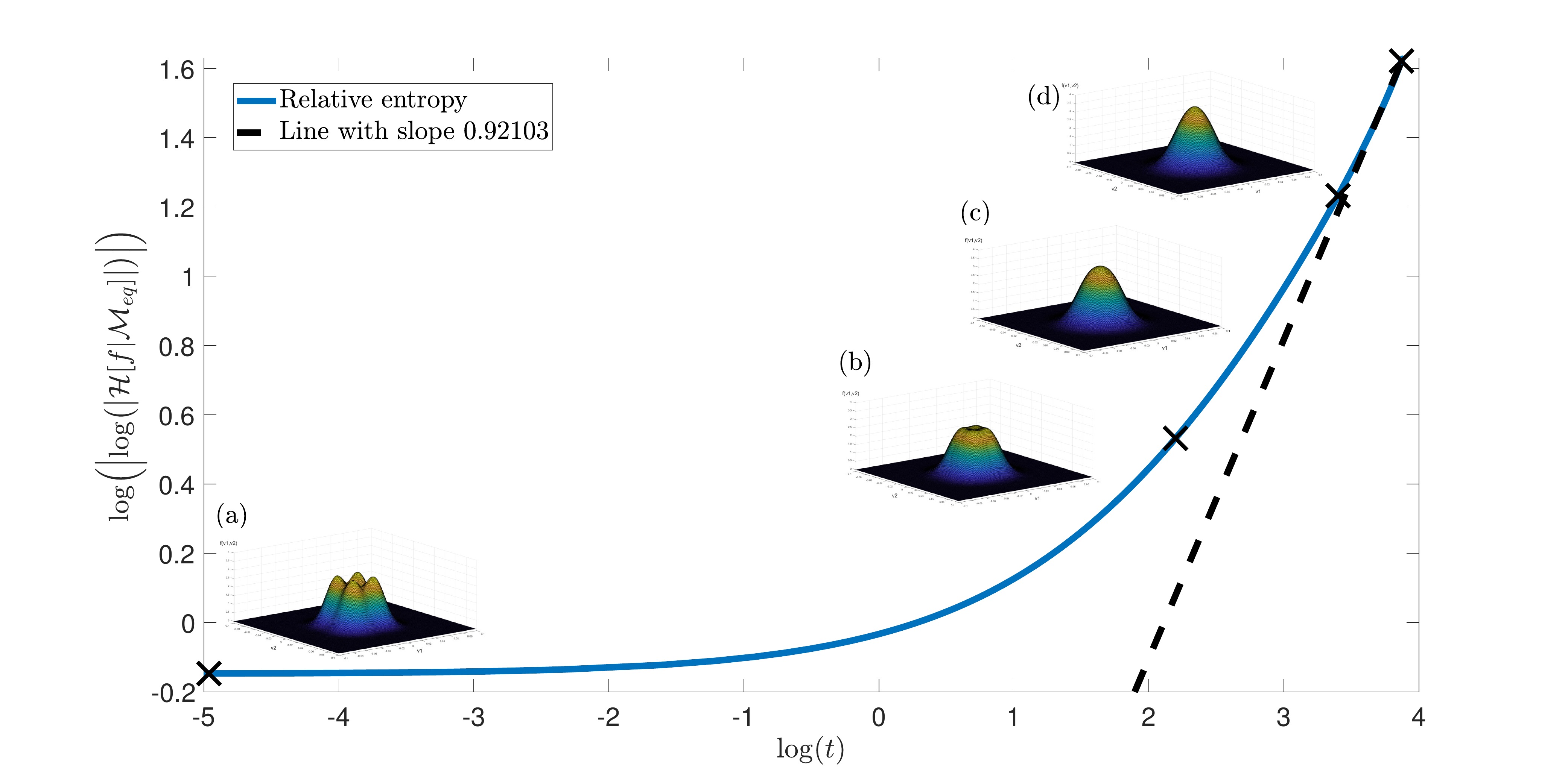} 
	\caption{Plot of $\ln\Bigl(\Bigl|\ln \bigl(\bigl|\mathcal{H}[f|\mathcal{M}_{eq}]\bigr|\bigr)\Bigr|\Bigr)$ against $\ln(t)$ for the numerical approximation $f$ to the space-homogeneous Fokker-Planck-Landau type equation with $\lambda = 1$ and weights calculated by the exact formulae in Section \ref{Fourier}, given initial condition (\ref{IC_Hardsphere}), with $T = 0.00015$, $\varepsilon = 20$, $L_v = 0.1$, $N = 32$ and $\Delta t = 0.1$, which has equilibrium solution given by a Maxwellian with temperature $T_{eq} = 0.000406$.  A straight line has been added to show that the slope near equilibrium is now approximately $0.92103$, slightly below the value of one expected for the existence of a spectral gap.  The ($v_1$, $v_2$)-marginals are included at times (a) $t = 0$, (b) $t = 9$, (c) $t = 30$ and (d) $t = 48$.}
	\label{EntropyPlot_Hardsphere_Exact}
\end{figure}

A plot of the relative entropy for hard sphere interactions on these scaled variables is shown in Fig.\ \ref{EntropyPlot_Hardsphere_Exact}.  When natural logarithms are taken, the curve is close to a straight line with slope $0.92103$ which is less than the slope of one that is expected for a spectral gap.  Nevertheless, this is still larger than the slope of two thirds for Coulomb interactions and the slope of one is merely an upper bound, so this result is still satisfactory.  Once again, by considering the marginals, when $t = 9$ (corresponding to position (b), or $t^H = 900$ in the original scaling), the curve is not yet straight but that is because the solution has too flat a peak and so is still relatively far from a Maxwellian.  At around $t = 30$ (corresponding to position (c), or $t^H = 3000$ in the original scaling), however, the shape of the marginal is closer to that of a Maxwellian and this is much more near to the part of the entropy plot which is a straight line.

\subsubsection{The Maxwell Molecule Case ($\lambda = 0$)}
When $\lambda = 0$, there is still a spectral gap for the Focker-Planck-Landau type equation but this can be seen as a borderline case before $\lambda$ drops below zero and starts to obey Strain and Guo's law of stretch-time exponential decay with exponent given by formula (\ref{S&G_decay}).  This means that a straight line in the relative entropy plot may be a little harder to detect.

\begin{figure}[!hbt]
	\centering
	\includegraphics[width=\linewidth]{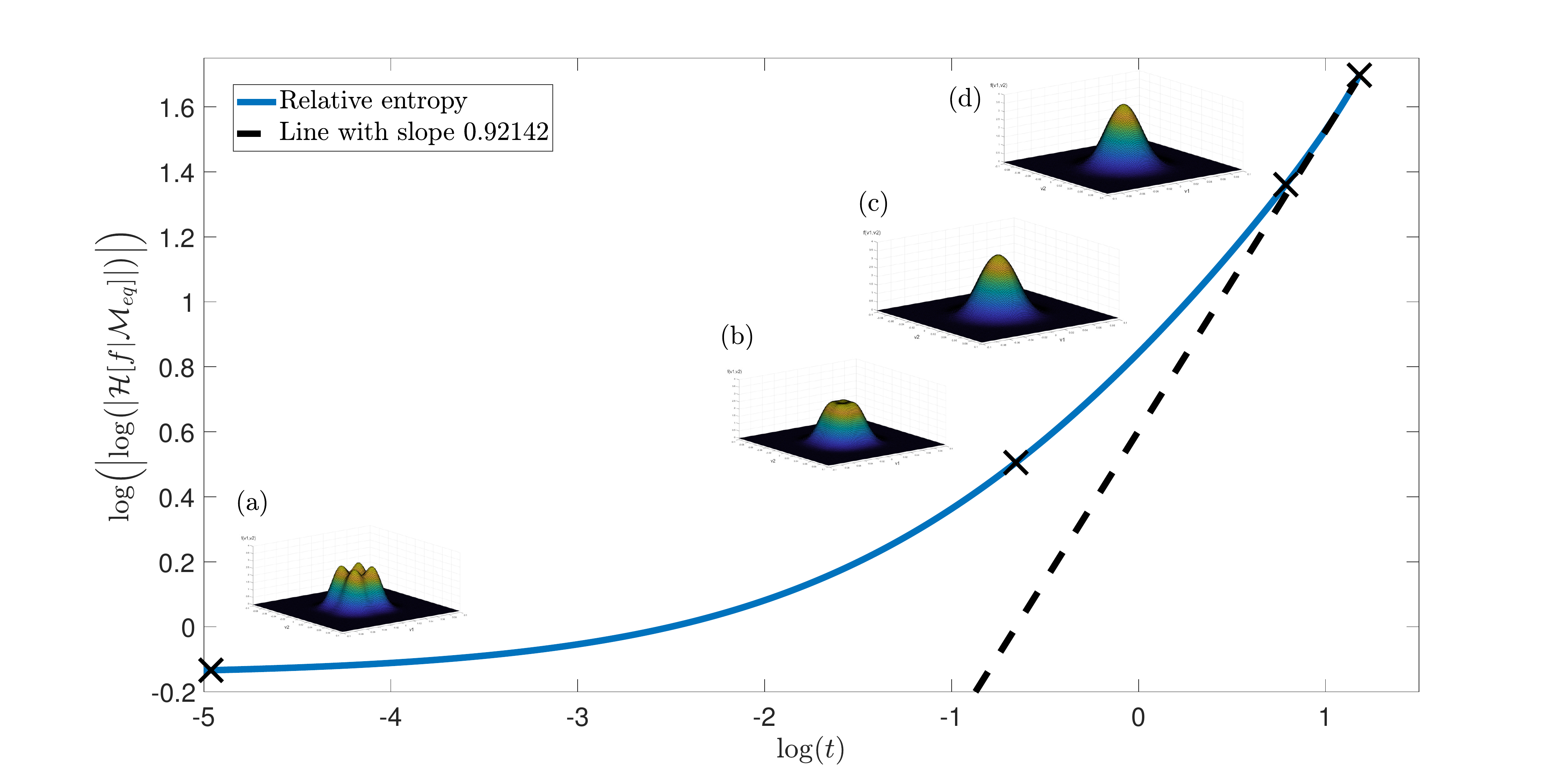} 
	\caption{Plot of $\ln\Bigl(\Bigl|\ln \bigl(\bigl|\mathcal{H}[f|\mathcal{M}_{eq}]\bigr|\bigr)\Bigr|\Bigr)$ against $\ln(t)$ for the numerical approximation $f$ to the space-homogeneous Fokker-Planck-Landau type equation with $\lambda = 0$ and weights calculated by the exact formulae in Section \ref{Fourier}, given initial condition (\ref{IC_Hardsphere}), with $T = 0.00015$, $\varepsilon = 20$, $L_v = 0.1$, $N = 32$ and $\Delta t = 0.1$, which has equilibrium solution given by a Maxwellian with temperature $T_{eq} = 0.000406$.  A straight line has been added to show that the slope near equilibrium is now approximately $0.92142$, slightly below the value of one expected for the existence of a spectral gap.  The ($v_1$, $v_2$)-marginals are included at times (a) $t = 0$, (b) $t = 0.52$, (c) $t = 2.2$ and $t = 4$.}
	\label{EntropyPlot_Maxwell}
\end{figure}

For the Maxwell type simulation, the same initial condition (\ref{IC_Hardsphere}) is used as for hard sphere interactions, with the same parameters $T = 0.00015$, $\rho_0 = 0.01$, $\varepsilon = 20$, $N = 32$, $L_v = 0.1$ and $\Delta t = 0.1$.  Due to the mass being smaller again, the same scaled variables for time and entropy are used, as in the discussion from the hard sphere results.  The relative entropy for this case is plotted in these scaled variables on a ln-ln scale in Fig.\ \ref{EntropyPlot_Maxwell} where there is still a straight line forming near the end of the simulation.  The slope of this line is approximately 0.92142, which is close to the value calculated for the hard sphere simulations.

\begin{figure}[!hbt]
	\centering
	\includegraphics[width=\linewidth]{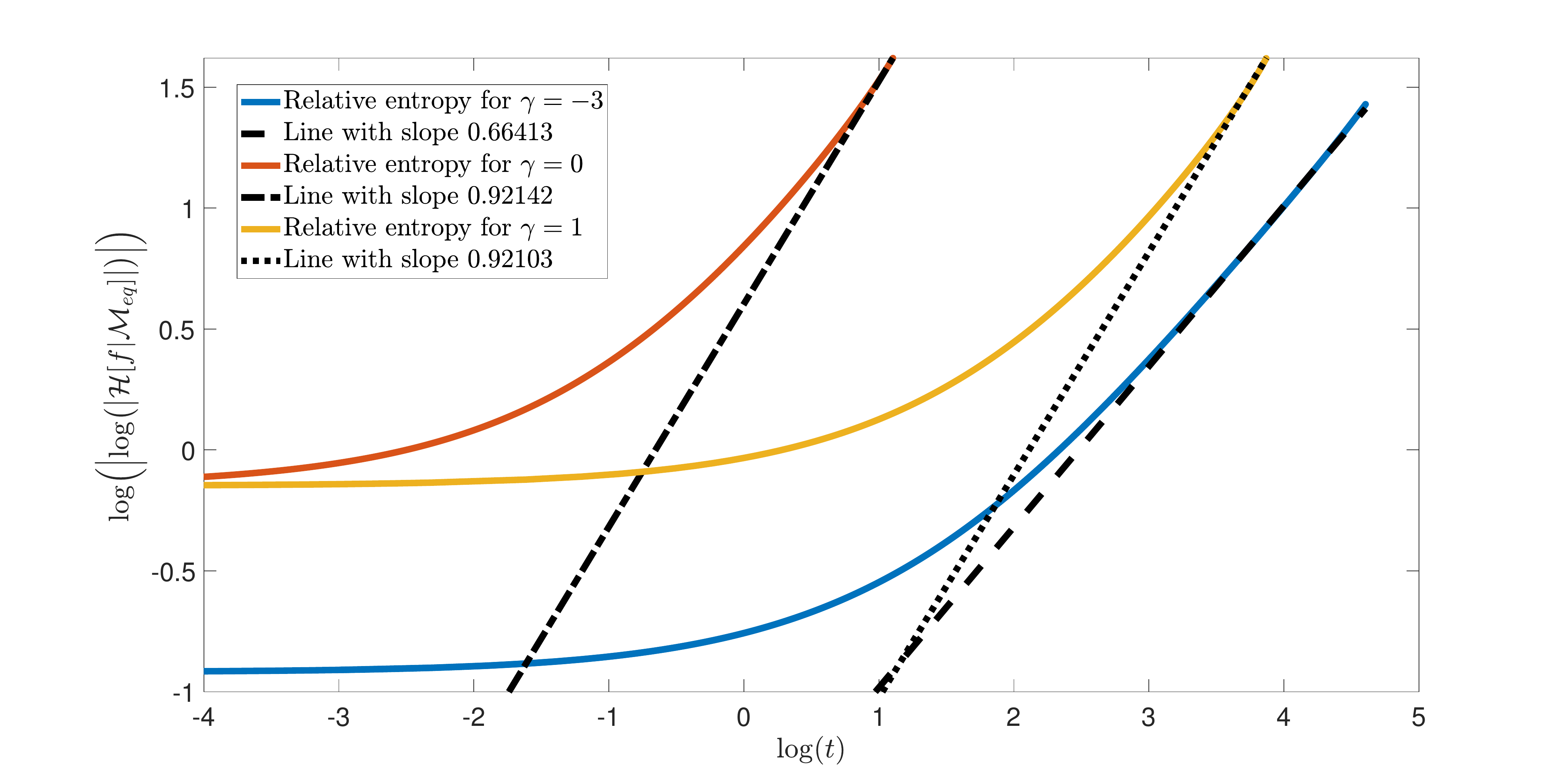} 
	\caption{Plot of $\ln\Bigl(\Bigl|\ln \bigl(\bigl|\mathcal{H}[f|\mathcal{M}_{eq}]\bigr|\bigr)\Bigr|\Bigr)$ against $\ln(t)$ for the numerical approximation $f$ to the space-homogeneous Fokker-Planck-Landau type equations with potentials $\lambda = -3$, 0 and 1, with weights calculated by the exact formulae in Section \ref{Fourier}.  The initial conditions and parameters used are the same as in Fig.\ \ref{EntropyPlot_Coulomb}-\ref{EntropyPlot_Maxwell}.  Straight lines are also added to show the decay rates approached by each simulation.}
	\label{EntropyPlot_Comparison}
\end{figure}

Finally, all three of the plots have been included on the same set of axes in Fig.\ \ref{EntropyPlot_Comparison}.  Here it can be seen that simulations associated to the Coulomb interactions (i.e.\ the Landau equation) give the strongest result.  Not only does the straight line persist for the longest time but the slope is closest to the predicted value.  This is perhaps indicative of the fact that the Focker-Planck-Landau type equation with Coulomb interactions is the most physically realisable case.  On the other hand, the fact that the slopes captured by both the hard sphere and Maxwell type simulations are so similar demonstrates that they are both capturing the same phenomenon, namely, the existence of the spectral gap.

\subsubsection{Results with Fewer Fourier Modes ($N = 24$)}
The results so far in this subsection have all used $N = 32$ Fourier modes.  This choice was used to push the method to greater accuracy and give more convincing results than the previous work by the current authors \cite{RGDPaper}.  If the computational power is not available to allow such a high choice of Fourier modes to be used in a reasonable amount of time, however, the method can still give impressive results without losing too much accuracy.  In particular, the most physically relevant case of Coulomb interactions still gives an accurate representation of the decay rate to equilibrium to two decimal places.  On the other hand, the value calculated for hard sphere and Maxwell type interactions does suffer more dramatically but it still remains larger than the rate of two thirds when there is no spectral gap.

\begin{figure}[!hbt]
	\centering
	\includegraphics[width=\linewidth]{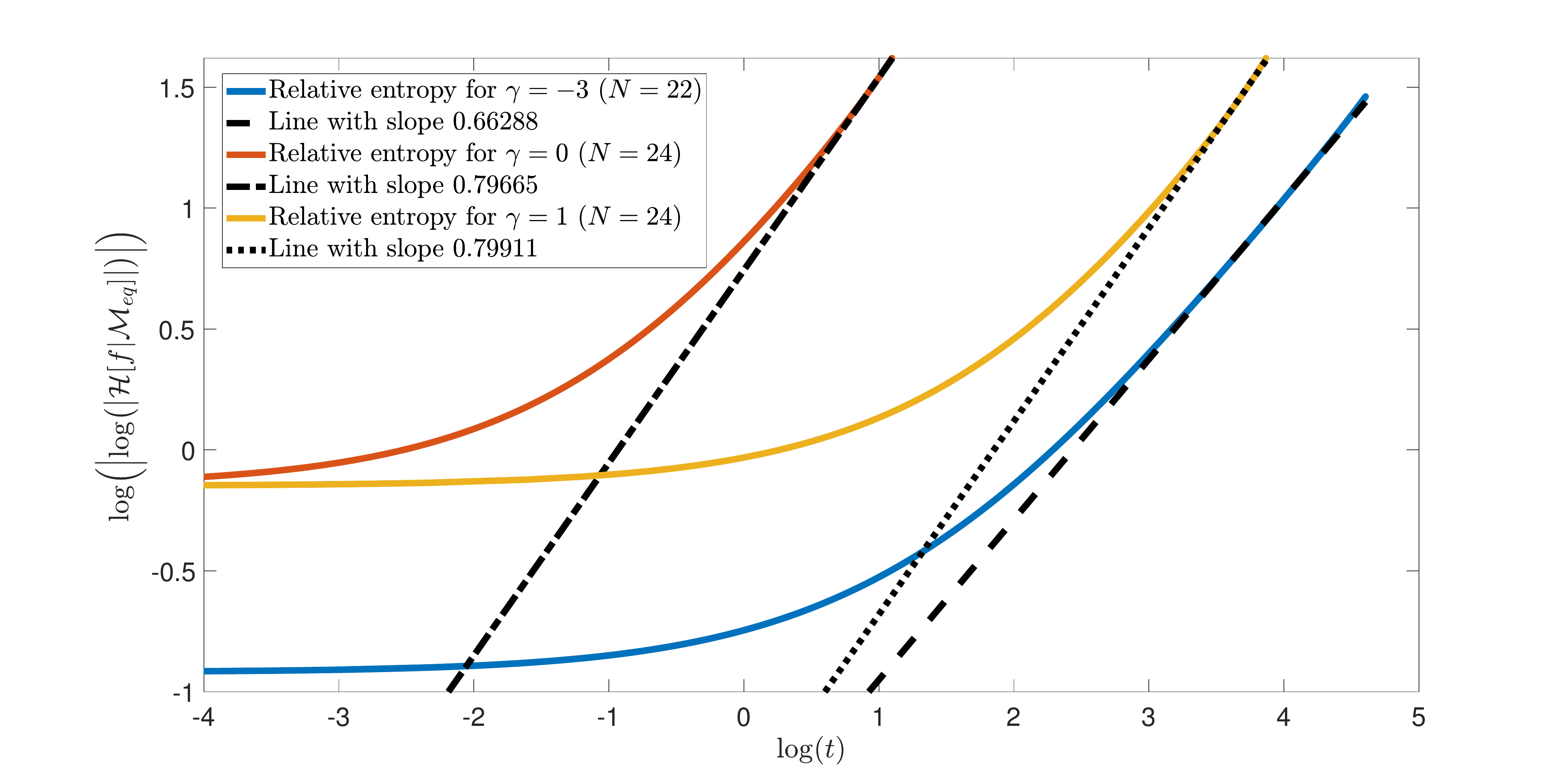} 
	\caption{Plot of $\ln\Bigl(\Bigl|\ln \bigl(\bigl|\mathcal{H}[f|\mathcal{M}_{eq}]\bigr|\bigr)\Bigr|\Bigr)$ against $\ln(t)$ for the numerical approximation $f$ to the space-homogeneous Fokker-Planck-Landau type equations with potentials $\lambda = -3$, 0 and 1, with weights calculated by the exact formulae in Section \ref{Fourier}.  The initial conditions and parameters used are all the same as in Fig.\ \ref{EntropyPlot_Coulomb}-\ref{EntropyPlot_Maxwell}, except for the choice of Fourier modes $N$.  For the Coulomb interactions plot, $N = 22$ but the other two use $N = 24$.  Straight lines are also added to show the decay rates approached by each simulation.}
	\label{EntropyPlot_Comparison_N24}
\end{figure}

As an example, the same simulations are run from Fig.\ \ref{EntropyPlot_Comparison} but with less Fourier modes.  The original hope was to use $N = 24$ Fourier modes (halfway between the choice of $N = 16$ in the previous work \cite{RGDPaper} and $N = 32$ used above) but, for some currently unknown reason, the simulation will not run Coulomb interaction simulations with $N = 24$ Fourier modes as the solution blows up after just one time-step.  There is known to be an issue with certain Fourier modes in the FFTW package, as discussed in the original work on this algorithm when applied to the Boltzmann equation by Gamba and Tharkabhushaman \cite{Harsha}, but this may not be the same problem here because this choice still runs for hard sphere and Maxwell type interactions.  For this reason, the Coulomb interactions are run with $N = 22$ Fourier modes and the others with $N = 24$.  A similar plot to Fig.\ \ref{EntropyPlot_Comparison}, on the same axis ranges, is shown in Fig.\ \ref{EntropyPlot_Comparison_N24} with these choices.

\subsection{Space-inhomogeneous Results for the Coulomb Case ($\lambda = -3$)}
\subsubsection{Results from $N=32$ Fourier Modes}
Trying to recover Strain and Guo's entropy decay rate of two thirds is a little more complicated in the space-inhomogeneous case, which appears to be a result of accumulating numerical error.  First of all, to alleviate these difficulties, a different form of initial condition is used from the four humps used in the space-homogeneous case.  In particular, a small perturbation of a Maxwellian by a cosine wave in space is chosen, which is the same used to demonstrate the phenomenon of Landau damping, namely
\begin{equation}
f_0(x, \boldsymbol{v}) = (1 + A \cos(k x)) \mathcal{M}_v (\boldsymbol{v}), ~~~\textrm{for } (x, \boldsymbol{v}) \in \Omega_x \times \Omega_{\boldsymbol{v}}, \label{IC_LD}
\end{equation}
again for the Maxwellian $\mathcal{M}_v (\boldsymbol{v}) = (2 \pi T)^{- \frac{3}{2}} e^{-\frac{|\boldsymbol{v}|^2}{2T}}$. Here, $T = 1.2$, $k = 0.5$ and $A = 0.05$ are used.  Additionally, the space domain is chosen as $\Omega_x = [0, L_x]$ with length $L_x = 4\pi$, so that there is exactly one period of the cosine wave and $\int_{0}^{L_x} f_0(x, \boldsymbol{v}) ~\textrm{d}x = \mathcal{M}_v (\boldsymbol{v})$.  This means the solution converges to the Maxwellian $\mathcal{M}_v (\boldsymbol{v})$, uniformly in space, as $t \to \infty$.

When the simulations were first run, it became clear that the choice of $N = 16$ Fourier modes and $N_v = 24$ velocity grid cells in each dimension used originally in the space-homogeneous case in \cite{RGDPaper} were not enough to accurately calculate the space-inhomogeneous entropy.  This problem can easily be fixed, however, by increasing the number of D.G.\ grid-cells in velocity spaces to $N_v = 48$ and the number of Fourier modes to $N = 32$, as in the space-homogeneous results in Sub-section \ref{Results_Homogeneous}.  These parameters are used along with the Knudsen number $\varepsilon = 20$; velocity domain width $L_v = 5.25$; $N_x = 24$ D.G.\ cells in space; and time-stepsize $\Delta t = 0.01$.  Note that increasing the number of D.G.\ cells in space has little effect on accuracy because the initial condition (\ref{IC_LD}) leads to simulations with very small variations in space.  This is why it is no problem to use as low a choice as $N_x = 24$ to speed up computations.

\begin{figure}[!hbt]
	\centering
	\includegraphics[width=\linewidth]{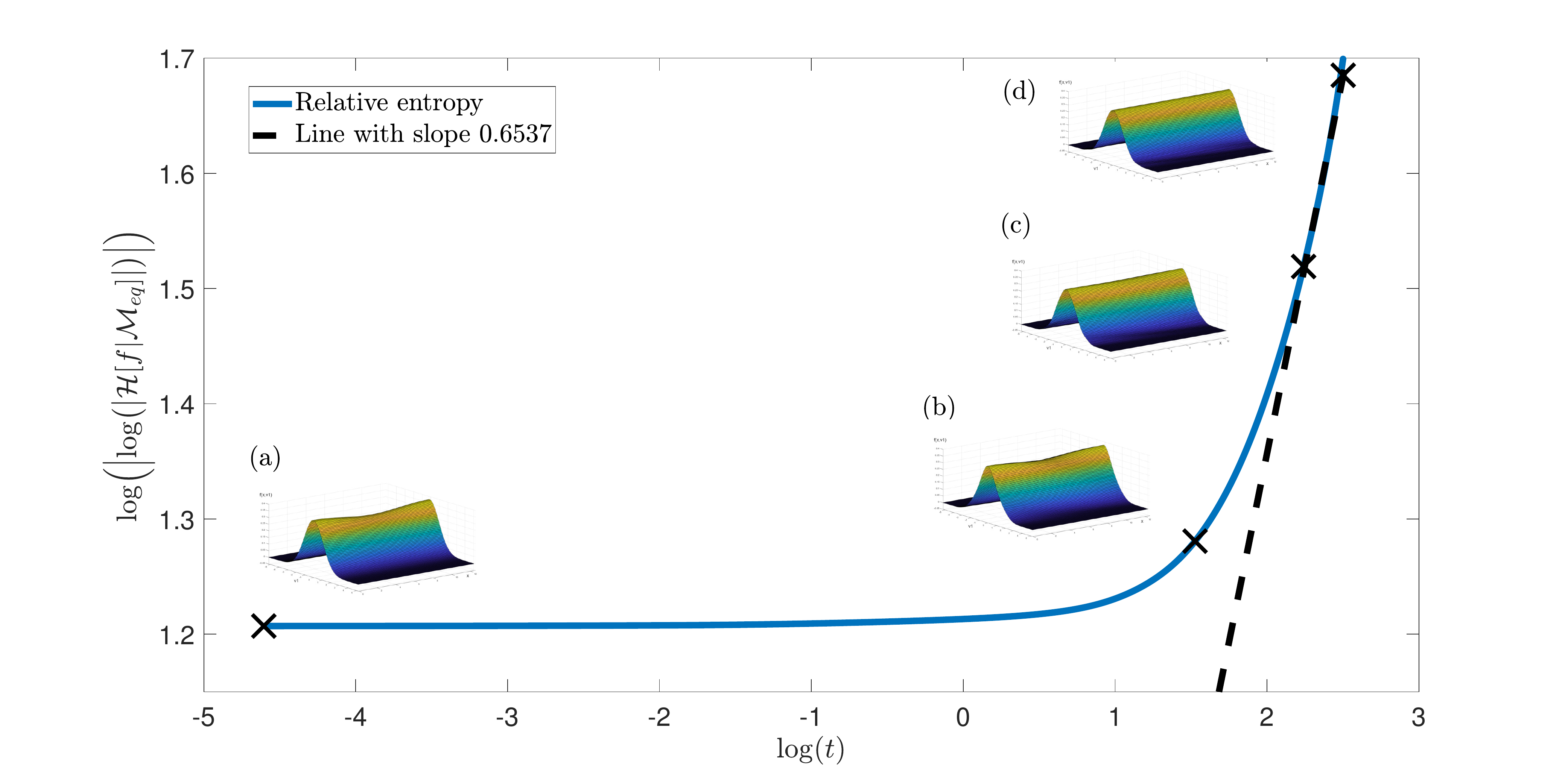} 
	\caption{Plot of $\ln\Bigl(\Bigl|\ln \bigl(\bigl|\mathcal{H}[f|\mathcal{M}_{eq}]\bigr|\bigr)\Bigr|\Bigr)$ against $\ln(t)$ for the numerical approximation $f$ to the space-inhomogeneous Landau equation, given initial condition (\ref{IC_LD}), with $T = 1.2$, $k = 0.5$, $A = 0.05$, $\varepsilon = 20$, $L_v = 5.25$, $N = 32$, $N_v = 48$, $N_x = 24$ and $\Delta t = 0.01$, which has equilibrium solution $\mathcal{M}_{eq}$ given by (\ref{M_eq_inhom_coulomb}).  A straight line has been added to show that the slope near equilibrium is close to two thirds, exhibiting the lack of spectral gap, but a degenerate spectrum corresponding to a stretch-time exponential decay given by $e^{\displaystyle -kt^p}$, with $p = \frac{2}{3}$ and some $k > 0$.  Marginals in $(x, v_1)$-space are also shown at times (a) $t = 0$ (b) $t = 4.6$, (c) $t = 9.4$ and (d) $t = 12$, to demonstrate that solution is only near equilibrium when close to the stretch-time exponential decay.}
	\label{EntropyPlot_InhomCoulomb}
\end{figure}

When the natural log of the relative entropy is then plotted on a ln-ln scale as a result of using these parameters, as in Fig.\ \ref{EntropyPlot_InhomCoulomb}, it once again approaches a straight line.  This time the slope of that line is approximately $0.6537$ which is again close to the slope of two thirds that is expected.  Some marginal plots are also included on this plot to show how the behaviour here is similar to that in the space-homogeneous case.  First, at $t = 4.6$ mean-free times, the ln-ln plot of relative entropy is not quite yet a straight line and it can be seen in the marginal plot at position (b) that the pdf is still taking a similar form to the initial condition in position (a).  As soon as the plot approaches the straight line, however, like at $t = 9.4$ mean-free times, the p.d.f.\ is starting to look more like the equilibrium solution (\ref{M_eq_inhom}), which is shown at positions (c) and (d).  

\subsubsection{Results from $N=16$ Fourier Modes}
To illustrate the issues when only $N = 16$ Fourier modes are used, first note that, for the current perturbation initial condition (\ref{IC_LD}), $\Phi_{eq}(x) = 0$ for all $x$ and $\rho_0 = L_x$ in the equilibrium Maxwellian (\ref{M_eq_inhom}) so that 
\begin{equation}
\mathcal{M}_{eq}(x, \boldsymbol{v}) = \frac{1}{(2\pi T_{eq})^{\frac{3}{2}}} e^{-\frac{|\boldsymbol{v}|^2}{2 T_{eq}}}. \label{M_eq_inhom_coulomb}
\end{equation}

As is shown in appendix \ref{Energy_calculations}, for $T = 1.2$ and $A = 0.05$, $T_{eq} = T + \frac{2}{3}A^2 = 1.2 + \frac{2}{3}(0.05)^2 = 1.201666\ldots$ and the equilibrium entropy evaluates to
\begin{equation}
\mathcal{H}[\mathcal{M}_{eq}] = -6\pi\left(\ln(2\pi (1.201666\ldots)) + 1\right) = -56.955565 ~~(\textrm{to 6 d.p.}). \label{Equilibrium_Entropy}
\end{equation}
When $N = 16$ Fourier modes, $N_v = 24$ velocity and $N_x = 24$ space D.G.\ grid cells are used in each dimension, however, the decreasing values of entropy pass the equilibrium value as early as the 342nd time-step, where it jumps from 
\begin{equation*}
\mathcal{H}[f](3.41) = -56.955547 ~~~~~\textrm{to} ~~~~~\mathcal{H}[f](3.42) = -56.955658.
\end{equation*}

\begin{figure}[!hbtp]
	\resizebox{\textwidth}{!}{
	\begin{tabular}{cc}
		\hspace{-15pt}\includegraphics[width=85mm]{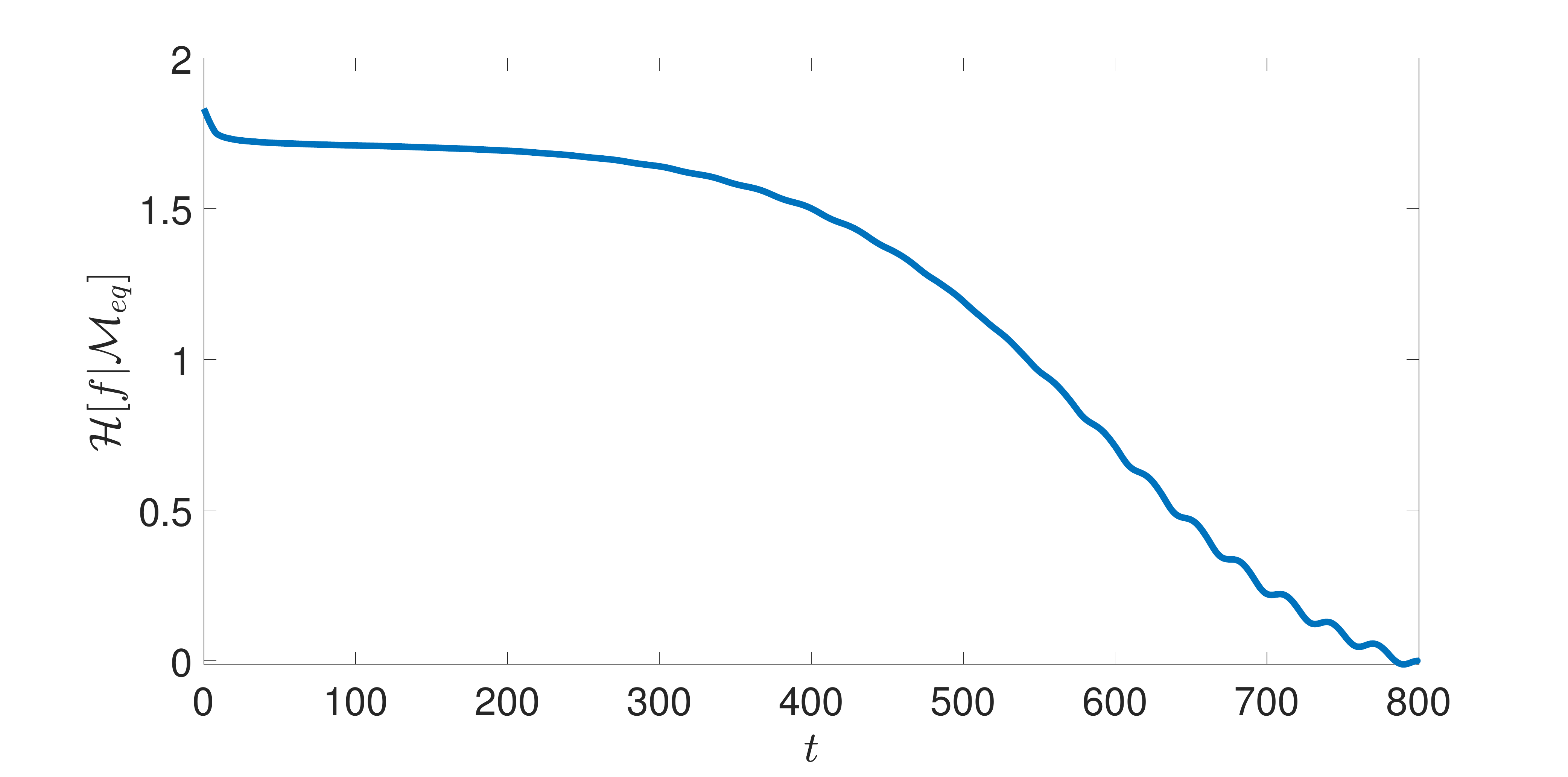} &   \includegraphics[width=85mm]{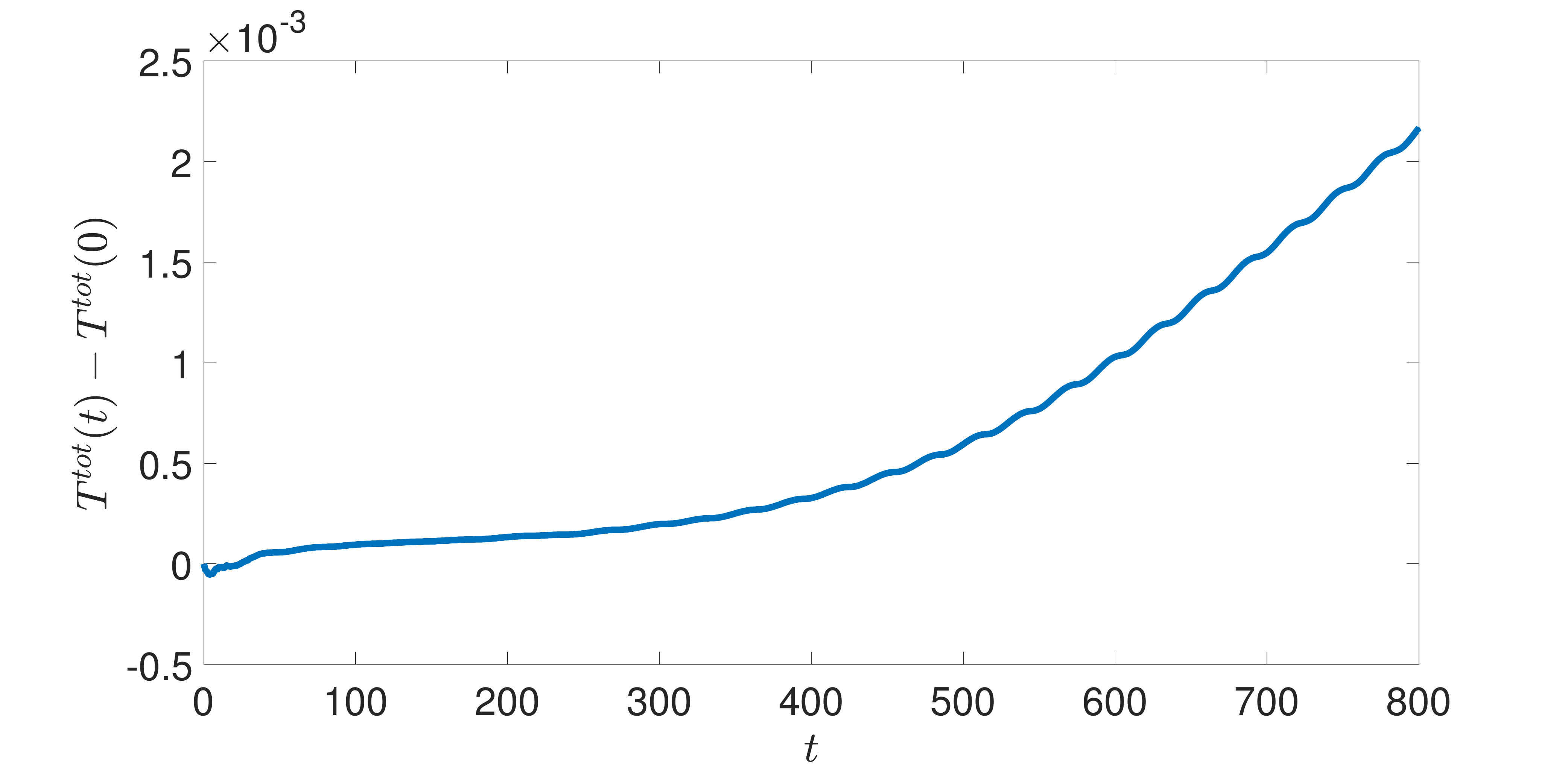} \\
		(a) & (b) \\[6pt]
	\end{tabular}
	}
	\caption{Results from a simulation of the space-inhomogeneous Landau equation, starting with the initial condition (\ref{IC_LD}), with $T = 1.2$, $k = 0.5$, $A = 0.05$, $\varepsilon = 20$, $L_v = 5.25$, $N = 16$, $N_v = 24$, $N_x = 24$ and $\Delta t = 0.01$. (a) Plot of $\mathcal{H}[f|\mathcal{M}_{eq}]$ for the numerical approximation $f$, which has equilibrium solution taken from the final time-step, namely $\mathcal{M}_{eq} = f(800)$. (b) Plot of the error in the total energy from the initial value $T^{tot}(0) = 1.201666...$.}
	\label{Instabilities}
\end{figure}

Initially, as a workaround for this issue, the idea was to run the simulation for long enough that the solution should reach a numerical approximation to equilibrium and then use the value of the entropy calculated from this long-time solution as the equilibrium entropy.  When running the simulations for so long, however, the numerical error begins to accumulate and an instability appears to be introduced.  Figure \ref{Instabilities}(a) shows that the entropy does seem to exhibit a type of exponential decay up until around $t = 200$ but then, instead of converging to some steady state value, decreases further and starts to oscillate.  A similar trend can be seen in the total energy.  This should be held constant throughout, but it is common for a slight deviation to occur in the energy of the space-inhomogeneous simulations of the order of roughly $10^{-4}$.  This can be seen up to about $t = 200$ in Fig.\ \ref{Instabilities}(b) and is expected to result from the time-splitting used.  What should not happen, however, is the faster increased deviation and oscillations that occur around the same time that the entropy is also oscillating.

\begin{figure}[!hbtp]
	\resizebox{\textwidth}{!}{
	\begin{tabular}{ccc}
		\hspace{-15pt}\includegraphics[width=55mm]{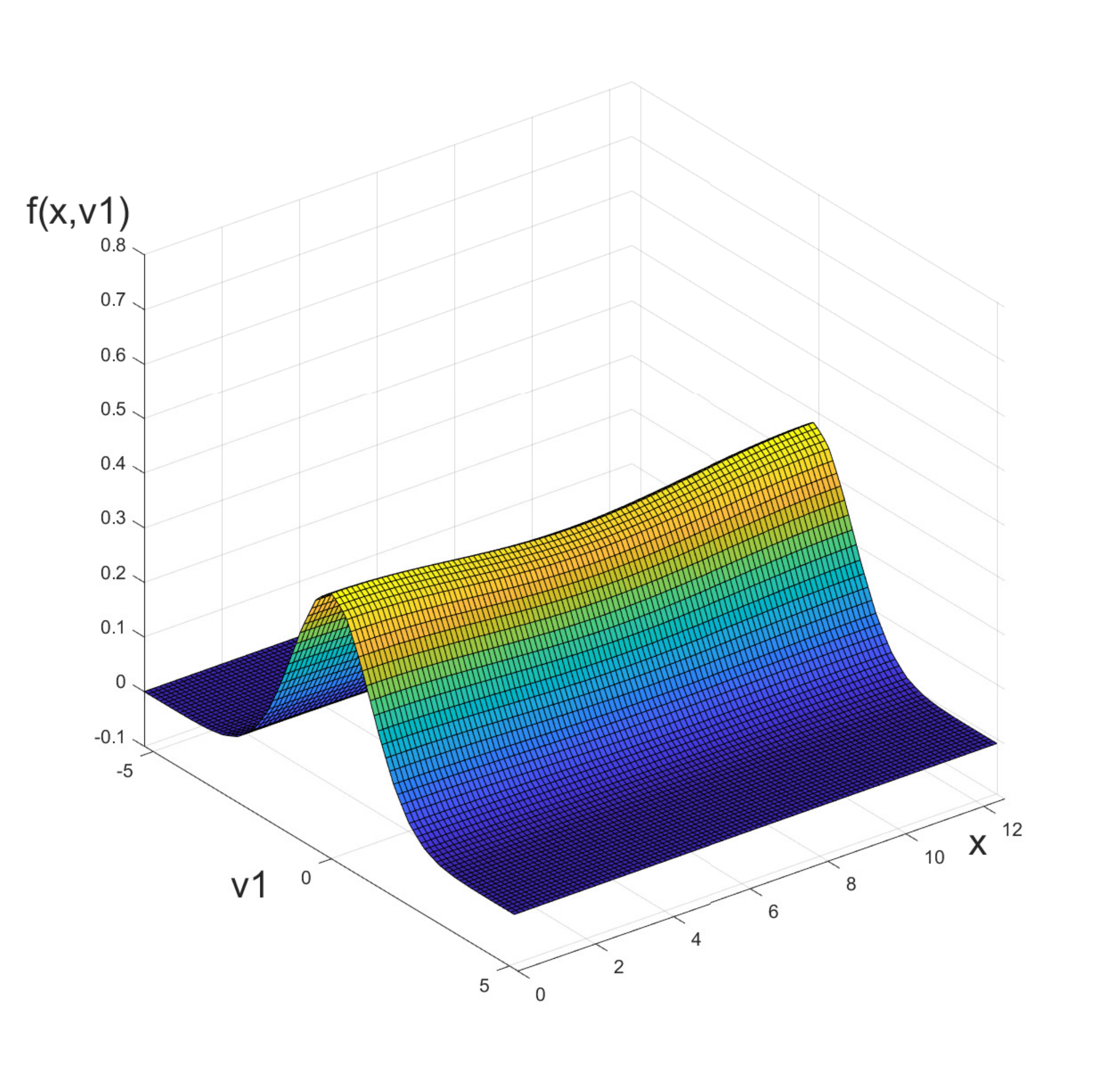} & \includegraphics[width=55mm]{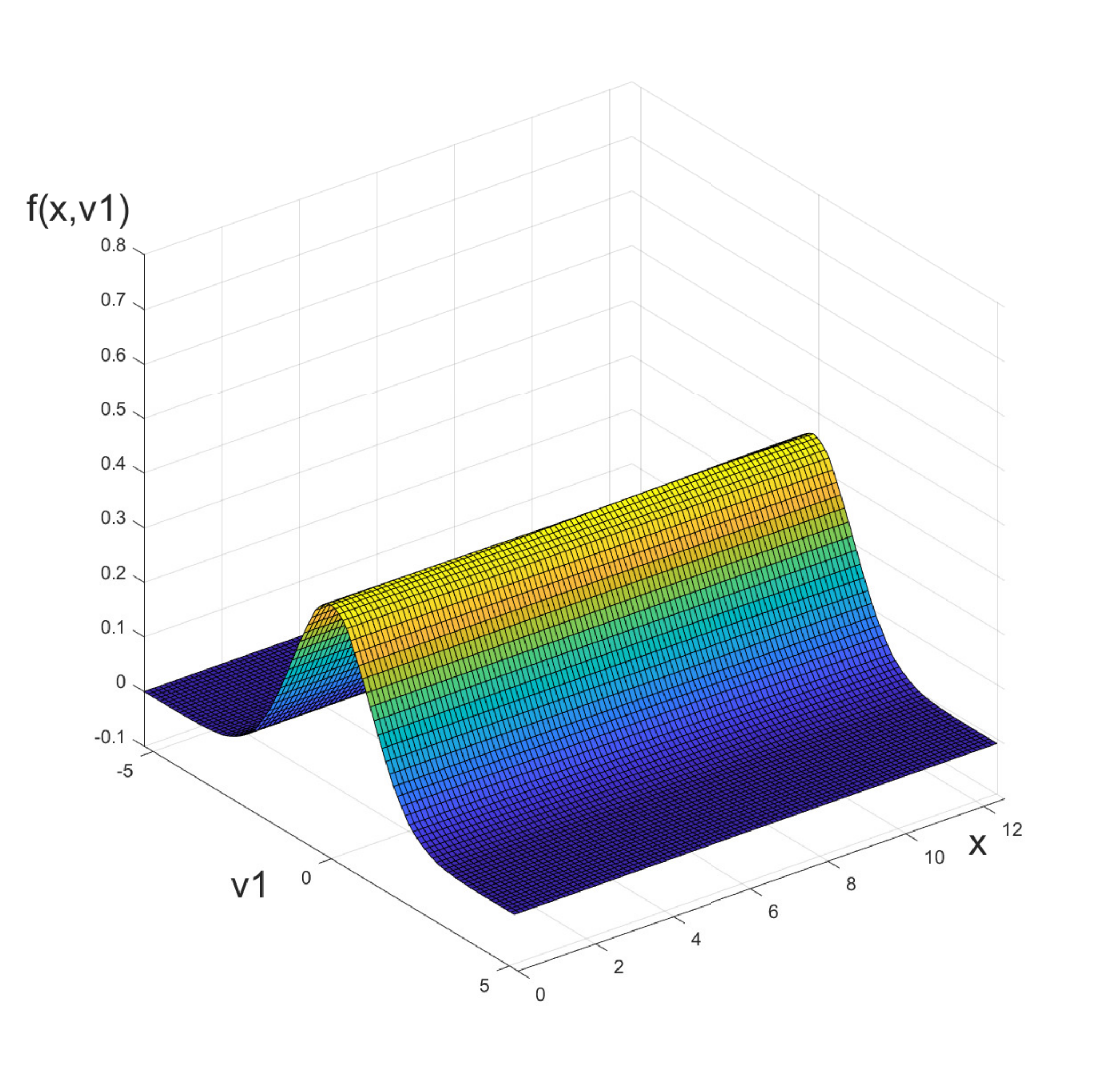} & \includegraphics[width=55mm]{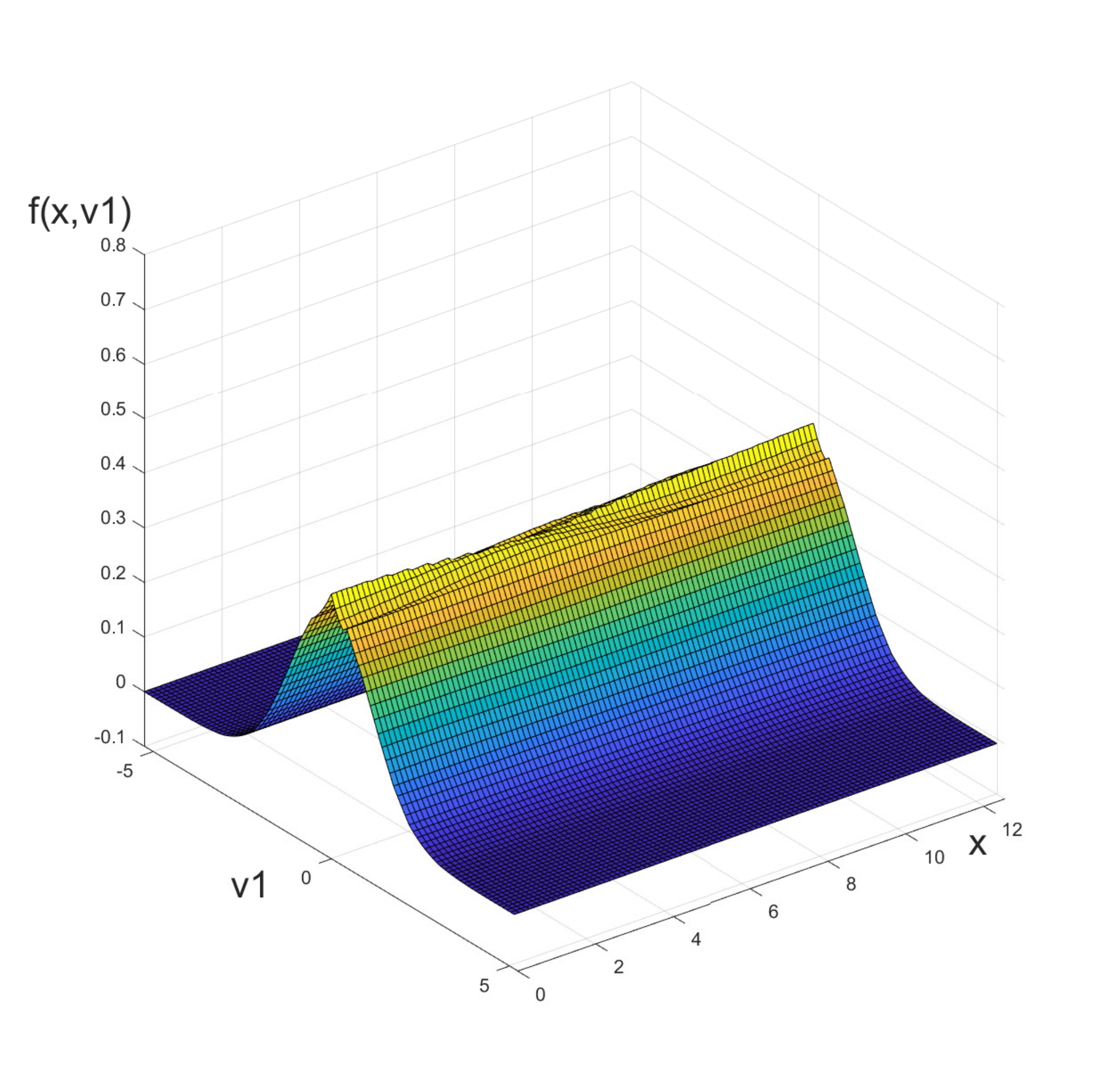} \\
		(a) & (b) & (c) \\[6pt]
	\end{tabular}
	}
	\caption{Marginals in $(x, v_1)$-space during a simulation of the space-inhomogeneous Landau equation, starting with the initial condition (\ref{IC_LD}), with $T = 1.2$, $k = 0.5$, $A = 0.05$, $\varepsilon = 20$, $L_v = 5.25$, $N = 16$, $N_v = 24$, $N_x = 24$ and $\Delta t = 0.01$ at times (a) $t = 0$ (the initial condition) (b) $t = 200$ (seemingly near equilibrium) and (c) $t = 700$ (unstable behaviour).}	\label{Marginal_Instabilities}
\end{figure}

It should also be noted that the instabilities here are different to those that arise in the Boltzmann and Landau equations associated with the issues discussed in Section \ref{Stability}, as the source of those errors are near the tails.  Here, the issue is close to the center of the Maxwellian, around $|\boldsymbol{v}| = 0$.  Figure \ref{Marginal_Instabilities}(a) shows a marginal in $(x, v_1)$-space of the initial condition and then in Fig.\ \ref{Marginal_Instabilities}(b) the marginal is shown at time $t = 200$ (the time up to which the entropy and total energy are behaving themselves in Fig.\ \ref{Instabilities}), where the approximation seems to be near the equilibrium.  Finally, Fig.\ \ref{Marginal_Instabilities}(c) shows an example of how the Maxwellian is contorting around $|\boldsymbol{v}| = 0$, with spikes appearing there at the $x$-boundaries and a kink in the middle of space.  This indicates some sort of instability interfering with the expected behaviour.

When using the values $N = 32$ and $N_v = 48$, however, the simulation does not reach the theoretical equilibrium entropy (\ref{Equilibrium_Entropy}) until after the 1525-th time-step, at which point the approximation appears to have reached the equilibrium solution.  This suggests that the solution to the Landau equation, starting with initial condition (\ref{IC_LD}), reaches equilibrium much faster than initially suspected and it was never necessary to push it to the point where the numerical error accumulates enough to cause an influence.

\section{Conclusion}
In this work, the conservative spectral method for solving Fokker-Planck-Landau type equations was expanded upon by extending the calculations to hard sphere and Maxwell type potentials.  Conditions for stability were then derived for each of the three cases.  Finally, examples of the numerical method for all three of these potentials were given in the space-homogeneous case, in addition to results for Coulomb interactions in the space-inhomogeneous setting, to show the power of the scheme.  In particular, the relative entropy during a simulation was shown to decay close to the correct rate for Coulomb interactions, in accordance with the rate of two thirds predicted by Strain and Guo.  This indicates that the current numerical scheme is an excellent model for the Landau equation.  When the model is applied to the Fokker-Planck-Landau type equation with hard sphere and Maxwell type interactions, the existence of the spectral gap is evident but the decay rate seen is slightly below the expected value of one.  Nevertheless, the decay rate captured was almost the same for both potentials and the effect responsible for this was the spectral gap.

The importance of the conservation routine was also emphasised by showing that the decay rate without it is less accurate.  Indeed, enforcing conservation does not preserve positivity but the regions in which the solution falls below zero are always near the tails and the solution is negligible at those locations anyway.  Clearly this is true as dropping those values in calculation of the entropy did not detract from the result, and so priority is given to conservation over positivity.

In addition to the numerical evidence provided here for the power of the conservative spectral method for Fokker-Planck-Landau equations, the current authors have proven analytically that the approximations from this scheme do indeed converge to the true solutions of the equation associated to hard potentials \cite{ErrorEstimates}.  It is hoped that this can then be extended to soft potentials.   Work is also underway to implement the present method in a multi-species setting, based on the calculations by Gamba et al.\ \cite{Liu} to develop an asymptotic preserving explicit-implicit numerical scheme for species with disparate masses.

\begin{appendices}
\section{Evaluating Integrals for $\hat{S}$} \label{S_calculations}
In general, to calculate an integral of the form $(2\pi)^{-\frac{3}{2}} \int_{B_R(\boldsymbol{0})}{f\left(\boldsymbol{u}\right)}e^{-i\boldsymbol{\omega } \cdot \boldsymbol{u}} ~\mathrm{d}\boldsymbol{u}$, first a substitution is made in order to reduce the scalar product in the exponential to a single multiplication.  To do this, note that the rotation matrix $A$ given by
\begin{equation*}
A=\left[ \begin{array}{c}
\frac{\omega_1 \omega_3}{\sqrt{\omega_1^2 + \omega_2^2}} \\ 
- \frac{\omega_2 |\boldsymbol{\omega}|}{\sqrt{\omega_1^2 + \omega_2^2}} \\ 
\omega_1 \end{array}
~~ \begin{array}{c}
\frac{\omega_2 \omega_3}{\sqrt{\omega_1^2 + \omega_2^2}} \\ 
\frac{\omega_1 |\boldsymbol{\omega}|}{\sqrt{\omega_1^2 + \omega_2^2}} \\ 
\omega_2 \end{array}
~~ \begin{array}{c}
- \sqrt{\omega_1^2 + \omega_2^2} \\ 
0 \\ 
\omega_3 \end{array}
\right]
\end{equation*} 
has the property that $A \boldsymbol{\omega} = (0, 0, |\boldsymbol{\omega}|)$.  Also, since $A$ is a rotation matrix, it is orthogonal and so $A^{-1} = A^T$ and $\det{A} = 1$.  

Then, changing variables via $\boldsymbol{u} = A^T \boldsymbol{v}$ and noting that $\boldsymbol{\omega} \cdot \boldsymbol{u} = \boldsymbol{\omega}^T A^T \boldsymbol{v} = (A \boldsymbol{\omega})^T \boldsymbol{v} = |\boldsymbol{\omega}| v_3$,
\begin{equation*}
(2\pi)^{-\frac{3}{2}} \int_{B_R(\boldsymbol{0})}{f\left(\boldsymbol{u}\right)}e^{-i\boldsymbol{\omega} \cdot \boldsymbol{u}} ~\mathrm{d}\boldsymbol{u} = (2\pi)^{-\frac{3}{2}} \int_{B_R(\boldsymbol{0})}{f\left(A^T \boldsymbol{v}\right)}e^{-i |\boldsymbol{\omega}| v_3} ~\mathrm{d}\boldsymbol{v}.
\end{equation*}
Finally, by changing to spherical coordinates via 
\begin{equation*}
\boldsymbol{v} = r \boldsymbol{\sigma} = r(\sin(\theta) \cos(\phi), \sin(\theta) \sin(\phi), \cos(\theta)), 
\end{equation*}
where $0 \leq r \leq R$, $-\pi \leq \phi \leq \pi$ and $0 \leq \theta \leq \pi$, 
\begin{align}
&(2\pi)^{-\frac{3}{2}} \int_{B_R(\boldsymbol{0})}{f\left(\boldsymbol{u}\right)}e^{-i\boldsymbol{\omega} \cdot \boldsymbol{u}} ~\mathrm{d}\boldsymbol{u} \nonumber \\
=& (2\pi)^{-\frac{3}{2}} \int_{0}^{R} \int_{-\pi}^{\pi} \int_{0}^{\pi} {f\left(r A^T \boldsymbol{\sigma}\right)}e^{-i r |\boldsymbol{\omega}| \cos(\theta)} r^2 \sin(\theta) ~\mathrm{d}\theta \mathrm{d}\phi \mathrm{d}r. \label{f_int}
\end{align}

Now, for $\hat{S}^1_{1, 1}(\boldsymbol{\omega})$, $f(\boldsymbol{u}) = |\boldsymbol{u}|^{\lambda + 2}$ and so $f\left(r A^T \boldsymbol{\sigma}\right) = r^{\lambda + 2}$.  By inserting this expression in the general integral formula (\ref{f_int}) and evaluating integrals with respect to $\phi$, this gives
\begin{equation*}
\hat{S}^1_{1, 1} (\boldsymbol{\omega}) = (2\pi)^{-\frac{1}{2}} \int_{0}^{R} r^{\lambda + 4} \int_{0}^{\pi} e^{-i r |\boldsymbol{\omega}| \cos(\theta)} \sin(\theta) ~\mathrm{d}\theta \mathrm{d}r.
\end{equation*}
It can be checked that 
\begin{equation*}
\int_{0}^{\pi} e^{-i r |\boldsymbol{\omega}| \cos(\theta)} \sin(\theta) ~\mathrm{d}\theta = \frac{2}{r |\boldsymbol{\omega}|} \sin(r |\boldsymbol{\omega}|)
\end{equation*}
and so
\begin{equation*}
\hat{S}^1_{1, 1} (\boldsymbol{\omega}) = \sqrt{\frac{2}{\pi}} \frac{1}{|\boldsymbol{\omega}|} \int_{0}^{R} r^{\lambda + 3} \sin(r |\boldsymbol{\omega}|) ~\mathrm{d}r.
\end{equation*}
Then, by evaluating this integral in each of the Coulomb, Maxwell type and hard sphere cases, if $|\boldsymbol{\omega}| \neq 0$, 
\begin{equation*}
\hat{S}^1_{1, 1} (\boldsymbol{\omega}) =
\left\{
\begin{aligned}
\displaystyle
\sqrt{\frac{2}{\pi}} \frac{1}{|\boldsymbol{\omega}|^2}&\Bigl(1 - \cos(R |\boldsymbol{\omega}|)\Bigr), &&\textrm{when } \lambda = -3, \\
\displaystyle
\sqrt{\frac{2}{\pi}} \frac{1}{|\boldsymbol{\omega}|^5}&\Bigl(-(R|\boldsymbol{\omega}|)^3 \cos(R|\boldsymbol{\omega}|) + \mathrlap{3 (R|\boldsymbol{\omega}|)^2 \sin(R|\boldsymbol{\omega}|)}\\
&~~~ + 6 (R|\boldsymbol{\omega}|) \cos(R|\boldsymbol{\omega}|) - 6 \sin(R|\boldsymbol{\omega}|) \Bigr), &&\textrm{when } \lambda = 0, \\
\displaystyle
\sqrt{\frac{2}{\pi}} \frac{1}{|\boldsymbol{\omega}|^6}&\Bigl(-(R|\boldsymbol{\omega}|)^4 \cos(R|\boldsymbol{\omega}|) + \mathrlap{4 (R|\boldsymbol{\omega}|)^3 \sin(R|\boldsymbol{\omega}|)} \\
&~~~ + \mathrlap{12 (R|\boldsymbol{\omega}|)^2 \cos(R|\boldsymbol{\omega}|) - 24 (R|\boldsymbol{\omega}|) \sin(R|\boldsymbol{\omega}|)} \\
&~~~~~~ - 24 \cos(R|\boldsymbol{\omega}|) + 24\Bigr), &&\textrm{when } \lambda = 1.
\end{aligned}
\right.
\end{equation*}

Next, for $\hat{S}^2_{3, 3}(\boldsymbol{\omega})$, $f(\boldsymbol{u}) = |\boldsymbol{u}|^{\lambda} u_3^2$ and so 
\begin{multline*}
f\left(r A^T \boldsymbol{\sigma}\right) = r^{\lambda + 2}\frac{1}{|\boldsymbol{\omega}|^2} \Bigl((\omega_1^2 + \omega_2^2) \sin^2(\theta) \cos^2(\phi) \\
- 2 \omega_3 \sqrt{\omega_1^2 + \omega_2^2} \sin(\theta) \cos(\theta) \cos(\phi) + \omega_3^2 \cos^2(\theta) \Bigr).
\end{multline*}
By inserting this expression in the general integral formula (\ref{f_int}) and evaluating integrals with respect to $\phi$ (noting that an integral of $\cos(\phi)$ over $-\pi \leq \phi \leq \pi$ returns zero), this gives
\begin{align*}
&\hat{S}^2_{3, 3} (\boldsymbol{\omega}) \\
=& (2\pi)^{-\frac{3}{2}} \int_{0}^{R} r^{\lambda + 4} \frac{1}{|\boldsymbol{\omega}|^2} \Bigl((\omega_1^2 + \omega_2^2) \pi \int_{0}^{\pi} (1 - \cos^2(\theta)) e^{-i r |\boldsymbol{\omega}| \cos(\theta)} \sin(\theta) ~\mathrm{d}\theta \\
&~~~~~~~~~~~~~~~~~~~~~~~~~~~~~~~~~~~~~~+ \omega_3^2 (2\pi) \int_{0}^{\pi} \cos^2(\theta) e^{-i r |\boldsymbol{\omega}| \cos(\theta)} \sin(\theta) ~\mathrm{d}\theta \Bigr) \mathrm{d}r.
\end{align*}
It can be checked that 
\begin{align*}
&\int_{0}^{\pi} \cos^2(\theta) e^{-i r |\boldsymbol{\omega}| \cos(\theta)} \sin(\theta) ~\mathrm{d}\theta \\
=& \frac{2}{(r |\boldsymbol{\omega}|)^3} \Bigl((r |\boldsymbol{\omega}|)^2 \sin(r |\boldsymbol{\omega}|) + 2 (r |\boldsymbol{\omega}|) \cos(r |\boldsymbol{\omega}|) - 2 \sin(r |\boldsymbol{\omega}|) \Bigr)
\end{align*}
and
\begin{equation*}
\int_{0}^{\pi} (1 - \cos^2(\theta)) e^{-i r |\boldsymbol{\omega}| \cos(\theta)} \sin(\theta) ~\mathrm{d}\theta = \frac{4}{(r |\boldsymbol{\omega}|)^3} \Bigl(\sin(r |\boldsymbol{\omega}|) - (r |\boldsymbol{\omega}|) \cos(r |\boldsymbol{\omega}|) \Bigr).
\end{equation*}
So,
\begin{multline*}
\hat{S}^2_{3, 3} (\boldsymbol{\omega}) = (2\pi)^{-\frac{3}{2}} \int_{0}^{R} r^{\lambda + 4} \frac{1}{|\boldsymbol{\omega}|^2} \biggl((\omega_1^2 + \omega_2^2) \frac{4 \pi}{(r |\boldsymbol{\omega}|)^3} \Bigl(\sin(r |\boldsymbol{\omega}|) - (r |\boldsymbol{\omega}|) \cos(r |\boldsymbol{\omega}|) \Bigr) \\
+ \omega_3^2 \frac{4 \pi}{(r |\boldsymbol{\omega}|)^3} \Bigl((r |\boldsymbol{\omega}|)^2 \sin(r |\boldsymbol{\omega}|) + 2 (r |\boldsymbol{\omega}|) \cos(r |\boldsymbol{\omega}|) - 2 \sin(r |\boldsymbol{\omega}|) \Bigr) \biggr) \mathrm{d}r.
\end{multline*}
The easiest way to calculate these integrals is to use a substitution of $u = r |\boldsymbol{\omega}|$, allowing $\hat{S}^2_{3, 3} (\boldsymbol{\omega})$ to be written as
\begin{multline*}
\hat{S}^2_{3, 3} (\boldsymbol{\omega}) = (2\pi)^{-\frac{3}{2}} \frac{4 \pi}{(|\boldsymbol{\omega}|)^{\lambda + 7}} \Biggl((\omega_1^2 + \omega_2^2) \biggl(\int_{0}^{R|\boldsymbol{\omega}|} u^{\lambda + 1} \sin(u) ~\mathrm{d}u \\
~~~~~~~~~~~~~~~~~~~~~~~~~~~~~~~~~~~~~~~~~~~~~~~~~~- \int_{0}^{R|\boldsymbol{\omega}|} u^{\lambda + 2} \cos(u) ~\mathrm{d}u \biggr) \\
~~~~~~~~~~~~~+ \omega_3 \biggl(\int_{0}^{R|\boldsymbol{\omega}|} u^{\lambda + 3} \sin(u) ~\mathrm{d}u - 2 \int_{0}^{R|\boldsymbol{\omega}|} u^{\lambda + 1} \sin(u) ~\mathrm{d}u \\
+ 2 \int_{0}^{R|\boldsymbol{\omega}|} u^{\lambda + 2} \cos(u) ~\mathrm{d}u \biggr) \Biggr).
\end{multline*}
Then, by evaluating these integrals in each of the Coulomb, Maxwell type and hardsphere cases, if $|\boldsymbol{\omega}| \neq 0$, 
\begin{equation*}
\hat{S}^2_{3, 3} (\boldsymbol{\omega}) =
\left\{
\begin{aligned}
\displaystyle
\sqrt{\frac{2}{\pi}} \frac{1}{|\boldsymbol{\omega}|^4}&\biggl(\Bigl(\omega_1^2 + \omega_2^2\Bigr) \frac{R|\boldsymbol{\omega}| - \sin(R|\boldsymbol{\omega}|)}{R|\boldsymbol{\omega}|} \\
&~~~- \mathrlap{\omega_3^2 \frac{R|\boldsymbol{\omega}| + (R|\boldsymbol{\omega}|) \cos(R|\boldsymbol{\omega}|) - 2 \sin(R|\boldsymbol{\omega}|)}{R|\boldsymbol{\omega}|}\biggr),} \\
&~~~~~~~~~~~~~~~~~~~~~~~~~~~~~~~~~~~~~~~~~~~~~~~~~~~~ &&\textrm{when } \lambda = -3, \\
\displaystyle
\sqrt{\frac{2}{\pi}} \frac{1}{|\boldsymbol{\omega}|^7}&\biggl(\Bigl(\omega_1^2 + \omega_2^2\Bigr)\Bigl(- (R|\boldsymbol{\omega}|)^2 \sin(R|\boldsymbol{\omega}|) - \mathrlap{3 (R|\boldsymbol{\omega}|) \cos(R|\boldsymbol{\omega}|)} \\
&~~~~~~~~~~~~~~~~~~ + 3 \sin(R|\boldsymbol{\omega}|)\Bigr) \\
&~~~+ \omega_3^2 \Bigl(- (R|\boldsymbol{\omega}|)^3 \cos(R|\boldsymbol{\omega}|) + \mathrlap{5 (R|\boldsymbol{\omega}|)^2 \sin(R|\boldsymbol{\omega}|)} \\
&~~~~~~~~~~~~~ + 12 (R|\boldsymbol{\omega}|) \cos(R|\boldsymbol{\omega}|) - \mathrlap{12 \sin(R|\boldsymbol{\omega}|)\Bigr)\biggr),} \\
&~~~~~~~~~~~~~~~~~~~~~~~~~~~~~~~~~~~~~~~~~~~~~~~~~~~~ &&\textrm{when } \lambda = 0, \\
\displaystyle
\sqrt{\frac{2}{\pi}} \frac{1}{|\boldsymbol{\omega}|^8}&\biggl(\Bigl(\omega_1^2 + \omega_2^2\Bigr)\Bigl(- (R|\boldsymbol{\omega}|)^3 \sin(R|\boldsymbol{\omega}|) - \mathrlap{4 (R|\boldsymbol{\omega}|)^2 \cos(R|\boldsymbol{\omega}|)} \\
&~~~~~~~~~~~~~~~~~~ + 8 (R|\boldsymbol{\omega}|) \sin(R|\boldsymbol{\omega}|) + \mathrlap{8 \cos(R|\boldsymbol{\omega}|) - 8\Bigr)} \\
&~~~+ \omega_3^2 \Bigl(- (R|\boldsymbol{\omega}|)^4 \cos(R|\boldsymbol{\omega}|) + \mathrlap{6 (R|\boldsymbol{\omega}|)^3 \sin(R|\boldsymbol{\omega}|)} \\
&~~~~~~~~~~~~~ + 20 (R|\boldsymbol{\omega}|)^2 \cos(R|\boldsymbol{\omega}|) - \mathrlap{40 (R|\boldsymbol{\omega}|) \sin(R|\boldsymbol{\omega}|)} \\
&~~~~~~~~~~~~~ - 40 \cos(R|\boldsymbol{\omega}|) + 40\Bigr)\biggr), &&\textrm{when } \lambda = 1.
\end{aligned}
\right.
\end{equation*}

Finally, for $\hat{S}^2_{1, 3}(\boldsymbol{\omega})$, $f(\boldsymbol{u}) = |\boldsymbol{u}|^{\lambda} u_1 u_3$ and so 
\begin{multline*}
f\left(r A^T \boldsymbol{\sigma}\right) = r^{\lambda + 2}\frac{1}{|\boldsymbol{\omega}|^2} \Bigl(- \omega_1 \omega_3 \sin^2(\theta) \cos^2(\phi) + \frac{\omega_1 \omega_3^2}{\sqrt{\omega_1^2 + \omega_2^2}} \sin(\theta) \cos(\theta) \cos(\phi) \\
~~~~~~~~~~~~~~~~~~~~~~~~~+ \omega_2 |\boldsymbol{\omega}| \sin^2(\theta) \sin(\phi) \cos(\phi) - \frac{\omega_2 \omega_3 |\boldsymbol{\omega}|}{\sqrt{\omega_1^2 + \omega_2^2}} \sin(\theta) \cos(\theta) \sin(\phi)\\
- \omega_1 \sqrt{\omega_1^2 + \omega_2^2} \sin(\theta) \cos(\theta) \cos(\phi) + \omega_1 \omega_3 \cos^2(\theta) \Bigr).
\end{multline*}
By inserting this expression in the general integral formula (\ref{f_int}) and evaluating integrals with respect to $\phi$ (noting that an integral of $\cos(\phi)$, $\sin(\phi)$ and $\sin(\phi)\cos(\phi)$ over $-\pi \leq \phi \leq \pi$ returns zero), this gives
\begin{multline*}
\hat{S}^2_{1, 3} (\boldsymbol{\omega}) = (2\pi)^{-\frac{3}{2}} \frac{\omega_1 \omega_3}{|\boldsymbol{\omega}|^2} \int_{0}^{R} r^{\lambda + 4} \Bigl(-\pi \int_{0}^{\pi} (1 - \cos^2(\theta)) e^{-i r |\boldsymbol{\omega}| \cos(\theta)} \sin(\theta) ~\mathrm{d}\theta \\
+ 2\pi \int_{0}^{\pi} \cos^2(\theta) e^{-i r |\boldsymbol{\omega}| \cos(\theta)} \sin(\theta) ~\mathrm{d}\theta \Bigr) ~\mathrm{d}r.
\end{multline*}
Using the results for the integrals with respect to $\theta$ from $\hat{S}^2_{3, 3}$,
\begin{multline*}
\hat{S}^2_{1, 3} (\boldsymbol{\omega}) = (2\pi)^{-\frac{3}{2}} 4\pi \frac{\omega_1 \omega_3}{|\boldsymbol{\omega}|^2} \int_{0}^{R} r^{\lambda + 4} \biggl(\Bigl(\frac{1}{(r |\boldsymbol{\omega}|)} - \frac{3}{(r |\boldsymbol{\omega}|)^3}\Bigr) \sin(r |\boldsymbol{\omega}|) \\
+ \frac{3}{(r |\boldsymbol{\omega}|)^2} \cos(r |\boldsymbol{\omega}|) \biggr) ~\mathrm{d}r.
\end{multline*}
Again, using a substitution of $u = r |\boldsymbol{\omega}|$, $\hat{S}^2_{1, 3} (\boldsymbol{\omega})$ can be written as
\begin{equation*}
\hat{S}^2_{1, 3} (\boldsymbol{\omega}) = (2\pi)^{-\frac{3}{2}} 4\pi \frac{\omega_1 \omega_3}{|\boldsymbol{\omega}|^{\lambda + 7}} \int_{0}^{R|\boldsymbol{\omega}|} \Bigl((u^{\lambda + 3} - 3u^{\lambda + 1}) \sin(u) + 3u^{\lambda + 2} \cos(u) \Bigr) ~\mathrm{d}u.
\end{equation*}
Then, by evaluating these integrals in each of the Coulomb, Maxwell type and hardsphere cases, if $|\boldsymbol{\omega}| \neq 0$, 
\begin{equation*}
\hat{S}^2_{1, 3} (\boldsymbol{\omega}) =
\left\{
\begin{aligned}
\displaystyle
-\sqrt{\frac{2}{\pi}} \frac{\omega_1 \omega_3}{|\boldsymbol{\omega}|^4} &\frac{2R|\boldsymbol{\omega}| + R|\boldsymbol{\omega}| \cos(R|\boldsymbol{\omega}|) - 3\sin(R|\boldsymbol{\omega}|)}{R|\boldsymbol{\omega}|}, &&\textrm{when } \lambda = -3, \\
\displaystyle
\sqrt{\frac{2}{\pi}} \frac{\omega_1 \omega_3}{|\boldsymbol{\omega}|^7}&\Bigl(- (R|\boldsymbol{\omega}|)^3 \cos(R|\boldsymbol{\omega}|) + \mathrlap{6 (R|\boldsymbol{\omega}|)^2 \sin(R|\boldsymbol{\omega}|)} \\
&~ + 15 (R|\boldsymbol{\omega}|) \cos(R|\boldsymbol{\omega}|) - 15 \sin(R|\boldsymbol{\omega}|)\Bigr), && \textrm{when } \lambda = 0, \\
\displaystyle
\sqrt{\frac{2}{\pi}} \frac{\omega_1 \omega_3}{|\boldsymbol{\omega}|^8}&\Bigl(- (R|\boldsymbol{\omega}|)^4 \cos(R|\boldsymbol{\omega}|) + \mathrlap{7 (R|\boldsymbol{\omega}|)^3 \sin(R|\boldsymbol{\omega}|)} \\
&~ + 24 (R|\boldsymbol{\omega}|)^2 \cos(R|\boldsymbol{\omega}|) - \mathrlap{48 (R|\boldsymbol{\omega}|) \sin(R|\boldsymbol{\omega}|)} \\
&~ - 48 \cos(R|\boldsymbol{\omega}|) + 48\Bigr), && \textrm{when } \lambda = 1.
\end{aligned}
\right.
\end{equation*}

\section{Calculating Bounds for $\hat{S}$} \label{S_bounds}
\subsection{The case $\lambda = -3$:}
First, by the triangle inequality and noting that $|\boldsymbol{\xi}_k| \geq h_{\xi}  = \frac{\pi}{L_v}$ when $|\boldsymbol{\xi}_k| \neq 0$,
\begin{equation*}
|\hat{S}^1_{1,1}(\boldsymbol{\xi}_k)| \leq \sqrt{\frac{2}{\pi}} \frac{1}{|\boldsymbol{\xi}_k|^2} (2) \leq \frac{2}{\pi^2} \sqrt{\frac{2}{\pi}} {L_v^2}.
\end{equation*}
Also, since $\hat{S}^1_{1,1}(\boldsymbol{0}) = \sqrt{\frac{1}{2\pi}}L_v^2$,
\begin{flalign*}
&& |\hat{S}^1_{1,1}(\boldsymbol{\xi}_k)| \leq \sqrt{\frac{1}{2\pi}}L_v^2, ~~~~~~~~\textrm{for any } k = 1, 2, \ldots, M.
\end{flalign*}

Then, when $|\boldsymbol{\xi}_k| \neq 0$,
\begin{align*}
|\hat{S}^2_{3,3}(\boldsymbol{\xi}_k)| &\leq \sqrt{\frac{2}{\pi}} \frac{1}{|\boldsymbol{\xi}_k|^4} \Biggl( |\boldsymbol{\xi}_k|^2 \biggl(1 + \frac{1}{L_v |\boldsymbol{\xi}_k|} \biggr) + |\boldsymbol{\xi}_k|^2 \biggl(1 + 1 + \frac{2}{L_v |\boldsymbol{\xi}_k|} \biggr) \Biggr) \\
&= 3\sqrt{\frac{2}{\pi}} \frac{1}{|\boldsymbol{\xi}_k|^2} \biggl(1 + \frac{1}{L_v} \frac{1}{|\boldsymbol{\xi}_k|} \biggr) \\
&\leq 3\sqrt{\frac{2}{\pi}} \frac{L_v^2}{\pi^2} \biggl(1 + \frac{1}{L_v} \frac{L_v}{\pi} \biggr)\\
&= \frac{3}{\pi^3} \Bigl(\pi + 1\Bigr) \sqrt{\frac{2}{\pi}} L_v^2.
\end{align*}
Also, since $\hat{S}^2_{3,3}(\boldsymbol{0}) = \frac{1}{3 \sqrt{2\pi}} L_v^2$,
\begin{flalign*}
&& |\hat{S}^2_{3,3}(\boldsymbol{\xi}_k)| \leq \frac{3}{\pi^3} \Bigl(\pi + 1\Bigr) \sqrt{\frac{2}{\pi}} L_v^2, ~~~~~\textrm{for any } k = 1, 2, \ldots, M.
\end{flalign*}

This then means that the diagonal terms satisfy, for each $i = 1, 2, 3$ and $k = 1, 2, \ldots, M$,
\begin{equation*}
|\hat{S}_{i,i}(\boldsymbol{\xi}_k)| \leq |\hat{S}^1_{i,i}(\boldsymbol{\xi}_k)| + |\hat{S}^2_{i,i}(\boldsymbol{\xi}_k)| = \Biggl(\sqrt{\frac{1}{2\pi}} + \frac{3}{\pi^3} \Bigl(\pi + 1\Bigr) \sqrt{\frac{2}{\pi}} \Biggr) L_v^2.
\end{equation*}

Similarly, when $|\boldsymbol{\xi}_k| \neq 0$,
\begin{align*}
|\hat{S}^2_{1,3}(\boldsymbol{\xi}_k)| &\leq \sqrt{\frac{2}{\pi}} \frac{|\boldsymbol{\xi}_k| |\boldsymbol{\xi}_k|}{|\boldsymbol{\xi}_k|^4} \biggl(2 + 1 + \frac{3}{L_v |\boldsymbol{\xi}_k|} \biggr) \\
&= 3\sqrt{\frac{2}{\pi}} \frac{1}{|\boldsymbol{\xi}_k|^2} \biggl(1 + \frac{1}{L_v} \frac{1}{|\boldsymbol{\xi}_k|} \biggr) \\
&\leq \frac{3}{\pi^3} \Bigl(\pi + 1\Bigr) \sqrt{\frac{2}{\pi}} L_v^2.
\end{align*}
Also, since $\hat{S}^2_{3,3}(\boldsymbol{0}) = 0$,
\begin{flalign*}
&& |\hat{S}^2_{1,3}(\boldsymbol{\xi}_k)| \leq \frac{3}{\pi^3} \Bigl(\pi + 1\Bigr) \sqrt{\frac{2}{\pi}} L_v^2, ~~~~~\textrm{for any } k = 1, 2, \ldots, M.
\end{flalign*}
This a bound for any off-diagonal term and so, since it is smaller than the bound for the diagonal terms, for each $i, j = 1, 2, 3$ and $k = 1, 2, \ldots, M$, when $\lambda = -3$,
\begin{equation*}
|\hat{S}_{i,j}(\boldsymbol{\xi}_k)| \leq |\hat{S}_{i,i}(\boldsymbol{\xi}_k)| \leq \Biggl(\sqrt{\frac{1}{2\pi}} + \frac{3}{\pi^3} \Bigl(\pi + 1\Bigr) \sqrt{\frac{2}{\pi}} \Biggr) L_v^2 \approx 0.719L_v^2 \leq L_v^2.
\end{equation*}

\subsection{The case $\lambda = 0$:}
Here, by factoring in the highest power of $\boldsymbol{\xi}_k$ appearing in brackets and by the triangle inequality, when $|\boldsymbol{\xi}_k| \neq 0$,
\begin{align*}
|\hat{S}^1_{1,1}(\boldsymbol{\xi}_k)| &\leq \sqrt{\frac{2}{\pi}} \frac{1}{|\boldsymbol{\xi}_k|^2} \biggl(L_v^3 + 3 L_v^2 \frac{1}{|\boldsymbol{\xi}_k|} + 6 L_v \frac{1}{|\boldsymbol{\xi}_k|^2} + 6 \frac{1}{|\boldsymbol{\xi}_k|^3}\biggr) \\
&\leq \sqrt{\frac{2}{\pi}} \frac{L_v^2}{\pi^2} \biggl(L_v^3 + 3 L_v^2 \frac{L_v}{\pi} + 6 L_v \frac{L_v^2}{\pi^2} + 6 \frac{L_v^3}{\pi^3}\biggr) \\
&= \sqrt{\frac{2}{\pi}} \frac{1}{\pi^5} \Bigl(\pi^3 + 3 \pi^2 + 6 \pi + 6 \Bigr) L_v^5.
\end{align*}
Similarly, when $|\boldsymbol{\xi}_k| \neq 0$,
\begin{align*}
|\hat{S}^2_{3,3}(\boldsymbol{\xi}_k)| &\leq \sqrt{\frac{2}{\pi}} \frac{1}{\pi^5} \biggl(\Bigl(\pi^2 + 3 \pi + 3 \Bigr) + \Bigl(\pi^3 + 5 \pi^2 + 12 \pi + 12 \Bigr) \biggr) L_v^5 \\
&=\sqrt{\frac{2}{\pi}} \frac{1}{\pi^5} \Bigl(\pi^3 + 6 \pi^2 + 15 \pi + 15 \Bigr) L_v^5
\end{align*}
\begin{flalign*}
\textrm{and } && |\hat{S}^2_{1,3}(\boldsymbol{\xi}_k)| \leq \sqrt{\frac{2}{\pi}} \frac{1}{\pi^5} \Bigl(\pi^3 + 6 \pi^2 + 15 \pi + 15 \Bigr) L_v^5. &&
\end{flalign*}
Also, since $\hat{S}^1_{1,1}(\boldsymbol{0}) = \frac{2}{5}\sqrt{\frac{1}{2\pi}}L_v^2$, $\hat{S}^2_{3,3}(\boldsymbol{0}) = \frac{2}{15 \sqrt{2\pi}} L_v^2$ and $\hat{S}^2_{3,3}(\boldsymbol{0}) = \boldsymbol{0}$, which are all less than the previous bounds, the above bounds are true for all $k = 1, 2, \ldots, M$. 

Again, since the bounds for $|\hat{S}^2_{1,3}(\boldsymbol{\xi}_k)|$ and $|\hat{S}^2_{3,3}(\boldsymbol{\xi}_k)|$ are the same, the off-diagonal terms are clearly bounded by a smaller value than the diagonal terms.  So, for each $i, j = 1, 2, 3$ and $k = 1, 2, \ldots, M$, when $\lambda = 0$,
\begin{align*}
|\hat{S}_{i,j}(\boldsymbol{\xi}_k)| &\leq |\hat{S}^1_{i,i}(\boldsymbol{\xi}_k)| + |\hat{S}^2_{i,i}(\boldsymbol{\xi}_k)| \\
&\leq \sqrt{\frac{2}{\pi}} \frac{1}{\pi^5} \Biggl(\Bigl(\pi^3 + 3 \pi^2 + 6 \pi + 6 \Bigr) + \Bigl(\pi^3 + 6 \pi^2 + 15 \pi + 15 \Bigr) \Biggr) L_v^5 \\
&= \sqrt{\frac{2}{\pi}} \frac{1}{\pi^5} \Bigl(2 \pi^3 + 9 \pi^2 + 21 \pi + 21 \Bigr) L_v^5 \\
& \approx 0.620L_v^5 \\
&\leq L_v^5.
\end{align*}

\subsection{The case $\lambda = 1$:}
Finally, by the same method as for $\lambda = 0$, when $|\boldsymbol{\xi}_k| \neq 0$,
\begingroup\belowdisplayskip=0pt
\begin{align*}
&|\hat{S}^1_{1,1}(\boldsymbol{\xi}_k)| \\
\leq &\sqrt{\frac{2}{\pi}} \frac{1}{|\boldsymbol{\xi}_k|^2} \biggl(L_v^4 + 4 L_v^3 \frac{1}{|\boldsymbol{\xi}_k|} + 12 L_v^2 \frac{1}{|\boldsymbol{\xi}_k|^2} + 24 L_v \frac{1}{|\boldsymbol{\xi}_k|^3} + 24 \frac{1}{|\boldsymbol{\xi}_k|^4} + 24 \frac{1}{|\boldsymbol{\xi}_k|^4}\biggr) \\
=& \sqrt{\frac{2}{\pi}} \frac{1}{\pi^6} \Bigl(\pi^4 + 4 \pi^3 + 12 \pi^2 + 24 \pi + 48 \Bigr) L_v^6,
\end{align*}
\endgroup
\begin{flalign*}
&& |\hat{S}^2_{3,3}(\boldsymbol{\xi}_k)| &\leq \sqrt{\frac{2}{\pi}} \frac{1}{\pi^6} \Bigl(\pi^4 + 7 \pi^3 + 24 \pi^2 + 48 \pi + 96 \Bigr) L_v^6 && \\
\textrm{and } && |\hat{S}^2_{1,3}(\boldsymbol{\xi}_k)| &\leq \sqrt{\frac{2}{\pi}} \frac{1}{\pi^6} \Bigl(\pi^4 + 7 \pi^3 + 24 \pi^2 + 48 \pi + 96 \Bigr) L_v^6. &&
\end{flalign*}
Also, since $\hat{S}^1_{1,1}(\boldsymbol{0}) = \frac{1}{3}\sqrt{\frac{1}{2\pi}}L_v^2$, $\hat{S}^2_{3,3}(\boldsymbol{0}) = \frac{1}{9 \sqrt{2\pi}} L_v^2$ and $\hat{S}^2_{3,3}(\boldsymbol{0}) = \boldsymbol{0}$, which are all less than the previous bounds, the above bounds are true for all $k = 1, 2, \ldots, M$. 

Again, since the bounds for $|\hat{S}^2_{1,3}(\boldsymbol{\xi}_k)|$ and $|\hat{S}^2_{3,3}(\boldsymbol{\xi}_k)|$ are the same, the off-diagonal terms are clearly bounded by a smaller value than the diagonal terms.  So, for each $i, j = 1, 2, 3$ and $k = 1, 2, \ldots, M$, when $\lambda = 1$,
\begin{align*}
|\hat{S}_{i,j}(\boldsymbol{\xi}_k)| &\leq |\hat{S}^1_{i,i}(\boldsymbol{\xi}_k)| + |\hat{S}^2_{i,i}(\boldsymbol{\xi}_k)| \\
&\leq \sqrt{\frac{2}{\pi}} \frac{1}{\pi^6} \Biggl(\Bigl(\pi^3 + 4 \pi^3 + 12 \pi^2 + 24 \pi + 48 \Bigr) \\
&~~~~~~~~~~~~~~~~~~+ \Bigl(\pi^4 + 7 \pi^3 + 24 \pi^2 + 48 \pi + 96 \Bigr) \Biggr) L_v^6 \\
&= \sqrt{\frac{2}{\pi}} \frac{1}{\pi^5} \Bigl(2 \pi^4 + 11 \pi^3 + 36 \pi^2 + 72 \pi + 144 \Bigr) L_v^6 \\
& \approx 1.047 L_v^6 \\
&\lesssim L_v^6.
\end{align*}

\section{Timescales for Simulations with Different Masses} \label{Timescales}
Consider two simulations of the space-homogeneous Fokker-Planck-Landau type equation (\ref{Landau_homo}), where the solution of one has mass a factor of $\tau > 0$ different to the other.  If the two solutions are denoted $f^a$ and $f^b$ then this means that $f^b = \tau f^a$.  Assume also that $f^a$ is modeled on time-scale $t^a$ and $f^b$ on time-scale $t^b$.  Then, the equations which the function $f^a$ and $f^b$ satisfy respectively are
\begin{flalign}
&& \frac{\partial f^a}{\partial t^a} &= \frac{1}{\varepsilon}Q(f^a,f^a) && \label{f_a}\\
\textrm{and }&& \frac{\partial f^b}{\partial t^b} &= \frac{1}{\varepsilon}Q(f^b,f^b). \label{f_b}&&
\end{flalign}
Now, using $f^b = \tau f^a$ in equation (\ref{f_b}) gives
\begin{flalign}
&& \frac{\partial \left(\tau f^a \right)}{\partial t^b} &= \frac{1}{\varepsilon}Q(\tau f^a, \tau f^a),~~~~~~~~~~~~~~~~~~~~~~~~~~~~~~~ && \nonumber \\
\textrm{which is equivalent to } && \tau \frac{\partial f^a}{\partial t^b} &= \frac{\tau^2}{\varepsilon}Q(f^a, f^a), && \label{f_b_scaled}
\end{flalign}
by considering the bilinear property of $Q$.

Then, if the timescales are chosen such that $t^a = \tau t^b$, $\frac{\partial f^a}{\partial t^b} = \tau \frac{\partial f^a}{\partial t^a}$ by the chain rule, and equation (\ref{f_b_scaled}) becomes
\begin{equation*}
\tau^2 \frac{\partial f^a}{\partial t^a} = \frac{\tau^2}{\varepsilon}Q(f^a,f^a),
\end{equation*}
which is equivalent to equation (\ref{f_a}) after dividing through by $\tau^2$.  This suggests that when $f^b$ is modeled on the time-scale $t^b = \frac{1}{\tau} t^a$ then any results will be comparable to that of modeling $f^a$ on time-scale $t^a$.

One final thing to notice here is that when $f^b = \tau f^a$ then the entropy of $f^b$ satisfies $\mathcal{H}^b[f^b](t) = \tau \mathcal{H}^a[f^a](t) + \tau \ln (\tau)$.  In this case, the equilibrium Maxwellians $\mathcal{M}^a_{eq}$ and $\mathcal{M}^b_{eq}$ approached by $f^a$ and $f^b$, respectively, satisfy $\mathcal{M}^b_{eq} = \tau \mathcal{M}^a_{eq}$ as well.  This means that the relative entropy is scaled as
\begin{equation*}
\mathcal{H}^b[f^b|\mathcal{M}^b_{eq}] = \left(\tau \mathcal{H}^a[f^a](t) + \tau \ln (\tau) \right) - \left(\tau \mathcal{H}^a[\mathcal{M}^a](t) + \tau \ln (\tau) \right) = \tau \mathcal{H}^a[f^a|\mathcal{M}^a_{eq}].
\end{equation*}

\section{Space-inhomoegeneous Equilibrium Energy Calculations} \label{Energy_calculations}
As explained in the introduction while discussing the space-inhomogeneous equilibrium Maxwellian (\ref{M_eq_inhom}), the equilibrium total energy $T^{tot}_{eq}$ satisfies $T^{tot}_{eq} = T^{tot}(0)$, for $T^{tot}$ calculated by expression (\ref{T_tot}).  Here, $T^K(0) = T$, for $T$ used in the Maxwellian in the initial condition (\ref{IC_LD}).  Also, it can easily be shown that the exact solution to Poisson's equation associated to the initial condition (\ref{IC_LD}) is $\Phi(x, 0) = 4A(1 - \cos(\frac{1}{2}x)) + C$ (for some constant $C$) which, when used in formula $\ref{T_E}$, gives $T^E(0) = A^2 L_x$.

By using expression (\ref{M_eq_inhom_coulomb}) for the equilibrium solution, as well as $\Phi_{eq}(x) = 0$, to calculate $T^{tot}_{eq} = \lim\limits_{t \to \infty} T^{tot}(t)$,
\begin{equation*}
T^{tot}_{eq} = \int_{0}^{L_x} \int_{\Omega_{\boldsymbol{v}}} \frac{1}{(2\pi T_{eq})^{\frac{3}{2}}} e^{-\frac{|\boldsymbol{v}|^2}{2 T_{eq}}} \left(\frac{1}{2} |\boldsymbol{v}|^2 + \frac{1}{2} |0|^2\right) ~\textrm{d}\boldsymbol{v} \textrm{d}x = \frac{3}{2}\rho_0 T_{eq}.
\end{equation*}
So, $T^{tot}_{eq} = T^{tot}(0)$ is equivalent to
\begin{equation*}
\frac{3}{2} \rho_0 T_{eq} = \frac{3}{2} \rho_0 T + A^2 L_x,
\end{equation*}
which gives
\begin{equation*}
T_{eq} = T + \frac{2}{3}A^2,
\end{equation*}
since $\rho_0 = L_x$ here.  
\end{appendices}

\section{ACKNOWLEDGMENTS}
The authors would like to thank Chenglong Zhang for help in understanding the code used to implement the conservative spectral method and being available to give advice on any developments.  Karl Schulz has also been extremely helpful in understanding modern high performance computing techniques used to improve the code structure.   Support from Oden Institute of Computational Engineering and Sciences at the University of Texas Austin is gratefully acknowledged by the first author. Both authors have been  partially funded by grants DMS-RNMS-1107291 (Ki-Net),  NSF DMS1715515 and  DOE DE-SC0016283 project \emph{Simulation Center for Runaway Electron Avoidance and Mitigation}.  This manuscript is based on a section of Clark Pennie's Ph.D.  Thesis Dissertation \cite{clark_pennie_phd_thesis}, written at The University of Texas at Austin, under the advising of the second author.
\bibliographystyle{acm}
\bibliography{Refs}

\begin{thebibliography}{10}

\bibitem{OpenMP}
Open{MP} {A}rchitecture {R}eview {B}oard, {O}pen{MP} {A}pplication {P}rogram
  {I}nterface {V}ersion 3.0.
\newblock http://www.openmp.org/mp-documents/spec30.pdf, May 2008.

\bibitem{TACC}
The {U}niversity of {T}exas at {A}ustin, {T}exas {A}dvanced {C}omputing
  {C}enter.
\newblock http://www.tacc.utexas.edu, TACC.

\bibitem{MHGM-jcp2014}
{\sc A.~Munafo, J.R.~Haack, I.~G., and Magin, T.}
\newblock A spectral-lagrangian boltzmann solver for a multi-energy level gas,.

\bibitem{Carrillo_etal_2020}
{\sc A.J.~Carrillo, J.~Hu, L.~W., and Hu, J.}
\newblock A particle method for the homogeneous {L}andau equation.
\newblock {\em J. Comput. Physics\/} (2020).

\bibitem{BoltzmannConvergence}
{\sc Alonso, R., Gamba, I., and Tharkabhushaman, S.}
\newblock Convergence and error estimates for the {L}agrangian based
  conservative spectral method for {B}oltzmann equations.
\newblock {\em SIAM Num. Anal. 56\/} (2018), 3534--3579.

\bibitem{Bobylev_Karpov_Potapenko-2012}
{\sc A.V.~Bobylev, S.~K., and Potapenko, I.}
\newblock Dsmc.

\bibitem{BobylevPotapenko}
{\sc Bobylev, A., and Potapenko, I.}
\newblock Monte {C}arlo methods and their analysis for {C}oulomb collisions in
  multicomponent plasmas.
\newblock {\em J. Comput. Phys. 246\/} (2013), 123--144.

\bibitem{Crouseilles_Spectral}
{\sc Crouseilles, N., and Filbet, F.}
\newblock Numerical approximation of collisional plasmas by high order methods.
\newblock {\em J. Comput. Physics 201\/} (2004), 546--572.

\bibitem{Degond&LD}
{\sc Degond, P., and Lucquin-Desreux, B.}
\newblock The {F}okker-{P}lanck asymptotics of the {B}oltzmann collision
  operator in the {C}oulomb case.
\newblock {\em Math. Models Meth. Appl. Sci. 2\/} (1992), 167--182.

\bibitem{Desvillettes92}
{\sc Desvillettes, L.}
\newblock On asymptotics of the {B}oltzmann equation when the collisions become
  grazing.
\newblock {\em Trans. Th. Stat. Phys. 21\/} (1992), 259--276.

\bibitem{Desvillettes15}
{\sc Desvillettes, L.}
\newblock Entropy dissipation estimates for the {L}andau equation in the
  {C}oulomb case and applications.
\newblock {\em J. Funct. Anal. 269}, 5 (2015), 1359--1403.

\bibitem{DesvillettesVillani}
{\sc Desvillettes, L., and Villani, C.}
\newblock On the spatially homogeneous {L}andau equation for hard potentials.
\newblock {\em Communications in Partial Differential Equations 25}, 1-2
  (2000), 179--298.

\bibitem{Filbet_InhomSpectral}
{\sc Filbet, F., and Pareschi, L.}
\newblock A numerical method for the accurate solution of the
  {F}okker-{P}lanck-{L}andau equation in the nonhomogeneous case.
\newblock {\em J. Comput. Physics 179\/} (2002), 1--26.

\bibitem{fftw3}
{\sc Frigo, M., and Johnson, S.}
\newblock The design and implementation of {FFTW}3.
\newblock Proceedings of the {IEEE}, 2005.

\bibitem{MPI}
{\sc Gabriel, E., Fagg, G., Bosilca, G., Angskun, T., Dongarr, J., Squyres, J.,
  Sahay, V., Kambadur, P., Barrett, B., Lumsdaine, A., Castain, R., Daniel, D.,
  Graham, R., and Woodall, T.}
\newblock Open {MPI}: {G}oals, {C}oncept, and {D}esign of a {N}ext {G}eneration
  {MPI} {I}mplementation.
\newblock Proceedings, 11th European PVM/MPI Users' Group Meeting, 2004.

\bibitem{Liu}
{\sc Gamba, I., Jin, S., and Liu, L.}
\newblock Asymptotic-preserving schemes for two-species binary collisional
  kinetic system with disparate masses {I}: time discretization and asymptotic
  analysis.
\newblock {\em to appear in C.M.S.\/} (2019).

\bibitem{BoltzmannSupport}
{\sc Gamba, I., Panferov, V., and Villani, C.}
\newblock Upper {M}axwellian bounds for the spatially homogeneous {B}oltzmann
  equation.
\newblock {\em Arch. Rational Mech. Anal. 194\/} (2009), 253--282.

\bibitem{Harsha}
{\sc Gamba, I., and Tharkabhushaman, S.}
\newblock Spectral-{L}agrangian based methods applied to computation of
  non-equilibrium statistical states.
\newblock {\em J. Comput. Physics 228\/} (2009), 2012--2036.

\bibitem{GambaHaack1}
{\sc Haack, J., and Gamba, I.}
\newblock Conservative deterministic spectral {B}oltzmann solver near the
  grazing collisions limit, 28th {R}arefied {G}as {D}ynamics conference.
\newblock AIP Conference Proceedings, 2012.

\bibitem{GambaHaack2}
{\sc Haack, J., and Gamba, I.}
\newblock A conservative spectral method for the {B}oltzmann equation with
  anisotropic scattering and the grazing collisions limit.
\newblock {\em J. Comput. Physics 270\/} (2014), 40--57.

\bibitem{LandauEq}
{\sc Landau, L.}
\newblock Kinetic equation for the case of {C}oulomb interaction.
\newblock {\em Phys. Zs. Sov. Union 10\/} (1936), 154--164.

\bibitem{Lebedev}
{\sc Lebedev, V.}
\newblock {\em How to solve stiff systems of differential equations by explicit
  methods; Numerical methods and applications \emph{(1994)}}, 1~ed.
\newblock CRC Revivals.

\bibitem{Pareschi_Spectral}
{\sc Pareschi, L., Russo, G., and Toscani, G.}
\newblock Fast spectral methods for the {F}okker-{P}lanck-{L}andau collision
  operator.
\newblock {\em J. Comput. Physics 165\/} (2000), 216--236.

\bibitem{clark_pennie_phd_thesis}
{\sc Pennie, C.}
\newblock Conservative spectral methods for {F}okker-{P}lanck-{L}andau type
  equations: Simulations, long-time behaviour and error estimates, the
  {U}niversity of {T}exas at {A}ustin, 2020.

\bibitem{RGDPaper}
{\sc Pennie, C., and Gamba, I.}
\newblock Decay of entropy from a conservative spectral method for
  {F}okker-{P}lanck-{L}andau type equations.
\newblock AIP Conference Proceedings, 2019.

\bibitem{Pennie_gamba_LandauConv}
{\sc Pennie, C., and Gamba, I.}
\newblock Convergence and error estimates for the conservative spectral method
  for fokker-planck-landau equations.

\bibitem{ErrorEstimates}
{\sc Pennie, C., and Gamba, I.}
\newblock Convergence and error estimates for the conservative spectral method
  for {F}okker-{P}lanck-{L}andau equations.
\newblock {\em arXiv:2009.10352\/} (2020).

\bibitem{Strain&Guo}
{\sc Strain, R., and Guo, Y.}
\newblock Exponential decay for soft potentials near {M}axwellian.
\newblock {\em Arch. Rational Mech. Anal. 187\/} (2008), 287--339.

\bibitem{VillaniLandau}
{\sc Villani, C.}
\newblock On a new class of weak solutions to the spatially homogeneous
  {B}oltzmann and {L}andau equations.
\newblock {\em Archive for Rational Mechanics and Analysis 143\/} (1998),
  273--307.

\bibitem{Chenglong}
{\sc Zhang, C., and Gamba, I.}
\newblock A conservative scheme for {V}lasov {P}oisson {L}andau modeling
  collisional plasmas.
\newblock {\em J. Comput. Physics 340\/} (2017), 470--497.

\end{thebibliography}
\end{document}